\documentclass[11pt,a4paper]{article}
\pdfoutput=1

\usepackage{jheppub_kim}
\usepackage{rotating}
\usepackage{graphicx,epsfig}
\usepackage{amsmath}
\usepackage {amssymb}
\usepackage{subfigure}

\usepackage{relsize}

\def\beq{\begin{equation}}
\def\eeq{\end{equation}}
\def\bea{\begin{eqnarray}}
\def\eea{\end{eqnarray}}

\begin{document}

\title{Observational Constraints on Oscillating Dark-Energy Parametrizations}

\author[a]{Supriya Pan}

\author[b,c,d]{Emmanuel N. Saridakis}
 
\author[e]{Weiqiang Yang}

\affiliation[a]{Department of Mathematics, Raiganj Surendranath Mahavidyalaya, 
Sudarshanpur, 
Raiganj, Uttar Dinajpur, West Bengla 733134, India}

\affiliation[b]{Chongqing University of Posts \& Telecommunications, Chongqing, 400065, 
China}

\affiliation[c]{Physics Division,
National Technical University of Athens, 15780 Zografou Campus,
Athens, Greece}

\affiliation[d]{CASPER, Physics Department, Baylor University, Waco, TX 76798-7310, USA}

\affiliation[e]{Department of Physics, Liaoning Normal University, Dalian, 116029, P. R. 
China}

\emailAdd{span@research.jdvu.ac.in}
\emailAdd{Emmanuel$_-$Saridakis@baylor.edu}
\emailAdd{d11102004@163.com}

\abstract{We perform a detailed confrontation of various oscillating dark-energy 
parame-trizations with the latest sets of observational data. In particular, we  use 
data from  Joint Light Curve analysis (JLA) sample from Supernoave Type Ia, Baryon 
Acoustic Oscillations (BAO) distance measurements, Cosmic Microwave Background (CMB) 
observations, redshift space distortion, weak gravitational lensing, Hubble parameter 
measurements from 
cosmic chronometers, and we impose constraints on 
four oscillating models. From the analyses we find that the best-fit characters of 
almost all models are bent towards the phantom region, 
nevertheless in all of them  the quintessential regime is also allowed within 
1$\sigma$ confidence-level. Furthermore, the deviations from $\Lambda$CDM cosmology  are 
not significant, however for two of the models they could be visible at large scales, 
through the 
impact on the temperature anisotropy of the CMB spectra and on the matter  power spectra. 
Finally, 
we peform the Bayesian analysis, which shows that the 
current observational data support the $\Lambda$CDM paradigm over this set of oscillating 
dark-energy parametrizations. }


\maketitle

\section{Introduction}

The universe acceleration at late times is one of the most interesting findings of modern 
cosmology, and thus there are two main directions that one could follow to explain it. 
The first way is to keep general relativity as the gravitational theory and introduce new 
components, that go beyond the Standard Model of Particle Physics, collectively known as
the dark energy sector \cite{Copeland:2006wr,Cai:2009zp}. The second way is to 
construct a modified gravitational theory, whose additional degrees of freedom  can drive 
the universe acceleration \cite{Nojiri:2006ri,Capozziello:2011et,Cai:2015emx}.

At the phenomenological level both the above approaches lead to a specific universe 
accelerated expansion, that can be quantified by the evolution of the (effective in the 
case of modified gravity) dark energy equation-of-state parameter. Hence, 
parametrizations of the dark energy fluid can lead to reconstructions of the universe 
late-time expansion. The basic idea relies on the fact that the dark energy 
equation-of-state 
parameter $w_x = p_x/ \rho_x$, with $\rho_x$  and  $p_x$ 
the dark energy energy density and pressure respectively, can be parametrized using 
different functional forms in terms of the cosmological 
redshift. 

In principle, there is not a theoretical guiding rule to select the best $w_x(z)$,   
however using   observational data it is possible to find viable  parametrizations. In 
the literature one can find many parametric dark energy models, that  have been 
introduced and fitted with observational data: (i) one-parameter family of dark energy 
models \cite{Gong:2005de} (ii) two-parameters 
family of dark energy parametrizations, namely, 
Chevallier-Polarski-Linder parametrization 
\cite{Chevallier:2000qy,Linder:2002et}, Linear parametrization 
\cite{Cooray:1999da,Astier:2000as,Weller:2001gf},
Logarithmic parametrization \cite{Efstathiou:1999tm}, Jassal-Bagla-Padmanabhan 
parametrization \cite{Jassal:2005qc}, 
Barboza-Alcaniz parametrization \cite{Barboza:2008rh}, 
etc  (see \cite{Ma:2011nc, Feng:2012gf,Feng:2011zzo, Pantazis:2016nky, 
DiValentino:2016hlg, Escamilla-Rivera:2016qwv,Zhao:2017cud, Yang:2017amu, 
DiValentino:2017zyq, DiValentino:2017gzb,Yang:2017alx}), (iii)
three-parameters family of dark energy parametrizations \cite{Linder:2005ne}, and   
(iv) four-parameters family of dark energy parametrizations \cite{Linder:2005ne, 
DeFelice:2012vd, Marcondes:2017vjw}. 

One of the interesting parametrizations is the class of models in which $w_x(z)$ exhibits 
oscillating behaviour  \cite{Sahni:1999qe, Dodelson:2001fq, 
Feng:2004ff,Xia:2004rw, Xia:2006rr,Zhao:2006qg, Nojiri:2006ww, Jain:2007fa, 
Lazkoz:2010gz, 
Pace:2011kb, Ma:2011nc, Zhang:2017idq}. The oscillating dark 
energy 
models are appealing and prove to lead to desirable cosmological behaviour. In 
particular, 
they  can alleviate  the coincidence problem, since they may lead to both accelerating 
and decelertaing phases in a periodic manner \cite{Dodelson:2001fq}, and thus to dark 
matter and dark energy density parameters of the same order. Furthermore, one can 
construct  oscillating dark energy models that can unify the current 
acceleration with the 
early-time inflationary phase \cite{Feng:2004ff}. 

The main question that arises naturally is whether oscillating dark-energy models 
are in agreement with the latest observational data. Although an early, basic 
fitting was performed in \cite{Nesseris:2004wj}, such an investigation has not been 
fulfilled in detail. In the present work we are interested in performing a complete 
observational confrontation, in order to examine whether oscillating dark energy models 
are in agreement with the latest data, namely: Joint Light Curve analysis sample 
from Supernoave Type Ia, Baryon Acoustic Oscillations (BAO) distance measurements, Cosmic 
Microwave Background (CMB) observations, redshift space distortion, weak gravitational 
lensing, Hubble parameter measurements from 
cosmic chronometers, and finally the local Hubble constant value.

The manuscript is organized in the following way. In Section \ref{sec-2} we present the 
cosmological equations for a dark energy model, both at background and perturbative 
levels. In Section \ref{sec-models} we introduce the oscillating dark energy models, 
through suitable parametrizations of the dark-energy equation-of-state parameter.  
In Section \ref{sec-data} we present the various observational data sets that we will use 
in our analysis, and in Section \ref{sec-results} we perform a detailed observational 
confrontation for various oscillating models. In Section \ref{sec-compare} we compare 
the results for all models, both amongst each other as well as relating to  $w$CDM and 
$\Lambda$CDM cosmology. Finally,   Section \ref{sec-discuss} is devoted to the 
Conclusions.

\section{Cosmological equations: Background and perturbations}
\label{sec-2}

In this section we provide the basic equations, both at the background and at the 
perturbation level, of a general cosmological scenario. Throughout the work we consider 
the homogeneous and isotropic Friedmann-Lema\^itre-Robertson-Walker (FLRW) metric of the 
form 
\begin{eqnarray}
{\rm d}s^2 = -{\rm d}t^2 + a^2(t) \left[ \frac{{\rm d}r^2}{1-kr^2} + r^2 \left ( 
d\theta^2 
+ \sin^2 
\theta d \phi^2   \right)  \right],
\end{eqnarray} 
where $a(t)$ is the scale factor and   $k = -1,+1,0$ corresponds respectively to
open, closed and flat geometry. For simplicity, in the following we focus on the flat 
geometry, as it is favored by 
observations, although the analysis can be straightforwardly extended to the non-flat 
case too. The
Friedmann equations are extracted as
\begin{equation}
H^2 + \frac{k}{a^2}=\frac{8\pi G}{3} \rho_{tot}, \label{f1}%
\end{equation}
\begin{equation}
2 \dot{H} + 3H^2 +  \frac{k}{a^2}=-8\pi G  p_{tot}, \label{f2}%
\end{equation}
where
$H=\dot{a}/a$ is the Hubble function and dots denote derivatives with respect to the 
cosmic time, $t$. In the above equations $\rho_{tot}$ and $p_{tot}$ are respectively the 
total energy density and pressure of the universe content, considered to be effectively 
described by perfect fluids. In particular, we consider that  the universe consists of 
radiation, baryonic matter, dark matter and (effective) dark energy,
and therefore the total energy density and the total 
pressure of the universe read as 
 $\rho_{tot}= 
\rho_{r}+\rho_{b}+\rho_{c}+\rho_{x}$ and $p_{tot}= 
p_{r}+p_{b}+p_{c}+p_{x}$, where  $\rho_{r}$, $p_r$ correspond to   radiation,  
$\rho_{b}$, $p_b$ to   baryonic sector,  $\rho_{c}$, $p_c$ to   dark matter, 
and 
$\rho_{x}$, $p_x$ to the dark energy sector. If we additionally assume that these sectors 
do 
not mutually interact, then each one is separately conserved, namely it satisfies 
\begin{equation}
\dot{\rho}_i+3H(1+w_i)\rho_i=0,
\end{equation}
where $w_i\equiv p_i/\rho_i$ is the $i$-th component's  equation-of-state parameter.
Since radiation has $w_r=1/3$, we obtain $\rho_{r}\propto\left( 
a/a_{0} \right) ^{-4}$. Similarly, since as usual the baryonic and dark matter sectors 
are considered to be pressureless, we obtain $\rho_{b}\propto\left( a/a_{0} \right) 
^{-3}$ 
and $\rho_{c}\propto\left( a/a_{0} \right) ^{-3}$, with $a_0$ the current 
value of the scale factor. Finally, since the dark energy fluid 
has a general 
equation-of-state parameter $w_x\equiv 
p_x/\rho_x$, its evolution equation 
leads to
\begin{eqnarray}\label{de-evol}
\rho_{x}=\rho_{x,0}\,\left(  \frac{a}{a_{0}}\right)  ^{-3}\,\exp\left(
-3\int_{a_{0}}^{a}\frac{w_{x}\left(  a'\right)  }{a'}\,da'
\right).
\end{eqnarray}
Hence, we can   see that the evolution of 
the dark energy fluid is obviously highly dependent on the form
of $w_x (a)$.

Let us now investigate the perturbations of the above general cosmological scenario. The 
perturbation equations of a general dark energy scenario have been explored  in detail in 
the literature \cite{Copeland:2006wr}. We choose the synchronous gauge, and thus the 
perturbed FLRW metric takes the form
\begin{eqnarray}
\label{perturbed-metric}
ds^2 = a^2(\tau) \left [-d\tau^2 + (\delta_{ij}+h_{ij}) dx^idx^j  \right], 
\end{eqnarray}
where $\tau$ is the conformal time, and where $\delta_{ij}$,  $h_{ij}$ are 
respectively the unperturbed and the perturbated metric  tensors. Now, for the 
perturbed 
FLRW metric (\ref{perturbed-metric}), using the conservation equation for the  
energy-momentum  tensor  of the $i$-th 
fluid, namely $T^{\mu \nu}_{; \nu}= 0$, one can conveniently write down the 
Einstein's equations in the two gauges, namely, the conformal Newtonian gauge or in the 
synchronous 
gauges of the Fourier space $\kappa$.  We choose the latter gauge, and in that gauge,  
one 
can 
obtain the continuity and the Euler equations as \cite{Mukhanov, Ma:1995ey, Malik:2008im}:
\begin{eqnarray}
\delta'_{i}  = - (1+ w_{i})\, \left(\theta_{i}+ \frac{h'}{2}\right) - 
3\mathcal{H}\left(\frac{\delta p_i}{\delta \rho_i} - w_{i} \right)\delta_i \nonumber\\- 9 
\mathcal{H}^2\left(\frac{\delta p_i}{\delta \rho_i} - c^2_{a,i} \right) (1+w_i) 
\frac{\theta_i}
{{\kappa}^2}, \label{per1} \\
\theta'_{i}  = - \mathcal{H} \left(1- 3 \frac{\delta p_i}{\delta 
\rho_i}\right)\theta_{i} 
+ \frac{\delta p_i/\delta \rho_i}{1+w_{i}}\, {\kappa}^2\, \delta_{i} 
-{\kappa}^2\sigma_i,\label{per2}
\end{eqnarray}
where the prime denotes   differentiation with respect to the conformal time $\tau$. In 
these equations $\delta_i = \delta \rho_i/\rho_i$ is the density perturbation, 
$\mathcal{H}= a^{\prime}/a$, is the conformal 
Hubble factor, $h = h^{j}_{j}$ is the trace of the metric perturbations $h_{ij}$, and
$\theta_{i}\equiv i \kappa^{j} v_{j}$ is the divergence of the $i$-th fluid
velocity. Additionally, $\sigma_i$ is the anisotropic stress of the $i$-th fluid, which 
will be neglected in our analysis. Finally,  $c_{a,i}^2 = \dot{p}_i/\dot{\rho}_i$ is 
the adiabatic speed of sound of the $i$-th fluid.
 As it is known, for an imperfect 
fluid the quantity $c^2_{s} = \delta p_i / \delta \rho_i$ is 
the sound  speed for the $i$-th fluid. Thus, the adiabatic sound speed is related to 
the sound speed through 
\begin{equation}
c^2_{a,i} =  w_i - \frac{w_i^{\prime}}{3\mathcal{H}(1+w_i)}.
\end{equation}
  We mention here that many dark energy models can be described through 
imperfect fluids,  
which   have $c_s^2 = 1$ while $c_a$ could be different 
\cite{Erickson:2001bq, Weller:2003hw, Hannestad:2005ak}. Hence, although there exist
models with $c_s^2 > 1$ (e.g in $k$-essence models), in our analysis we fix 
this qunatity to be unity.

\section{Oscillating Dark-Energy models}
\label{sec-models}

In this section we consider dark energy parametrizations that exhibit oscillating 
behaviour with the evolution of the universe. Our primary intention is to investigate 
these models with current cosmological data.

For convenience we 
will use as independent variable the redshift, defined as $z=\frac{a_0}{a}-1$, with 
 the current scale factor $a_0$ set to $1$ for simplicity. 
 We will study the following four models:

\begin{itemize}
\item{Model I:} The first model in this class is
\begin{eqnarray}\label{model1-current}
 w_x(z) = w_0 + b \left\{  1- \cos \left[\ln (1+z)  \right] \right\},
\end{eqnarray}
where $w_0$ is the current value of $w_x (z)$ and $b$ is the model parameter.
The free parameter $b$  quantifies the dynamical character of the model. For $b =0$ we 
acquire $w_x (z) = w_0$, while any nonzero value of $b$ corresponds to a deviation 
of the model from the constant dark-energy equation-of-state parameter. Let us 
note that the generalized version of (\ref{model1-current}) can be found in 
\cite{Feng:2004ff}, which however allows a large number of parameters in terms of the 
frequency or 
period of the oscillations. The inclusion of several free parameters $-$ such as 
the 
frequency or period of the oscillations $-$ may add different aspects and richer behavior 
to the oscillating 
dark energy models, 
however, from 
the statistical point of view,
the presence of a large number of free parameters in a dark energy model
increases the degeneracy amongst them. The two-parameters models, on 
the other hand, are able to retain the oscillating features of the parametric 
dark energy models, whose study is the field of interest of the present work, and 
qualitatively they look similar to
the four-parameters models \cite{Feng:2004ff}.  
Thus, although the four-parameters oscillating dark energy models are the general ones,
here we restrict to models with only 
two free parameters in order to examine if an oscillating behavior is 
allowed in the  dark-energy equation of 
state, and quantitatively confront it with the 
observational data. This   may serve as a good starting point towards the 
analysis of the most general oscillating dark energy models.

\item {Model II:} In similar lines we introduce another oscillating function as 
\begin{eqnarray}\label{model2-current}
  w_x(z) = w_0 + b \sin \left[   \ln (1+z)  \right],
\end{eqnarray}
with $w_0$ and $b$ the model parameters as described for Model 
I. A general version of the above model can be found in 
\cite{Xia:2006rr,Zhao:2006qg} in which 
the authors have considered the period of oscillations along with other free parameters, 
thus
leading to an extended parameter space. Since a large number of parameters generally 
leads to degeneracy amongst them, in this work we consider the two-parameter 
model.

\item {Model III:} Another oscillatory dark energy parametrization is  \cite{Ma:2011nc}
\begin{eqnarray}\label{model3}
 w_x(z) = w_0 + b \left[  \frac{\sin (1+z)}{1+z} - \sin 1\right],
\end{eqnarray}
with $w_0$ and $b$ the model parameters with as described for Model 
I.

\item {Model IV:} Finally,  we consider a new model
\begin{eqnarray}\label{model4}
 w_x(z) = w_0 + b \left[ \frac{z}{1+z} \right] \cos(1+z),
\end{eqnarray} 
where $w_0$, $b$ are the free parameters as described above. 
One may 
note that the above model might be connected with the CPL model 
\cite{Chevallier:2000qy,Linder:2002et} for very low redshifts. 

\end{itemize}

\section{Observational Data}
\label{sec-data}

In this section we provide the various data sets that we will incorporate in the 
observational fittings. We will use 
data from the following probes:

\begin{enumerate}

\item \textit{Supernovae Type Ia:} We include the latest Joint Light Curve analysis 
sample \cite{Betoule:2014frx} from  Supernovae Type Ia, one of the cosmological 
data sets to probe the nature of dark energy. The sample contains 740 number of 
Supernovae Type Ia data, distributed in the 
redshift interval $z \in [0.01, 1.30]$. The $\chi^2$ function for this sample becomes

\begin{eqnarray}
\chi^2_{\rm JLA}=(\hat{\mu}-\hat{\mu}^{m})^T C^{-1}
(\hat{\mu}-\hat{\mu}^{m}),
\label{eq:JLA}
\end{eqnarray}
where $\hat{\mu}$ is the vector of effective absolute magnitudes,
$C$ is the covariance metrix of $\hat{\mu}$ quantifying the statistical and systematic 
errors (see \cite{Betoule:2014frx} for details), and 
$\hat{\mu}^{m}(z)= 5 \log_{10} \left(\frac{D_L (z)}{10\mbox{pc}}\right)$
is the distance modulus at redshift $z$ for the model in which $D_L (z) $ is the 
luminosity 
distance \cite{Riess:1998cb}.

\item \textit{Baryon Acoustic Oscillations (BAO) distance measurements:} 
For the BAO data, we use the ratio of $r_s/D_V$ acting as a ``standard ruler'' in 
which  the 
quantity $r_s$ refers to the comoving sound horizon at the baryon drag epoch and $D_V$ 
refers to 
the effective distance determined by $D_A$. The angular diameter distance  and the 
Hubble parameter 
$H$   are related through the following equation  \cite{Eisenstein:2005su}
\begin{eqnarray}
D_V(z)=\left[(1+z)^2D_A(a)^2\frac{z}{H(z)}\right]^{1/3}.
\label{eq:Dv}
\end{eqnarray}
We include  four  measurements of $r_s/D_V$ at four different redshifts, 
namely from the 6dF Galaxy Survey (6dFGS) measurement at 
$z_{\emph{\emph{eff}}}=0.106$
\cite{Beutler:2011hx}, from the Main Galaxy Sample of Data Release 7 of Sloan
Digital Sky Survey (SDSS-MGS) at $z_{\emph{\emph{eff}}}=0.15$
\cite{Ross:2014qpa}, and from the CMASS and LOWZ samples from the latest Data
Release 12 (DR12) of the Baryon Oscillation Spectroscopic Survey (BOSS) at
$z_{\mathrm{eff}}=0.57$    and at $z_{\mathrm{eff}%
}=0.32$ \cite{Gil-Marin:2015nqa}. The   likelihood  for BAO is given by
\begin{eqnarray}
\chi^2_{\rm BAO} = \sum_{i} \frac{\left[r_{\rm BAO,i}^{\rm obs} - r_{\rm BAO,i}^{\rm th} 
\right]^2}{
\sigma_i^2},
\label{eq:BAO}
\end{eqnarray}
where $r_{\rm BAO}= r_s(z_d)/D_V$ and $\sigma_i$'s are the uncertainties in the 
measurements 
for each data point $i = 1, 2, 3, 4$, respectively correspond to
$z_{\emph{\emph{eff}}}=0.106$ \cite{Beutler:2011hx}, $z_{\emph{\emph{eff}}}=0.15$ 
\cite{Ross:2014qpa}, $z_{\mathrm{eff}}=0.57$ \cite{Gil-Marin:2015nqa} and 
$z_{\mathrm{eff}}=0.32$ \cite{Gil-Marin:2015nqa}.

\item \textit{Cosmic Microwave Background (CMB) data:} 
We incorporate the CMB temperature and polarization anisotropies with their 
cross-correlations from the Planck Probe \cite{Adam:2015rua}. Specifically, we use the 
combinations of high- and low-$\ell$ 
TT likelihoods (overall multiple range $2\leq \ell \leq 2508$) as well as the 
combinations 
of the high- and low-$\ell$ polarization likelihoods \cite{Aghanim:2015xee},  which 
are notationally referred to as Planck TT, TE, EE+lowTEB. In order to analyze the data we 
use the publicly available Planck 
likelihood \cite{Aghanim:2015xee}, which eventually marginalizes over several nuisance 
parameters associated with the measurements. For a detailed study and the implementation 
of the CMB data, we refer the reader to \cite{Adam:2015rua, Aghanim:2015xee}.

\item \textit{Redshift space distortion:} We include two 
redshift space distortion (RSD) data from CMASS and LowZ galaxy samples. The
CMASS sample consists of 777202 galaxies with an effective redshift of $%
z_{\mathrm{eff}}=0.57$ \cite{Gil-Marin:2016wya}, 
while the LOWZ sample contains 361762 galaxies with an effective redshift of $z_{%
\mathrm{eff}}=0.32$~\cite{Gil-Marin:2016wya}. 
The data-vector containing the cosmological parameters of interest, namely 
$f(z)\sigma_8(z)$, $H(z) r_s(z_
d)$ (in $10^3 {\rm km}s^{-1}$ units) and $D_A(z)/r_s(z_d)$, reads as
\begin{equation}
 \label{data1}
D_{\rm data}(z) =
 \begin{pmatrix}
  f(z)\sigma_8(z)  \\
  H(z) r_s(z_d) \\
  D_A(z)/r_s(z_d)
 \end{pmatrix}.
 \end{equation}
 The data-vectors for the samples LOWZ and CMASS can be formed as   
  (see Table 3 of \cite{Gil-Marin:2016wya}):
 \begin{equation}
 \label{data_lowz}
D_{\rm data}(z_{\mathrm{eff}}=0.32) =
 \begin{pmatrix}
   0.45960 \\
  11.753 \\
  6.7443
 \end{pmatrix},
 \end{equation}
 from the LOWZ sample at $k_{\rm max}=0.18\,h{\rm Mpc}^{-1}$, and
  \begin{equation}
 \label{data_cmass}
D_{\rm data}(z_{\mathrm{eff}}=0.57) =
 \begin{pmatrix}
   0.41750 \\
  13.781 \\
  9.3276
 \end{pmatrix},
 \end{equation}
 from the CMASS sample at $k_{\rm max}=0.22\,h{\rm Mpc}^{-1}$. The covaiance
 matrices for the above two samples are given in \cite{Gil-Marin:2016wya}. In 
particular, the covariance matrix 
for the LOWZ sample at $k_{\rm max}=0.18\,h{\rm Mpc}^{-1}$ is
\begin{equation}
\label{cov1}
C^{{\rm LOWZ}} = 10^{-3}
 \begin{pmatrix}
 5.0837 & 23.818 & 10.490  \\
  &  300.30 & 73.448 \\
  &  & 47.493
 \end{pmatrix},
 \end{equation}
while the covariance matrix for the CMASS sample at $k_{\rm max}=0.22\,h{\rm 
Mpc}^{-1}$ is
\begin{equation}
\label{cov2}
C^{{\rm CMASS}}= 10^{-3}
 \begin{pmatrix}
1.3046 & 4.6434 & 3.5329  \\
  & 77.713 & 22.773 \\
  &  & 21.700
 \end{pmatrix}.
 \end{equation}

Now, the corresponding likelihood of any cosmological model is   given by
\begin{align} \label{likelihood}
\mathcal{L}\propto e ^{\left[ -(D_{\rm data}-D_{\rm model})^T {C}^{-1} (D_{\rm 
data}-D_{\rm model}) 
\right]},
\end{align}
where $D_{\rm model}$ represents the vector with the model prediction for the same 
cosmological
 parameters as $D_{\rm data}$ and $C^{-1}$ is the inverse of the covariance matrix.
Lastly, we mention that when these two RSD data are considered, the 
BOSS DR12 results will not be considered.

\item \textit{Weak lensing data:}  We consider the weak gravitational lensing 
from the Canada-France-Hawaii Telescope
Lensing Survey (CFHTLenS) \cite{Heymans:2013fya,Asgari:2016xuw}. 
The CFHTLenS
is the largest weak lensing survey at present and  
spans 154 square degrees in five optical bands. We use 
the tomographic CFHTLenS blue galaxy sample for the analysis. In particular, 
we note that the survey \cite{Heymans:2013fya}, which we follow in  this work, uses   
21 sets of 
cosmic shear correlation functions associated with six redshift bins. The weak 
correlations between 
the observed shapes of distant galaxies are generally induced due to the weak 
gravitational lensing 
by large scale structure. 
The cosmological information can be extracted through the two-point shear correlation 
function,
which is related to convergence power spectrum
\begin{eqnarray}
P^{ij}_{K}(l)=\int^{\eta_H}_0 d\eta 
\frac{q_i(\eta)q_j(\eta)}{[f_K(\eta)]^2}P_{\delta}\left(\frac{l}
{f_K(\eta)};\eta\right),
\end{eqnarray}
where $\eta$ is the comoving distance, $\eta_{H}$ is the horizon distance, and
$f_K(\eta)$ is the 
angular diameter distance out to $\eta$.  The quantity  $f_K(\eta)$ depends on the 
curvature scalar 
($k$) of spacetime, and $q_j$ is the lensing
efficiency function for the redshift bin $j$ (see \cite{Heymans:2013fya,Asgari:2016xuw} 
for more 
details).  
The tomographic correlation functions measured from the blue galaxy sample is consistent 
with zero 
intrinsic alignment nuisance parameter $A$. We use the likelihood analysis of the 
CFHTLenS 
data, 
where the true inverse covariance matrix is given by 
$\mathbf{C}^{-1}=\alpha_A\hat{\mathbf{C}}^{-1}
$. Here $\alpha_A=(n_{\mu}-p-2)/(n_{\mu}-1)$ and $\hat{\mathbf{C}}$ is the measured 
covariance 
matrix in which $p$ is the total number of data points, that is calculated in 
\cite{Heymans:2013fya} 
as follows: for $N_t$ tomographic redshift bins and $N_{\theta}$ angular scales, and 
considering 
the shear correlation functions
$\xi_{+}^{ij}$ and $\xi_{-}^{ij}$ (see \cite{Heymans:2013fya,Asgari:2016xuw} for more 
details) 
between the redshift bins $i$, $j$, we have $p = 
N_{\theta} N_{t} 
(N_{t}+ 1)$. The quantity $n_{\mu}$ refers to the total number of simulations.  The 
$\chi^2$ 
function for this data set is given by 
\begin{eqnarray}
\chi^2_{\ WL} = [\hat{d} - d(\pi)]^{T} C^{-1} [\hat{d} - d(\pi)],
\end{eqnarray}
where $\hat{d}$ is the vector of measured data points, and $d(\pi)$ represents 
the vector carrying the model parameters.

\item \textit{Cosmic Chronometers (CC) data:} 
In our analysis we consider the Hubble 
parameter values at different redshifts, using the massive and passively evolving
galaxies in our universe, known as cosmic chronometers. The 
measurements of the Hubble parameter values follow the spectroscpic
method with high accuracy, and moreover the technique of measurements
is model independent \cite{Nunes:2016dlj,Anagnostopoulos:2017iao}. The CC (or $H(z)$) 
data 
are 
compiled in \cite{Moresco:2016mzx},    and they contain 30 measurements 
distributed in the interval
$0< z< 2$. The $\chi^2$-statistics for the cosmic 
chronometers data is 
given by
\begin{eqnarray}
\chi^2_{\rm CC} = \sum_{i= 1}^{30} \frac{\left( H(z_i) - 
H_{th}(z_i)\right)^2}{\sigma_i^2},
\end{eqnarray}
where each $z_i$ with its corresponding uncertainty $\sigma_i$ can be found in Table 
4 of \cite{Moresco:2016mzx}.

\end{enumerate}

\section{Observational constraints}
\label{sec-results}

In this section we proceed to the detailed confrontation of the above oscillating dark 
energy models with observational data. We perform a combined analysis JLA $+$ BAO $+$ 
Planck TT, TE, EE $+$ LowTEB (CMB) $+$ RSD $+$ WL$+$ CC
to constrain the proposed oscillating dark energy 
models (\ref{model1-current}), (\ref{model2-current}),
(\ref{model3}) and (\ref{model4}). Our analysis follows the 
likelihood $\mathcal{L} \propto \exp \left(- \chi^2/2 \right)$,
where 
\begin{equation}
\chi^2 = \chi^2 _{JLA} +\chi^2 _{BAO}+ \chi^2_{CMB} +\chi^2_{RSD}+
\chi^2_{WL} + \chi^2_{CC}. 
\end{equation}
  The main statistical analysis is based on the ``Code for Anisotropies in the Microwave 
Background'' (CAMB) \cite{Lewis:2002ah}, a publicly available code. For each of the 
studied  models we modify the code accordingly, and then we additionally use CosmoMC, a 
Markov Chain Monte Carlo (MCMC) simulation, in order to extract the cosmological 
constraints 
for the oscillating dark energy models.

In summary,  we analyze the following 
eight-dimensional parameters space:
\begin{align}
\mathcal{P}_1 \equiv\Bigl\{\Omega_bh^2, \Omega_{c}h^2, 100 \theta_{MC}, \tau, w_0, b, 
n_s, 
log[10^{
10}A_s]\Bigr\},
\label{eq:parameter_space1}
\end{align}
where $\Omega_bh^2$, $\Omega_{c}h^2$ are respectively the baryon   
and the cold dark matter density    parameter, 
$100 \theta_{MC}$ and $\tau$ refer respectively  to the ratio of the sound horizon to the 
angular diameter distance and to the optical depth, 
$n_s$ and $A_s$ are respectively the scalar spectral index and the amplitude of the 
initial power spectrum \cite{Adam:2015rua}, and  $w_0$ and $b$ are the free 
parameters of the oscillating dark energy models.
 Additionally, the priors on the 
cosmological parameters used in the analysis are displayed in Table \ref{tab:priors}. 
Lastly, in the following the subscript ``0'' denotes the value of a quantity at present.

\begin{table}  [ht]
\centering
\begin{tabular}{cc}     
\hline\hline                                                                     
  $\ $    Parameter    $\ $                & Prior\\
\hline 
$\Omega_{c}h^2$              & $[0.01,0.99]$\\
$\Omega_{b} h^2$             & $[0.005,0.1]$\\
$100\theta_{MC}$             & $[0.5,10]$\\ 
$\tau$                       & $[0.01,0.8]$\\
$n_s$                        & $[0.5, 1.5]$\\
$\log[10^{10}A_{s}]$         & $[2.4,4]$\\
$w_0$                        & $[-2, 0]$\\
$b$                          & $[-3, 3]$ \\
\hline\hline                                                                             
\end{tabular}                                                                         
\caption{The flat priors on the cosmological parameters for the CosmoMC analysis. 
}\label{tab:priors}                                     
 \end{table}    
 \begin{figure*}[ht]
\includegraphics[width=0.24\textwidth]{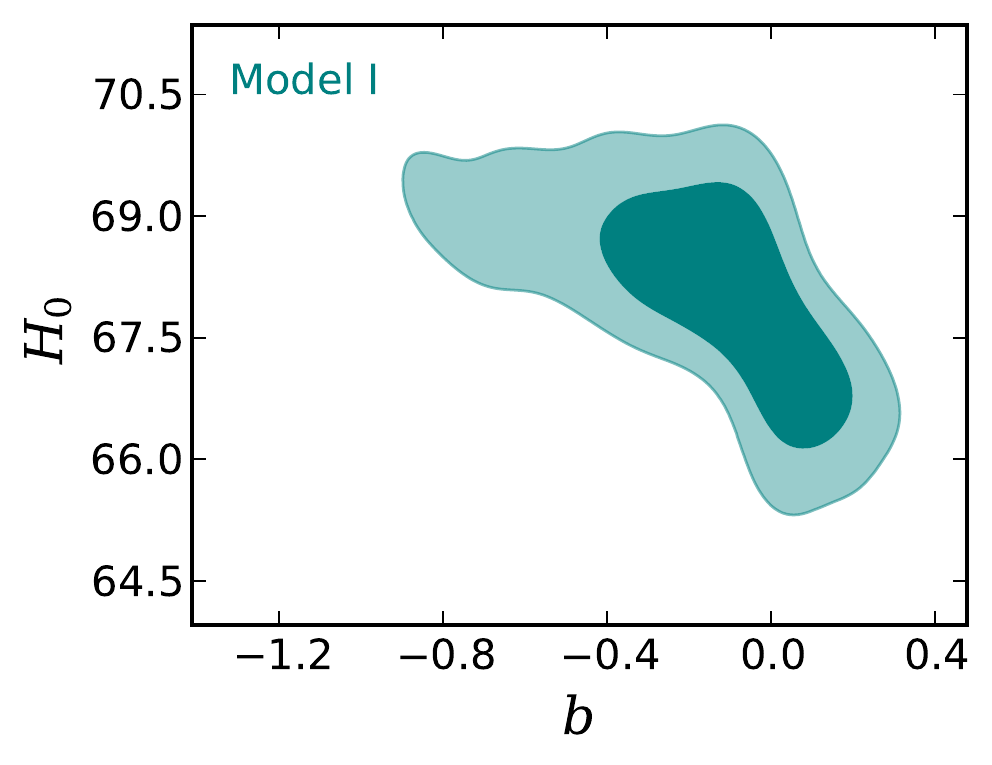}
\includegraphics[width=0.24\textwidth]{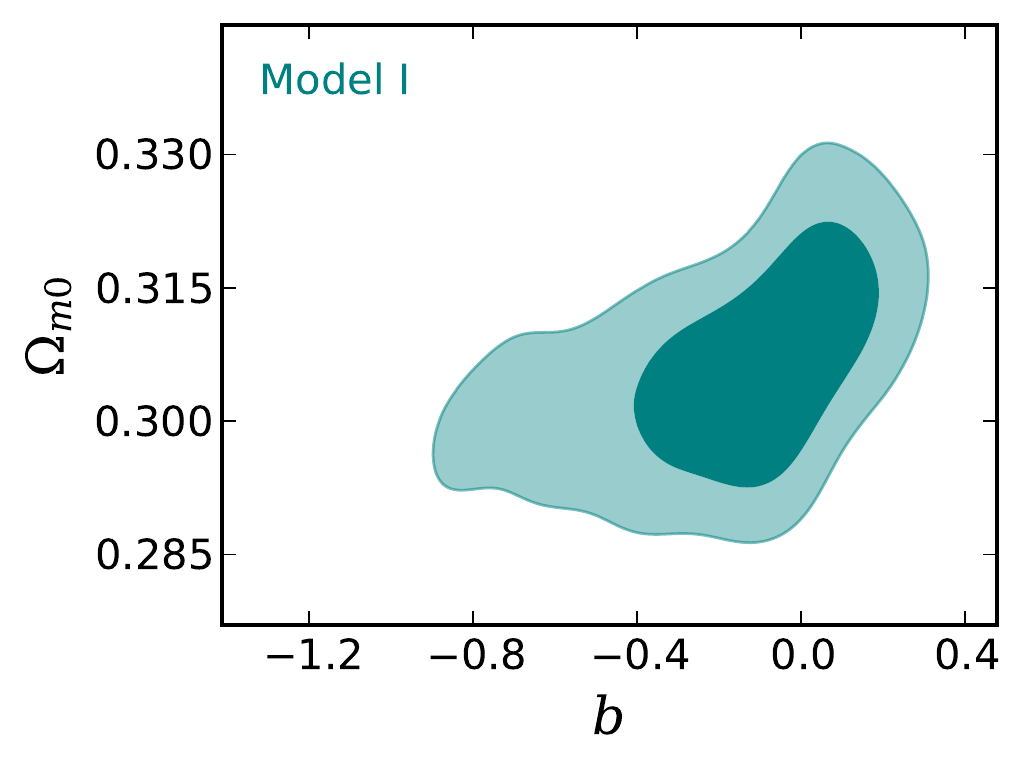}
\includegraphics[width=0.24\textwidth]{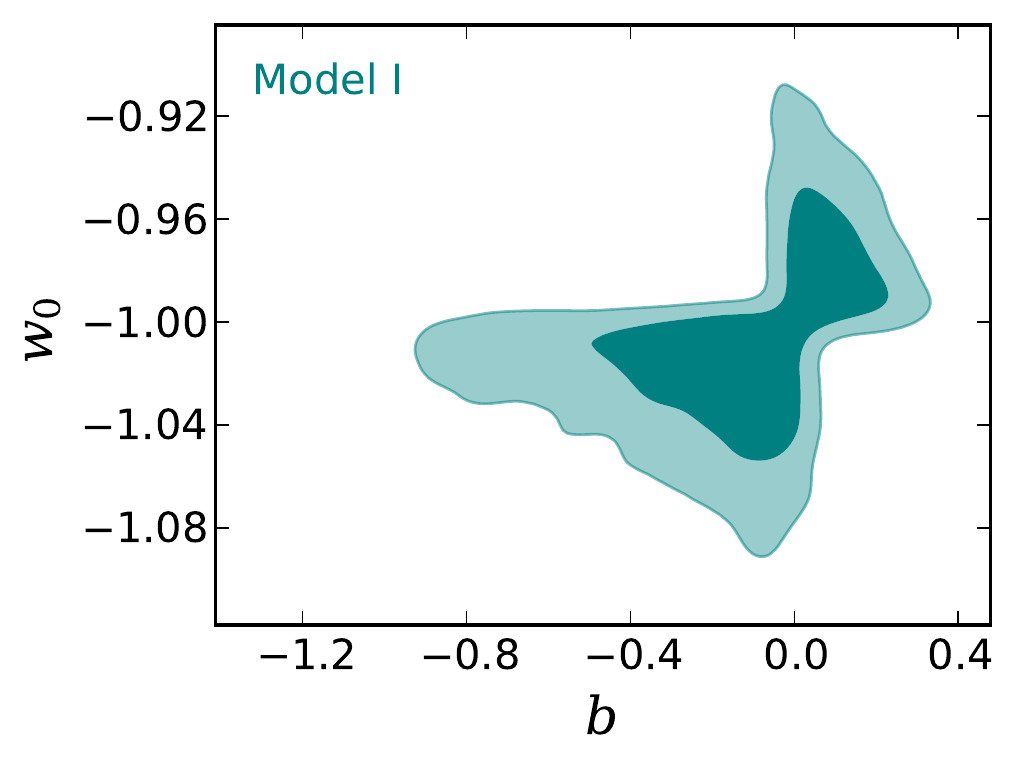}
\includegraphics[width=0.24\textwidth]{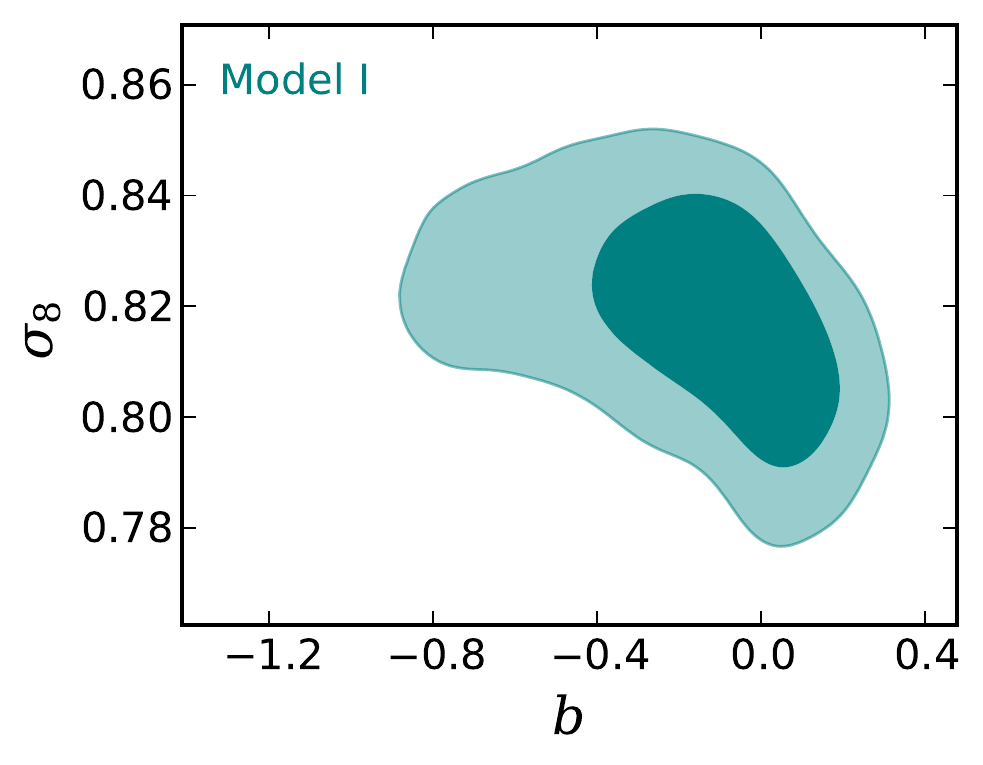}\\
\includegraphics[width=0.24\textwidth]{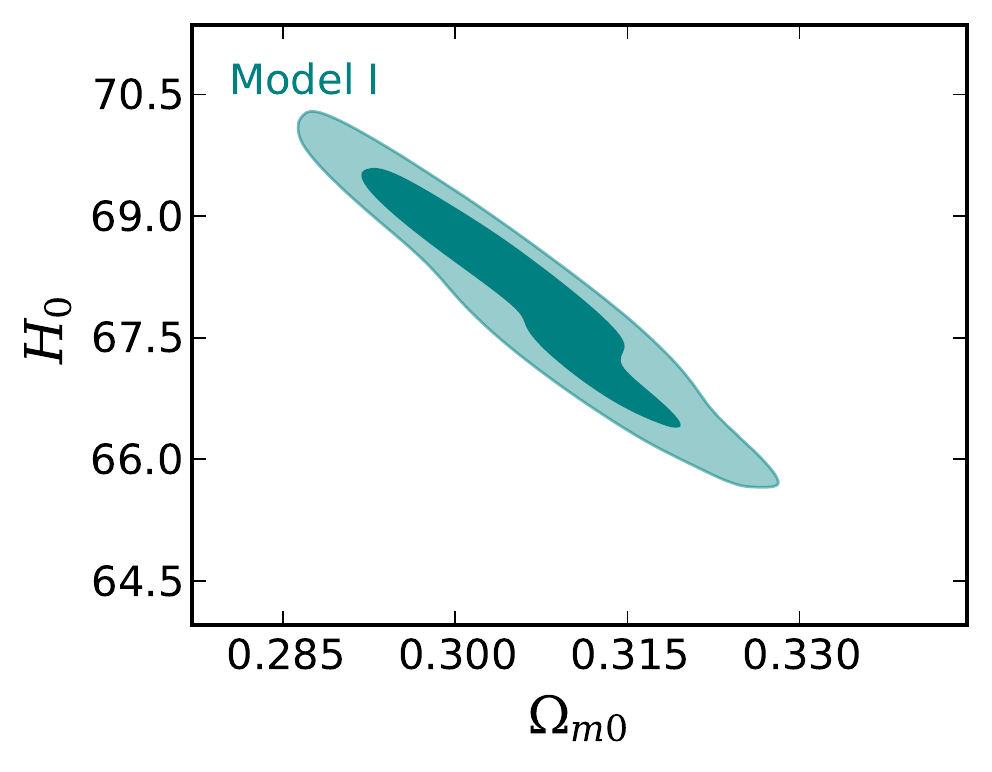}
\includegraphics[width=0.24\textwidth]{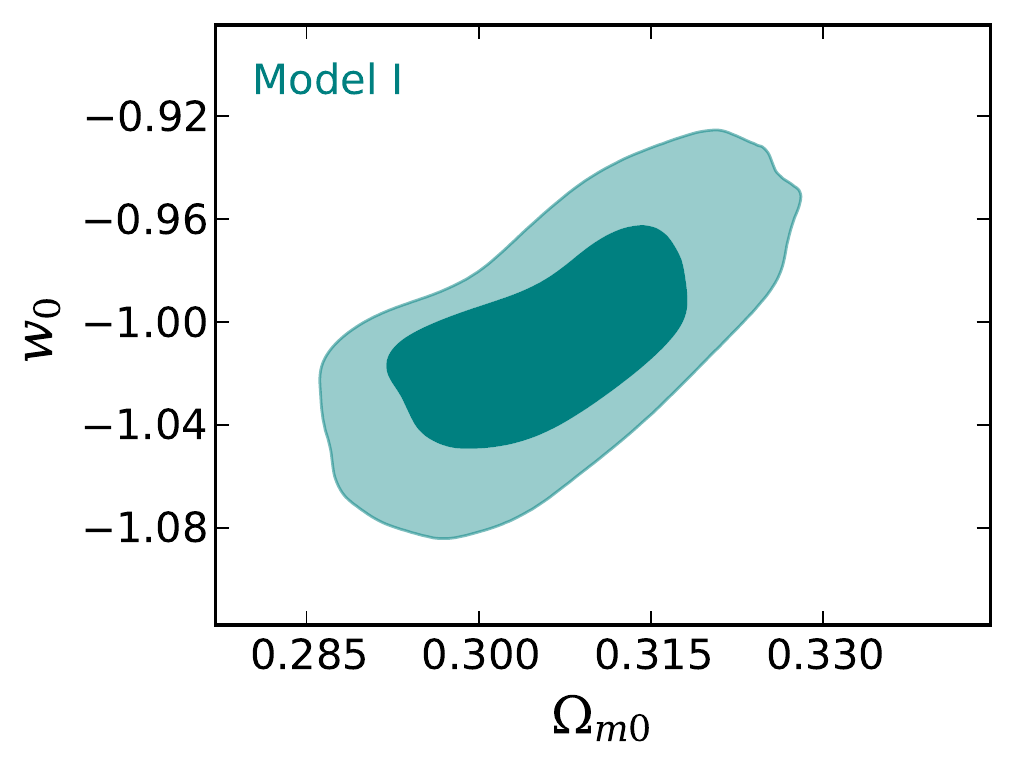}
\includegraphics[width=0.24\textwidth]{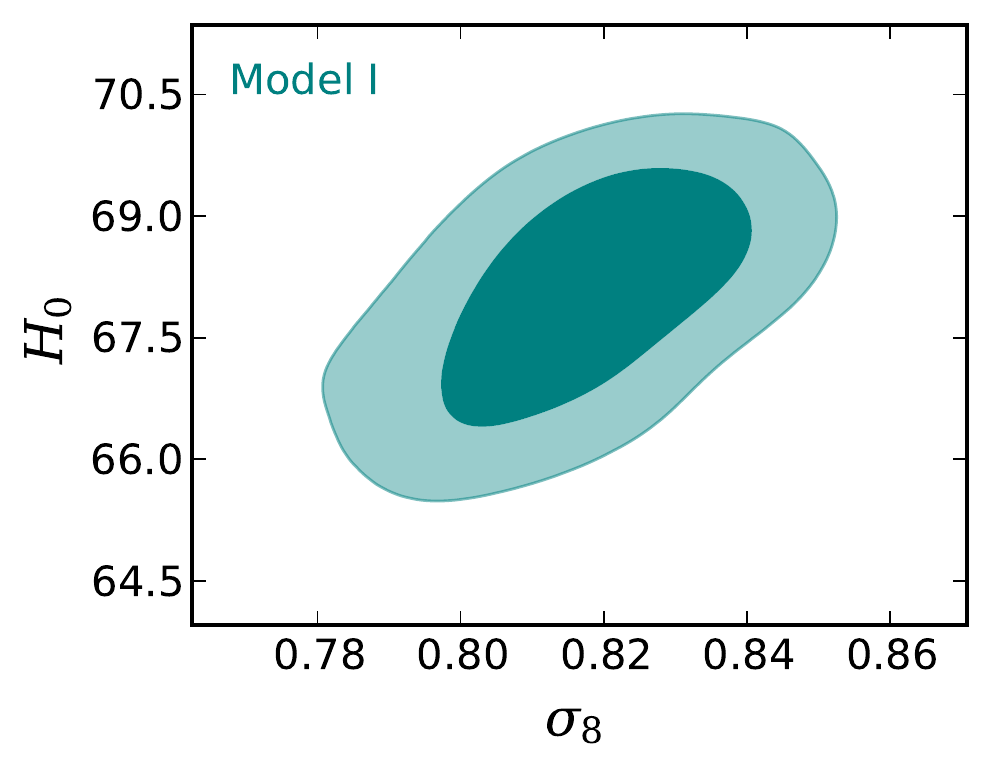}
\includegraphics[width=0.24\textwidth]{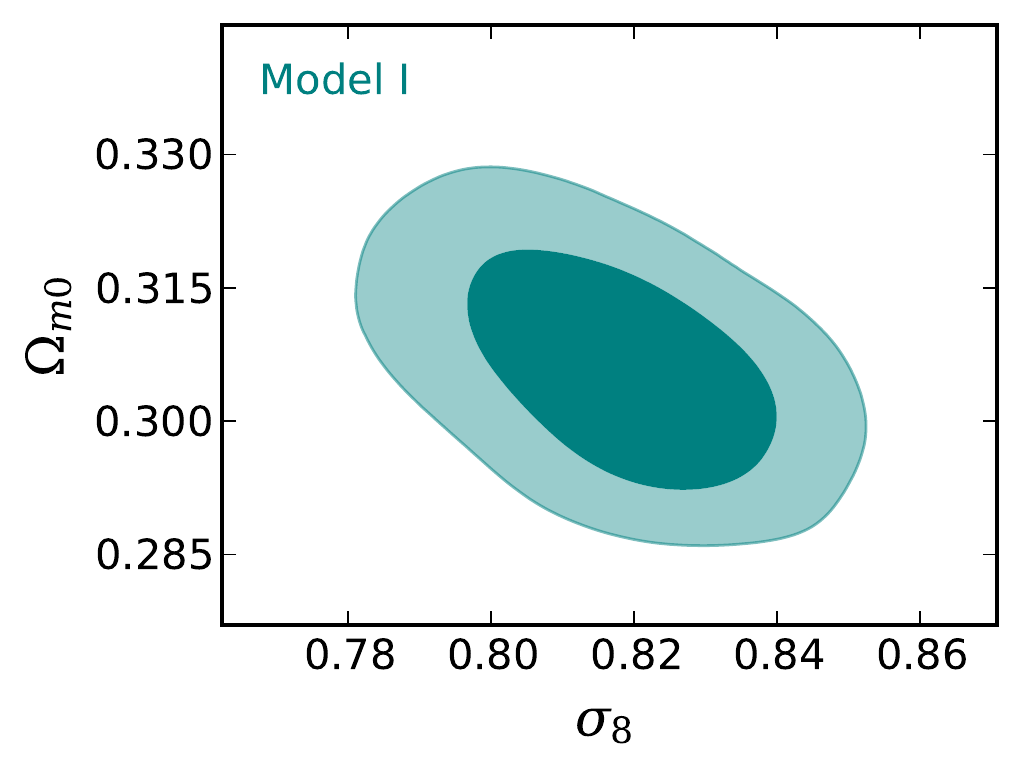}
\caption{\textit{1$\sigma$ (68.3\%) and 2$\sigma$ (95.4\%) confidence level contour plots 
for 
different combinations of the model parameters of Model I  (\ref{model1-current}), 
for the combined observational data JLA $+$ BAO $+$ Planck TT, TE, EE $+$ LowTEB 
$+$ RSD 
$+$ WL$+$ CC. We have defined, $\Omega_{m0} = \Omega_{c0}+\Omega_{b0}$.}}
\label{fig:contour-ModelI}
\end{figure*}  
\begin{figure*}
\includegraphics[width=0.19\textwidth]{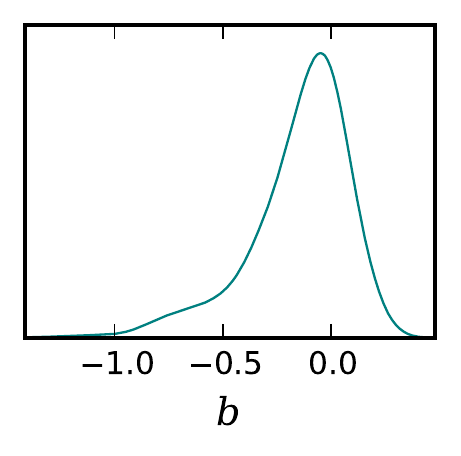}
\includegraphics[width=0.19\textwidth]{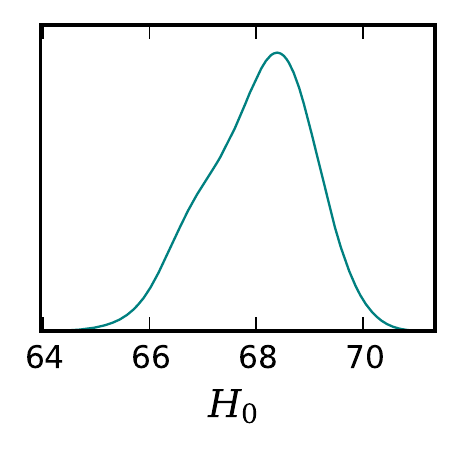}
\includegraphics[width=0.19\textwidth]{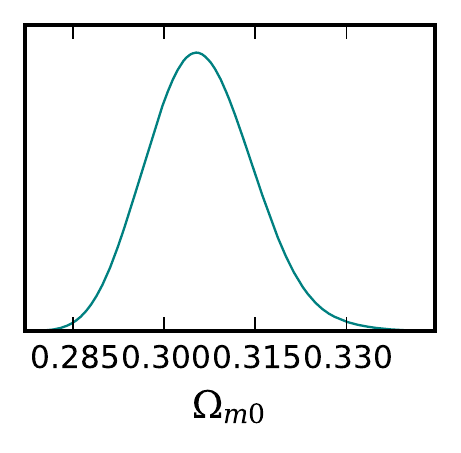}
\includegraphics[width=0.19\textwidth]{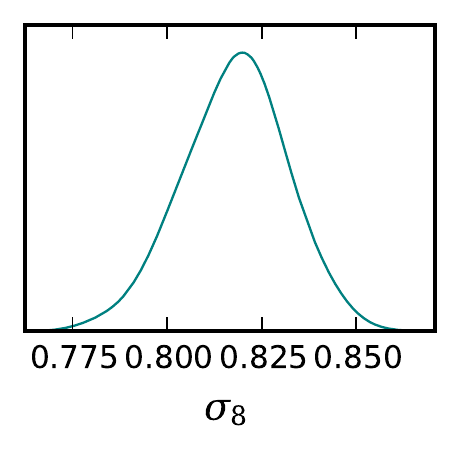}
\includegraphics[width=0.19\textwidth]{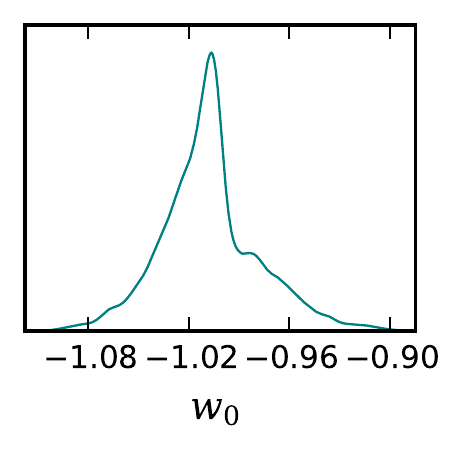}
\caption{\textit{The  marginalized 1-dimensional posterior distributions for the model 
parameters of Model I of (\ref{model1-current}), for the combined observational data JLA 
$+$ BAO $+$ Planck TT, TE, EE $+$ LowTEB 
$+$ RSD  $+$ WL$+$ CC.}}
\label{fig:ModelI-posterior}
\end{figure*}  
\begin{figure*}[!]
\includegraphics[width=0.5\textwidth]{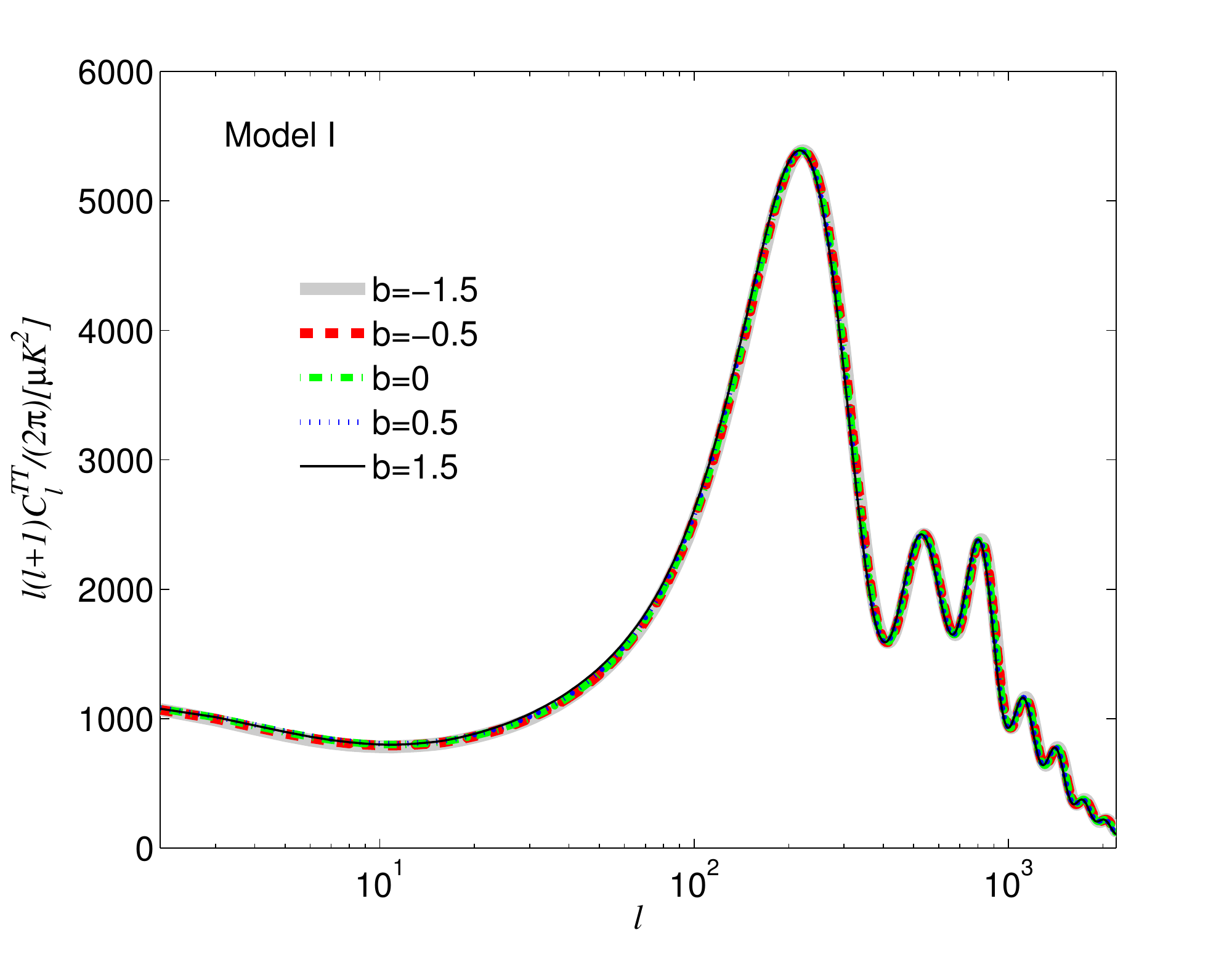}
\includegraphics[width=0.5\textwidth]{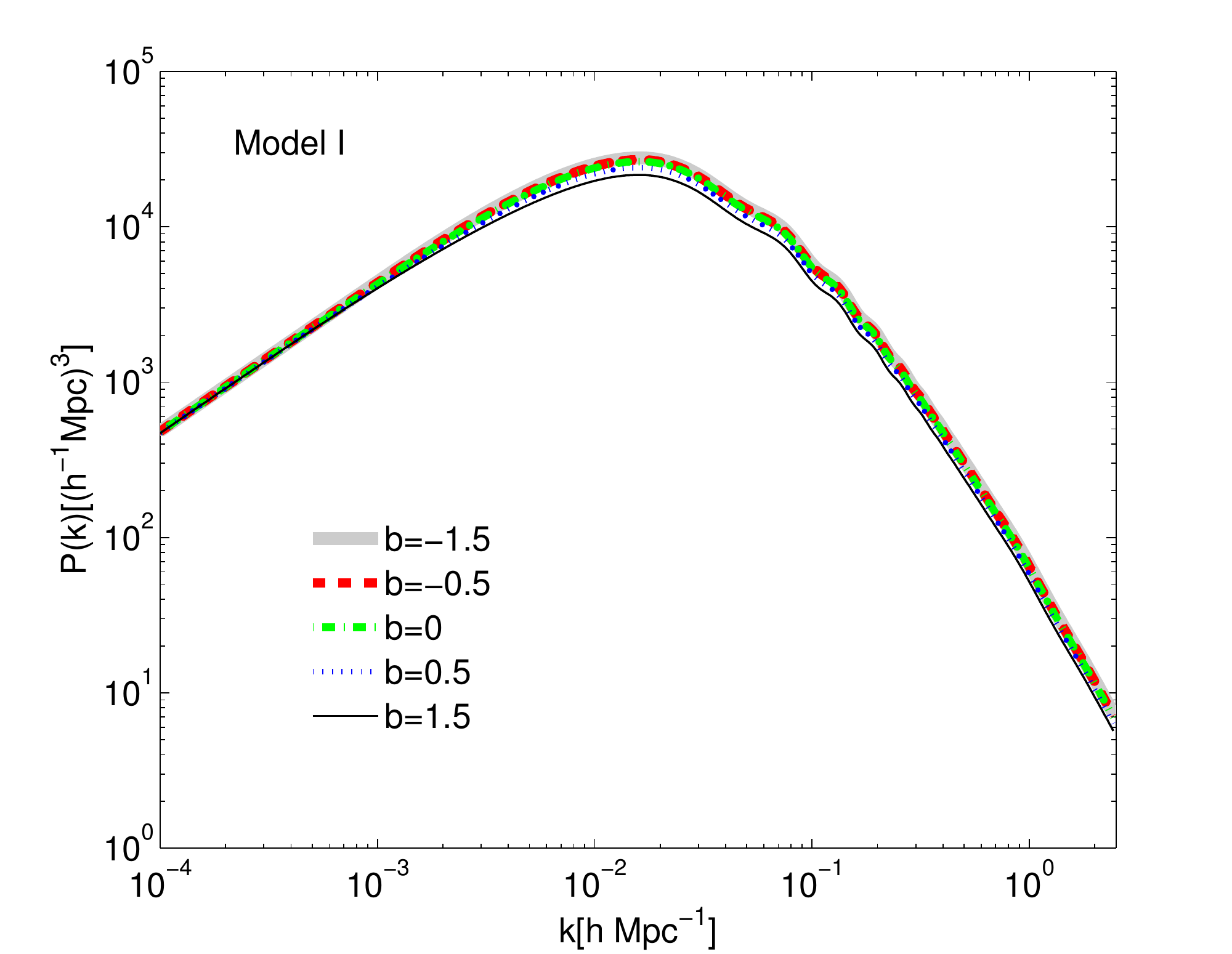}
\caption{\textit{The temperature anisotropy 
in the CMB spectra (left panel) and the matter power spectra (right panel),
for Model I of (\ref{model1-current}), for different values of 
the parameter $b$. }}
\label{fig:CMB-Matter-ModelI}
\end{figure*}

In the next subsections we describe the 
obtained results on each model from this combined analysis.

\subsection{Model I}

\begin{table}[ht]
\centering
\begin{tabular}{ccc}                                                                      
\hline\hline                                                                     
Parameters & Mean $\pm$  1$\sigma$ $\pm$ 2$\sigma$ $\pm$ 3$\sigma$ & Best fit \\ \hline
$\Omega_c h^2$ & $    0.1188_{-    0.0013-    0.0027-    0.0035}^{+    0.0015+    0.0025+ 
   0.0031}
$ & $    0.1187$\\
$\Omega_b h^2$ & $    0.02225_{-    0.00015-    0.00029-    0.00038}^{+    0.00015+    
0.00031+    
0.00038}$ & $    0.02218$\\
$100\theta_{MC}$ & $    1.04055_{-    0.00033-    0.00065-    0.00085}^{+    0.00033+    
0.00064+   
 0.00086}$ & $    1.04054$\\
$\tau$ & $    0.065_{-    0.020-    0.038-    0.048}^{+    0.019+    0.038+    0.050}$ & 
$    0.065$\\
$n_s$ & $    0.9749_{-    0.0044-    0.0085-    0.0107}^{+    0.0045+    0.0088+    
0.0120}$ & $    
0.9726$\\
${\rm{ln}}(10^{10} A_s)$ & $    3.069_{-    0.036-    0.070-    0.091}^{+    0.036+    
0.072+    0.094}$ & $    3.071$\\
$w_0$ & $   -1.0078_{-    0.032-    0.059-    0.080}^{+    0.023+    0.068+    0.094}$ & 
$   -1.0031$\\
$b$ & $   -0.1468_{-    0.142-    0.555-    0.803}^{+    0.275+    0.431+    0.511}$ & $  
 -0.1127$\\
$\Omega_{m0}$ & $    0.306_{-    0.009-    0.017-    0.020}^{+    0.008+ 0.017 +0.025}$ & 
$    0.
308$\\
$\sigma_8$ & $    0.818_{-    0.014-    0.029-    0.040}^{+    0.015+    0.027+    
0.033}$ & $    0.
817$\\
$H_0$ & $   68.05_{-    0.90-    2.02-    2.68}^{+    1.20+    1.77+    2.25}$ & $   
67.84$\\
\hline\hline                                                                             
\end{tabular}                                                                           
\caption{ Summary of the observational constraints on Model I of
(\ref{model1-current}), using 
the observational data JLA $+$ BAO $+$ Planck TT, TE, EE $+$ LowTEB $+$ RSD $+$ WL$+$ CC. 
We define $\Omega_{m0} = \Omega_{c0}+\Omega_{b0}$ and we use $H_0$ to denote the current 
value of the Hubble function.}\label{table-ModelI}                                     
 \end{table}                                                                             
 We perform the above combined analysis for the Model I of  (\ref{model1-current}), and 
in 
Table \ref{table-ModelI} we summarize the main observational constraints. Furthermore, in 
Fig.  \ref{fig:contour-ModelI} we present the 1$\sigma$ 
and 2$\sigma$ confidence-level contour  plots for several combinations of the model 
parameters and of the derived parameters. Similarly, in Fig. \ref{fig:ModelI-posterior} 
we display the marginalized
one-dimensional posterior distributions for the involved quantities.

Our analysis reveals that both the best-fit and the mean values of the dark energy 
equation-of-state parameter at present ($w_0$) exhibit phantom behaviour although very 
close to the cosmological constant boundary, however, as one can see from Table 
\ref{table-ModelI}, within 1$\sigma$ confidence-region the quintessential character of 
$w_0$ is not excluded.

Additionally, we analyze the behaviour of Model I at large scales 
through its impact on the temperature anisotropy of the CMB spectra and on the matter 
power spectra, shown respectively in the upper and lower panel of the Fig. 
\ref{fig:CMB-Matter-ModelI}, and moreover we compare the results  with   $w$CDM 
cosmology (obtained for $b = 0$). We find that for several values of $b$ we do not find a 
remarkable behaviour in the CMB spectra. On the other hand, from the matter power spectra 
we can see that for large positive $b$ values the model has a deviating nature from   
$w$CDM cosmology.

In summary, from the observational constraints we deduce that the model is close to   
$w$CDM cosmology, and hence to the $\Lambda$CDM paradigm too.

\subsection{Model II}
                                                                 
\begin{figure*}
\includegraphics[width=0.24\textwidth]{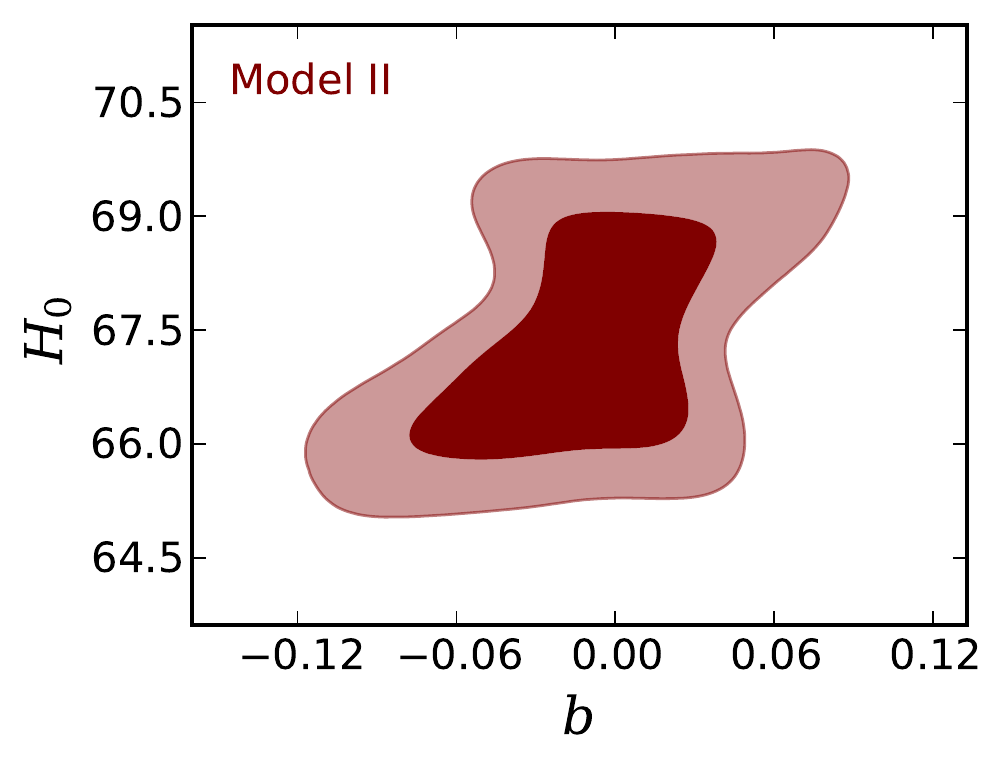}
\includegraphics[width=0.24\textwidth]{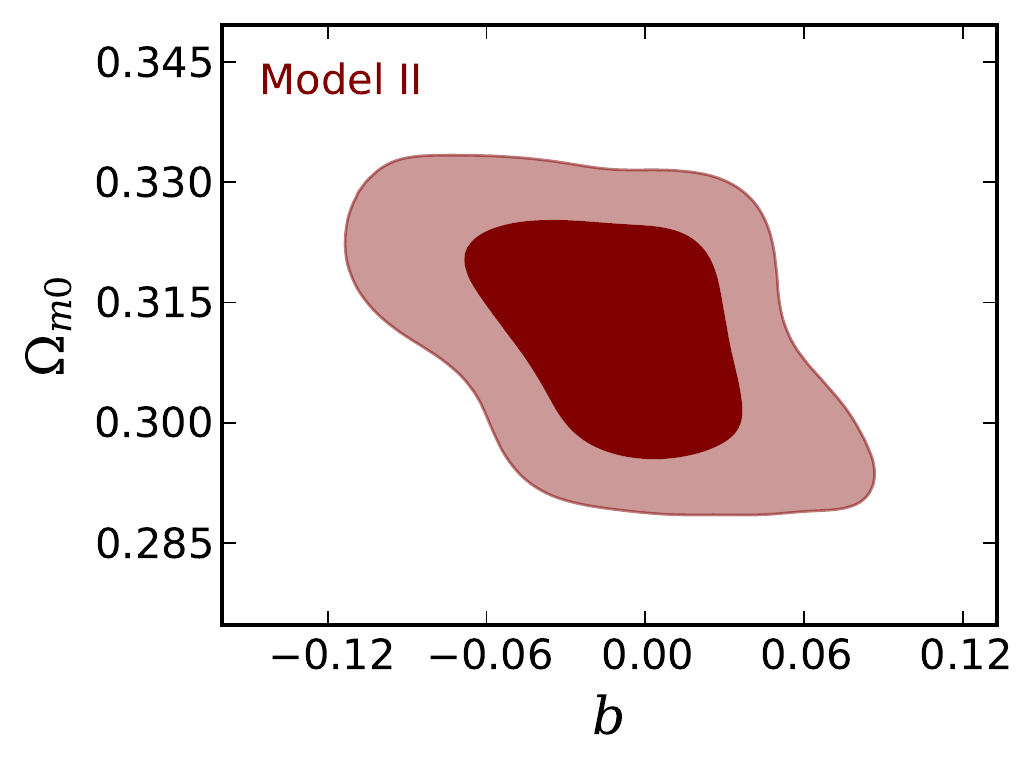}
\includegraphics[width=0.24\textwidth]{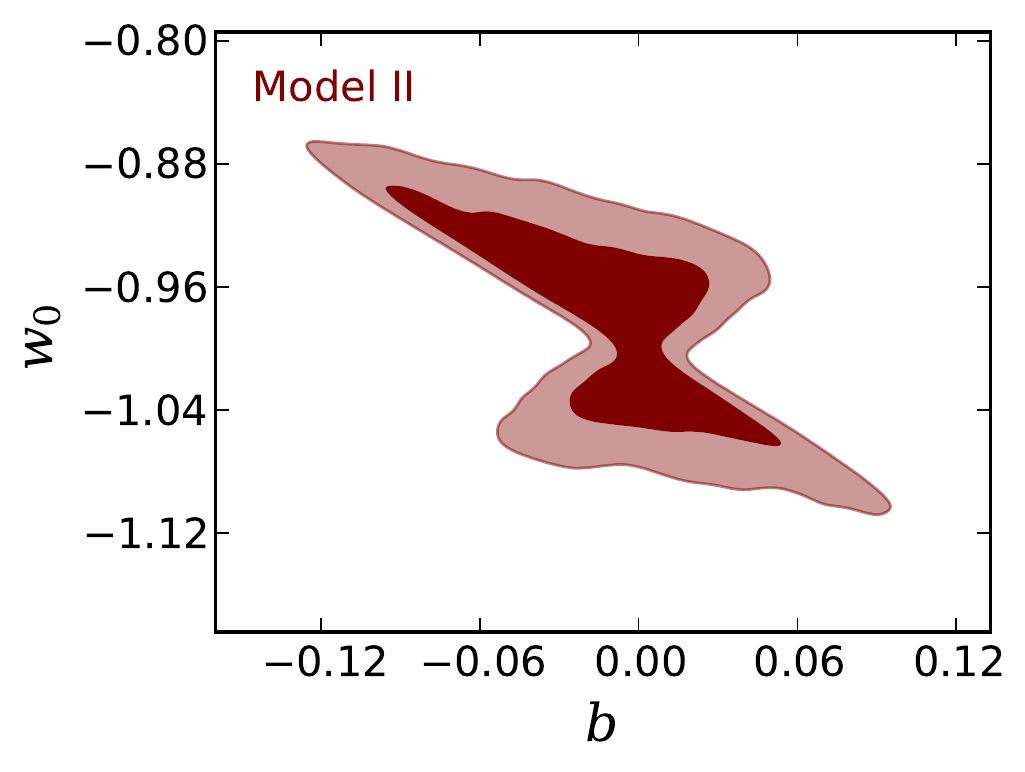}
\includegraphics[width=0.24\textwidth]{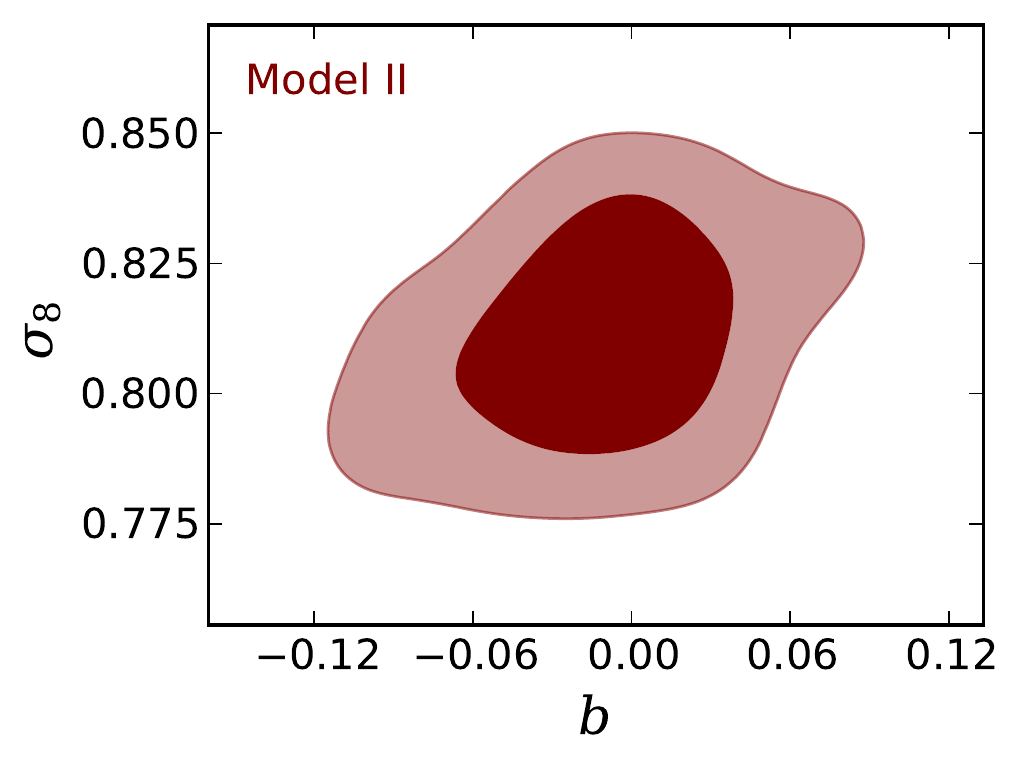}\\
\includegraphics[width=0.24\textwidth]{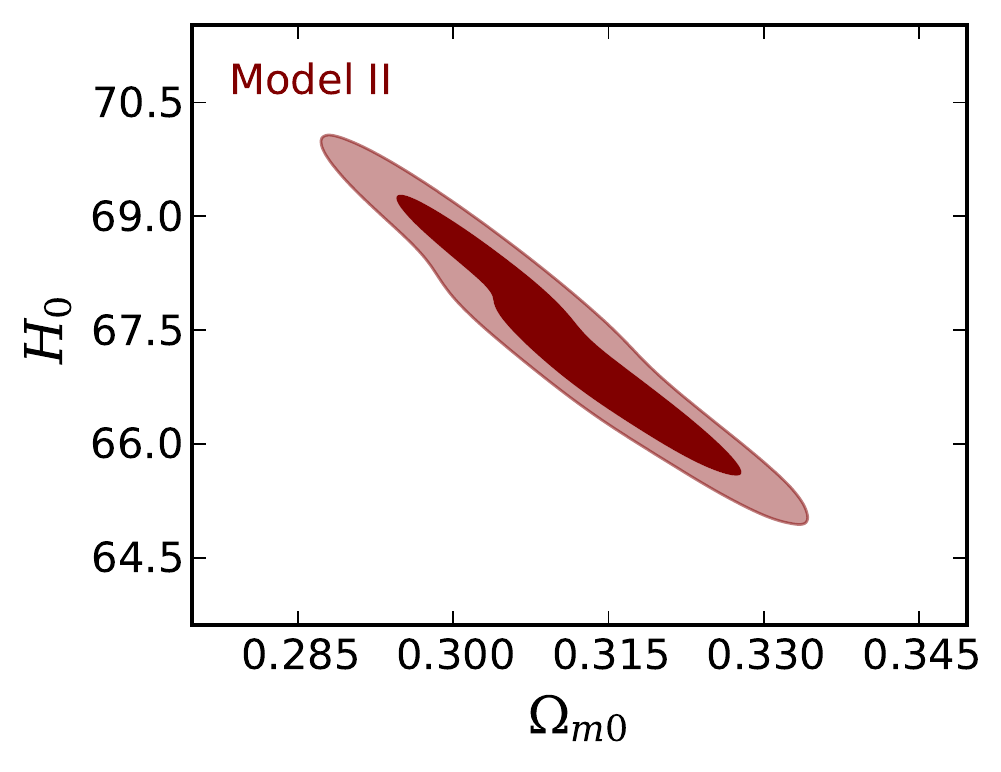}
\includegraphics[width=0.24\textwidth]{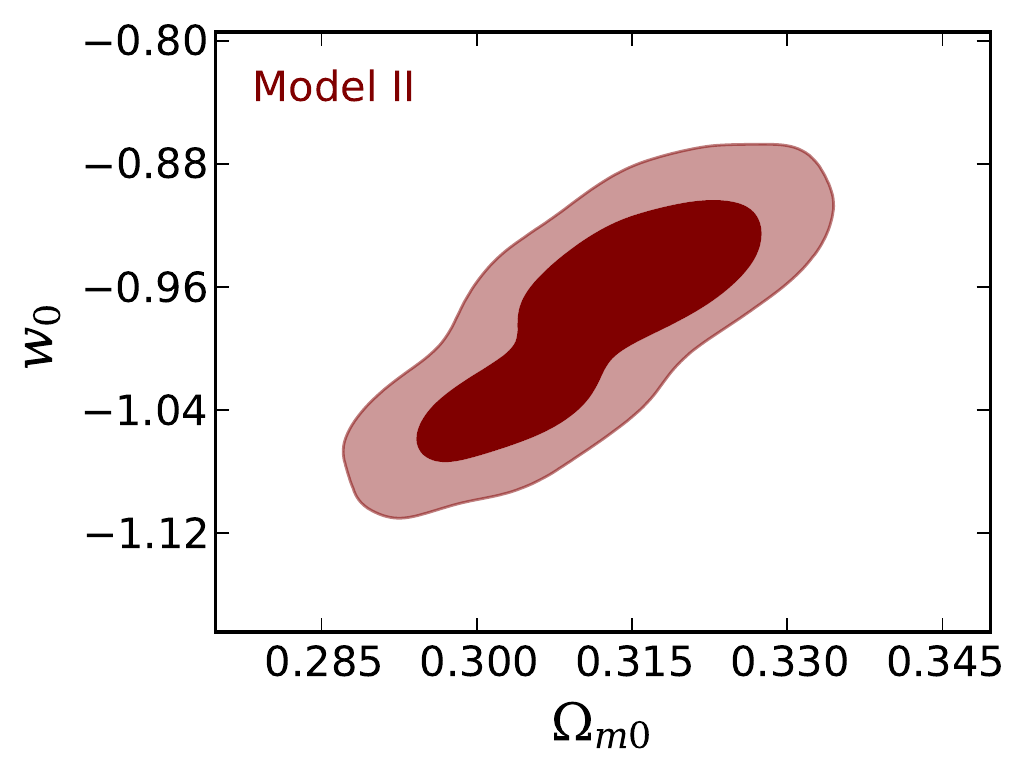}
\includegraphics[width=0.24\textwidth]{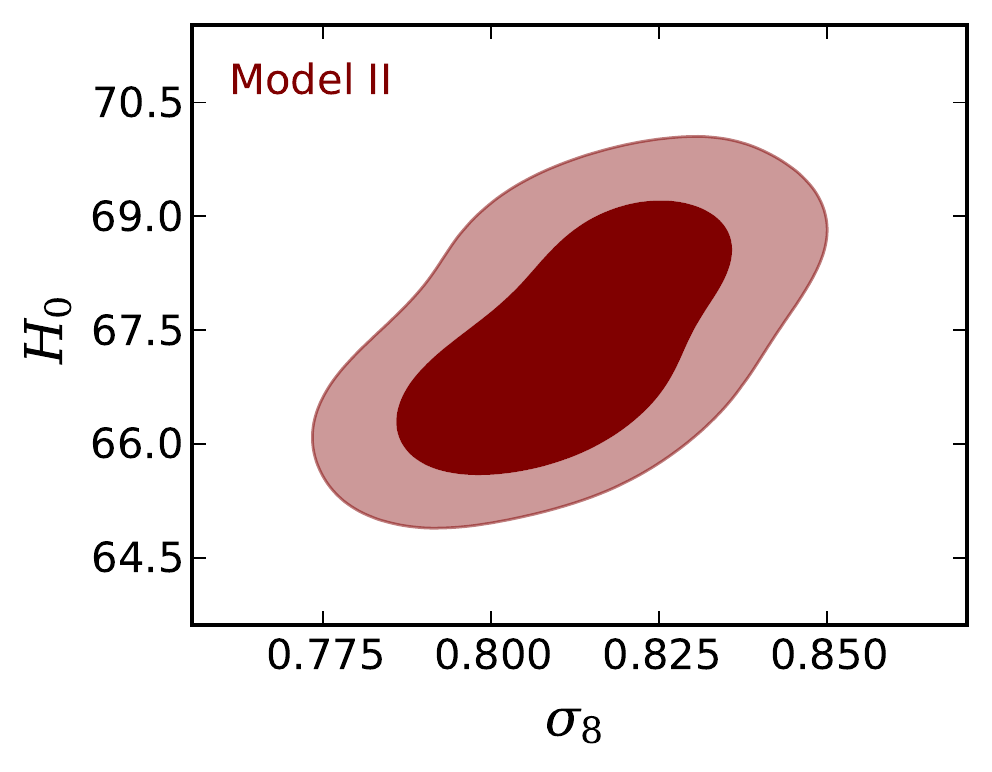}
\includegraphics[width=0.24\textwidth]{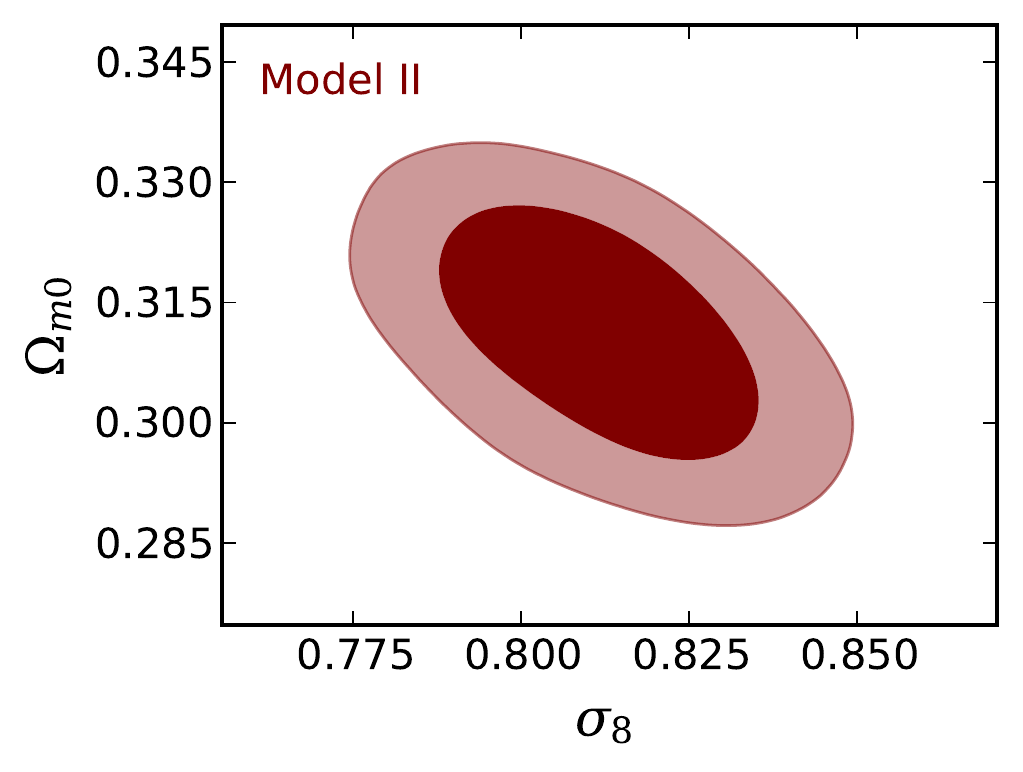}
\caption{\textit{1$\sigma$ (68.3\%) and 2$\sigma$ (95.4\%) confidence level contour plots 
for different combinations of the model parameters of Model II of 
(\ref{model2-current}), for the combined observational data JLA $+$ BAO $+$ Planck TT, 
TE, EE $+$ LowTEB 
$+$ RSD $+$ WL$+$ CC. }}
\label{fig:contour-ModelII}
\end{figure*} 
\begin{figure*}
\includegraphics[width=0.19\textwidth]{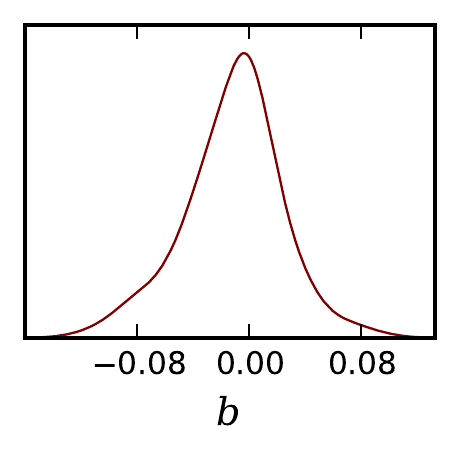}
\includegraphics[width=0.19\textwidth]{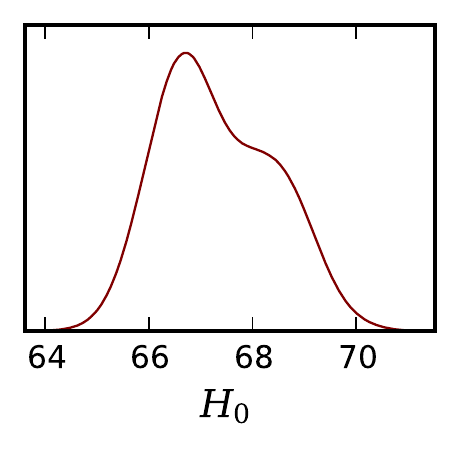}
\includegraphics[width=0.19\textwidth]{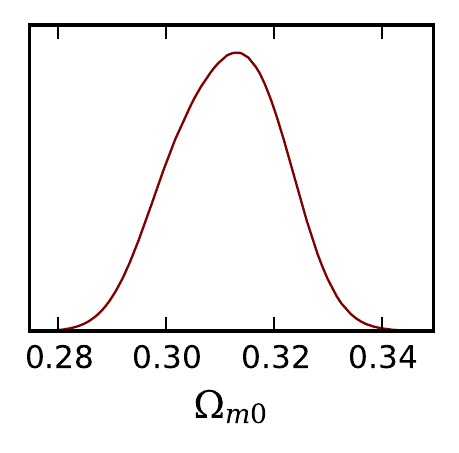}
\includegraphics[width=0.19\textwidth]{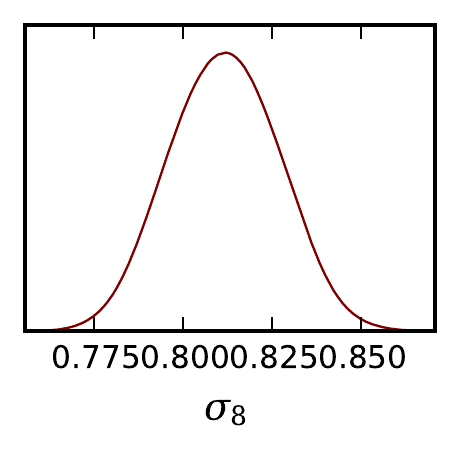}
\includegraphics[width=0.19\textwidth]{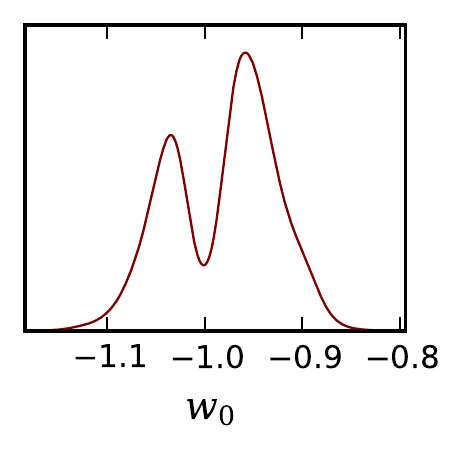}
\caption{\textit{The  marginalized 1-dimensional posterior distributions for the model 
parameters of Model II of (\ref{model2-current}), for the combined observational data JLA 
$+$ BAO $+$ Planck TT, TE, EE $+$ LowTEB 
$+$ RSD $+$ WL$+$ CC.  }}
\label{fig:ModelII-posterior}
\end{figure*} 
\begin{figure*}[!]
\includegraphics[width=0.5\textwidth]{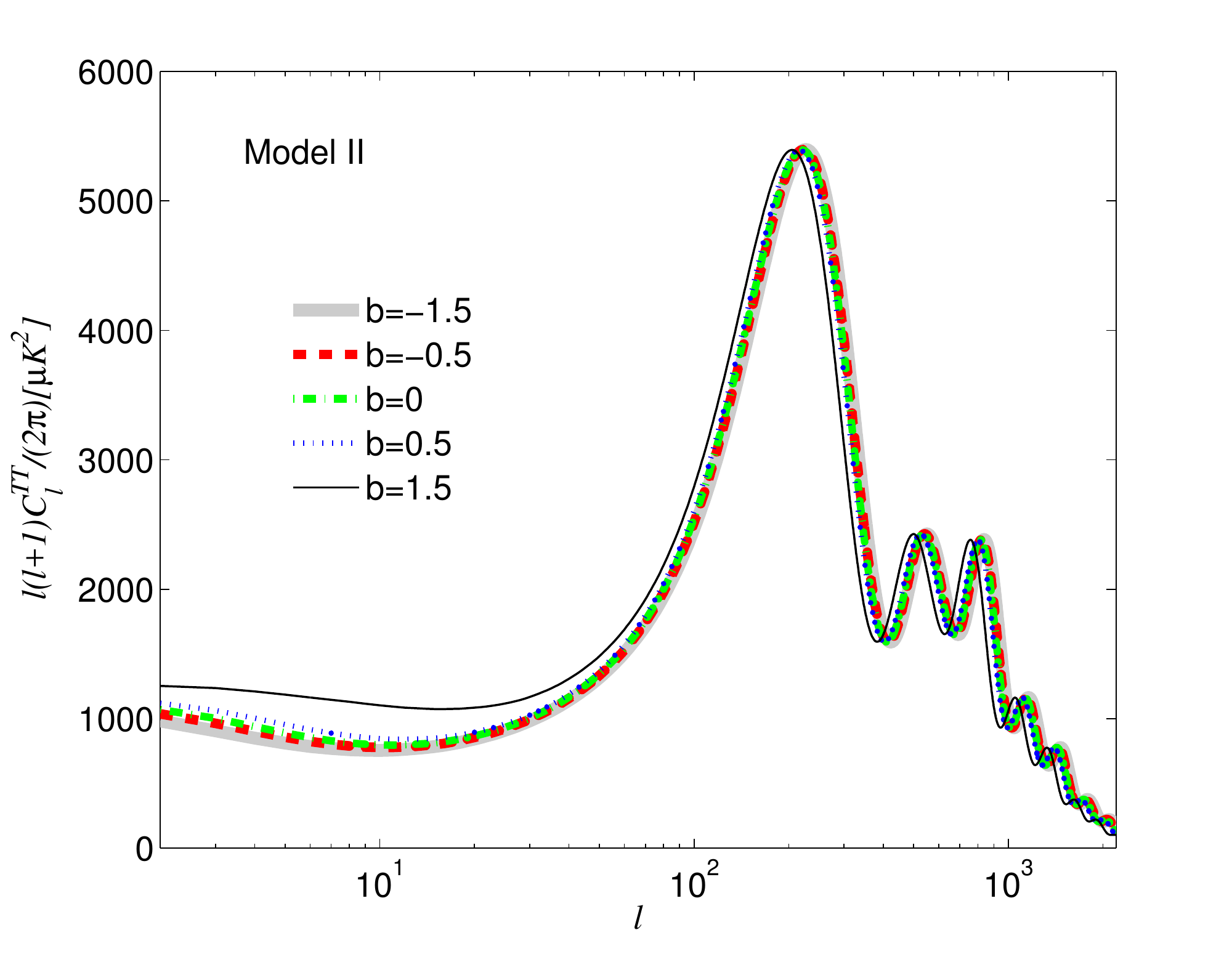}
\includegraphics[width=0.5\textwidth]{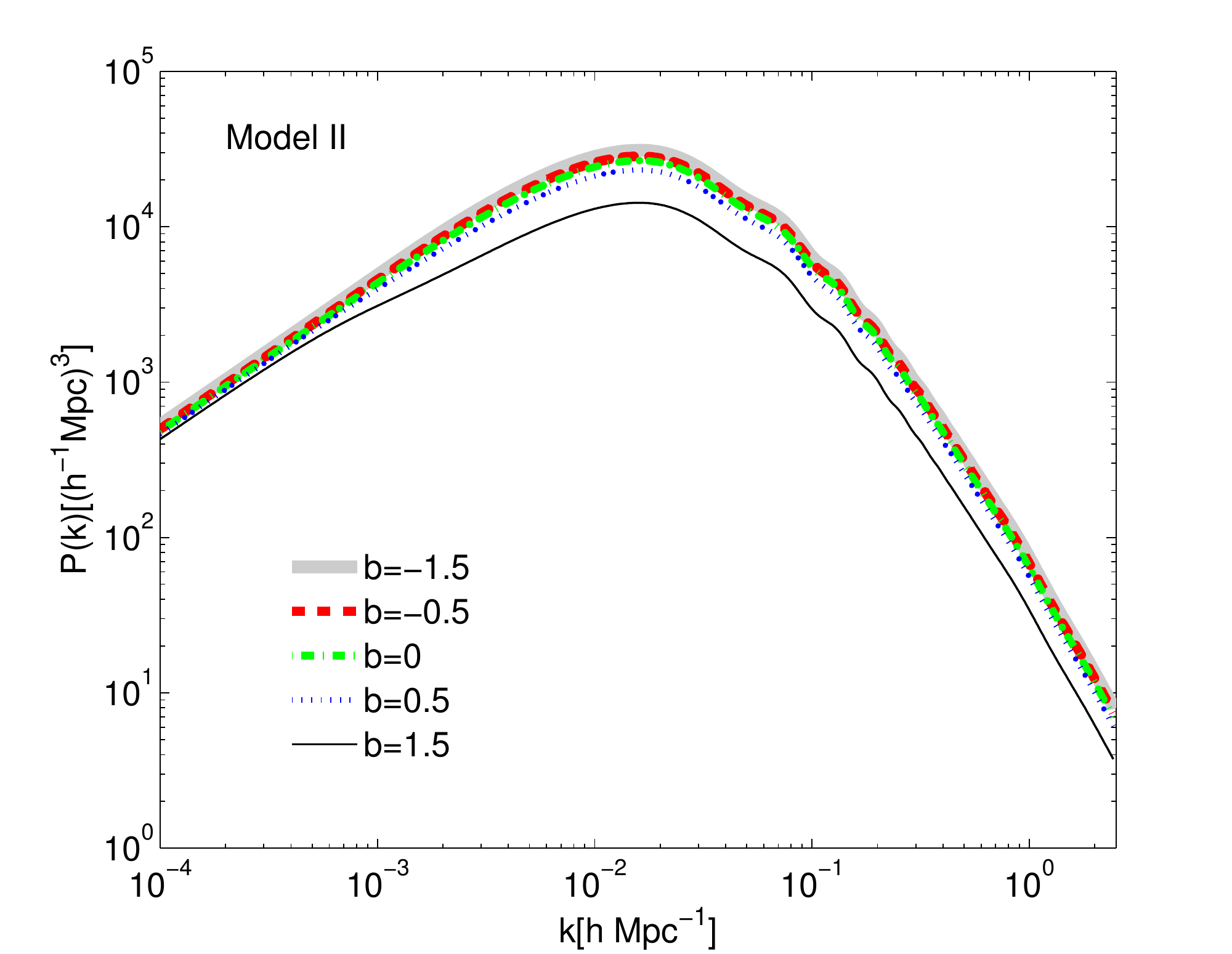}
\caption{\textit{The temperature anisotropy 
in the CMB spectra (left panel) and the matter power spectra (right panel),
for Model II of (\ref{model2-current}), for different values of 
the parameter $b$.  }}
\label{fig:CMB-Matter-ModelII}
\end{figure*}     
We perform the   combined analysis for the Model II of  (\ref{model2-current}), and in 
Table \ref{table-ModelII} we summarize the main observational constraints. Additionally, 
in Fig. \ref{fig:contour-ModelII} we depict the 1$\sigma$ 
and 2$\sigma$ confidence-level contour  plots for several combinations of the model 
parameters and the derived parameters, while in Fig. \ref{fig:ModelII-posterior} we 
display the corresponding 
marginalized one-dimensional posterior distributions for the involved quantities. 
\begin{table}[ht]
\centering
\begin{tabular}{ccc}            
 \hline\hline                                                                             
 Parameters & Mean $\pm$  1$\sigma$ $\pm$ 2$\sigma$ $\pm$ 3$\sigma$ & Best fit 
\\ \hline
$\Omega_c h^2$ & $    0.1180_{-    0.0014-    0.0024-    0.0031}^{+    0.0013+    0.0025+ 
   0.0033}
$ & $    0.1190$\\
$\Omega_b h^2$ & $    0.02230_{-    0.00016-    0.00029-    0.00039}^{+    0.00014+    
0.00030+    
0.00048}$ & $    0.02218$\\
$100\theta_{MC}$ & $    1.04064_{-    0.00033-    0.00062-    0.00081}^{+    0.00032+    
0.00060+   
 0.00079}$ & $    1.04048$\\
$\tau$ & $    0.073_{-    0.0187-    0.0358-    0.0438}^{+    0.0185+    0.0344+    
0.0450}$ & $    
0.060$\\
$n_s$ & $    0.9768_{-    0.0043-    0.0085-    0.0107}^{+    0.0043+    0.0085+    
0.0111}$ & $    
0.9736$\\
${\rm{ln}}(10^{10} A_s)$ & $    3.084_{-    0.035-    0.068-    0.085}^{+    0.035+    
0.066+    0.
087}$ & $    3.062$\\
$w_0$ & $   -0.9817_{-    0.0616-    0.1032-    0.1390}^{+    0.0535+    0.0938+    
0.1175}$ & $   -
1.0444$\\
$b$ & $   -0.0114_{-    0.0319-    0.0809-    0.1071}^{+    0.0378+    0.0739+    
0.1001}$ 
& $   -0.0144$\\
$\Omega_{m0}$ & $    0.311_{-    0.010-    0.019-    0.024}^{+    0.011+    0.019+    
0.023}$ & $   
 0.300$\\
$\sigma_8$ & $    0.812_{-    0.016-    0.029-    0.039}^{+    0.016+    0.029+    
0.039}$ 
& $    0.823$\\
$H_0$ & $   67.32_{-    1.39-    1.95-    2.37}^{+    1.09+    2.22+    2.86}$ & $   
68.74$\\
\hline\hline                                                                       
\end{tabular}                 
\caption{Summary of the observational constraints on Model II of (\ref{model2-current}) 
using the 
observational data JLA $+$ BAO $+$ Planck TT, TE, EE $+$ LowTEB $+$ RSD $+$ WL$+$ CC. }
\label{table-ModelII}                                                                    
\end{table}

The joint analysis on Model II shows that the best-fit 
value of the dark energy equation-of-state parameter  $w_0$  lies in the phantom 
regime, while the mean value of $w_0$ exhibits   quintessential character. We note that 
within $1\sigma$ confidence-region $w_0$ can exhibit phantom behaviour, too. 
However, as we 
can see from Table \ref{table-ModelII}, $w_0$ is close to the cosmological constant 
boundary. Additionally, from the temperature anisotropy 
in the CMB spectra and the matter power spectra depicted in Fig. 
\ref{fig:CMB-Matter-ModelII}, we can observe that at large 
scales this model exhibits a clear deviation from $w$CDM cosmology (and thus 
$\Lambda$CDM cosmology too) for large positive values of the parameter $b$.

\subsection{Model III}

\begin{figure*}
\includegraphics[width=0.24\textwidth]{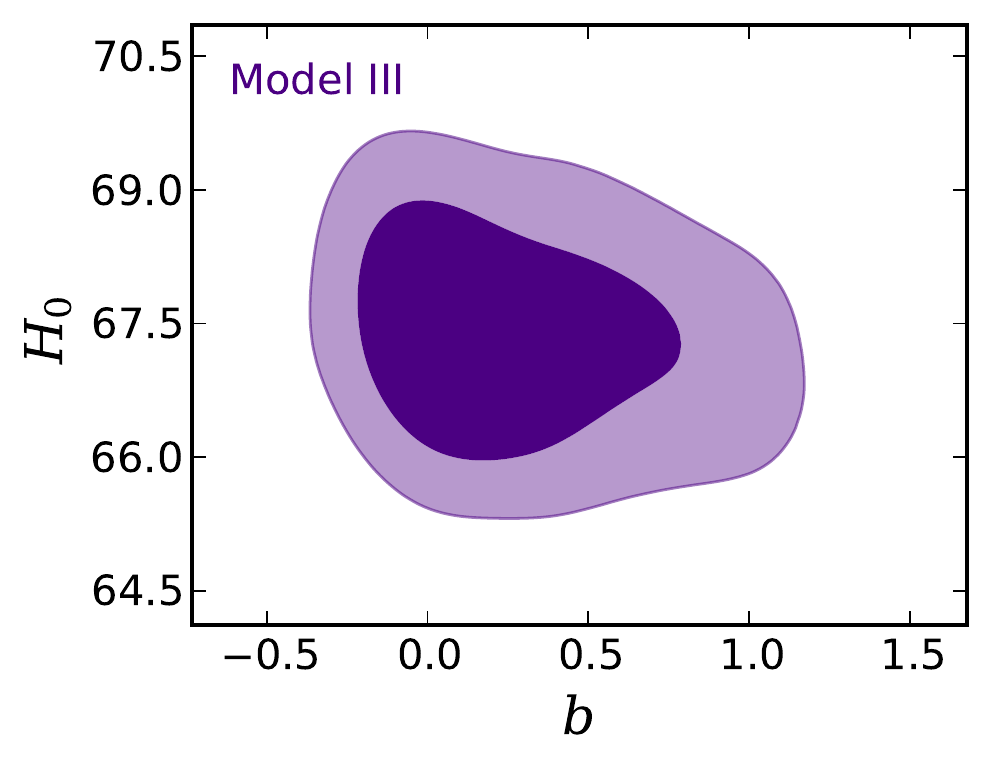}
\includegraphics[width=0.24\textwidth]{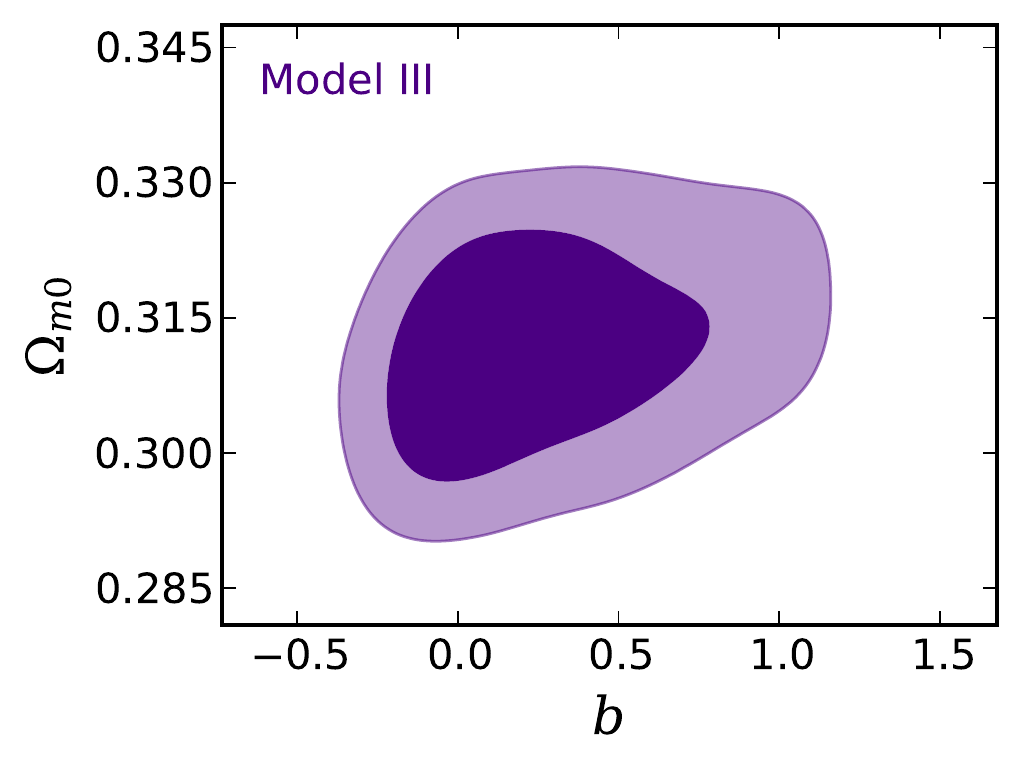}
\includegraphics[width=0.24\textwidth]{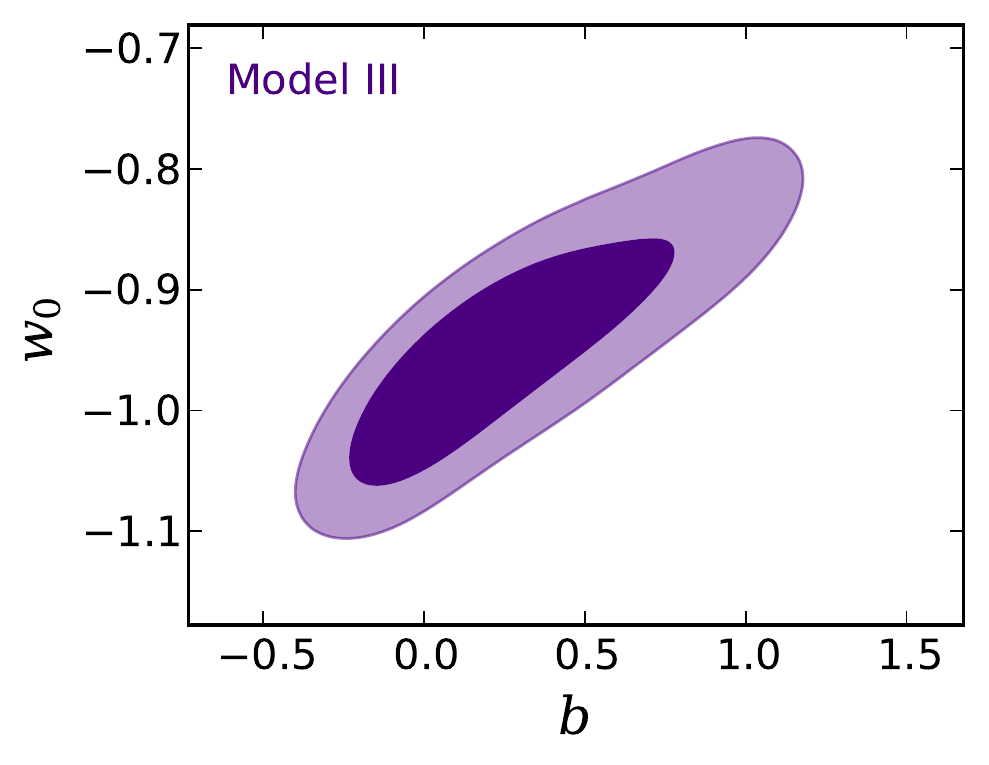}
\includegraphics[width=0.24\textwidth]{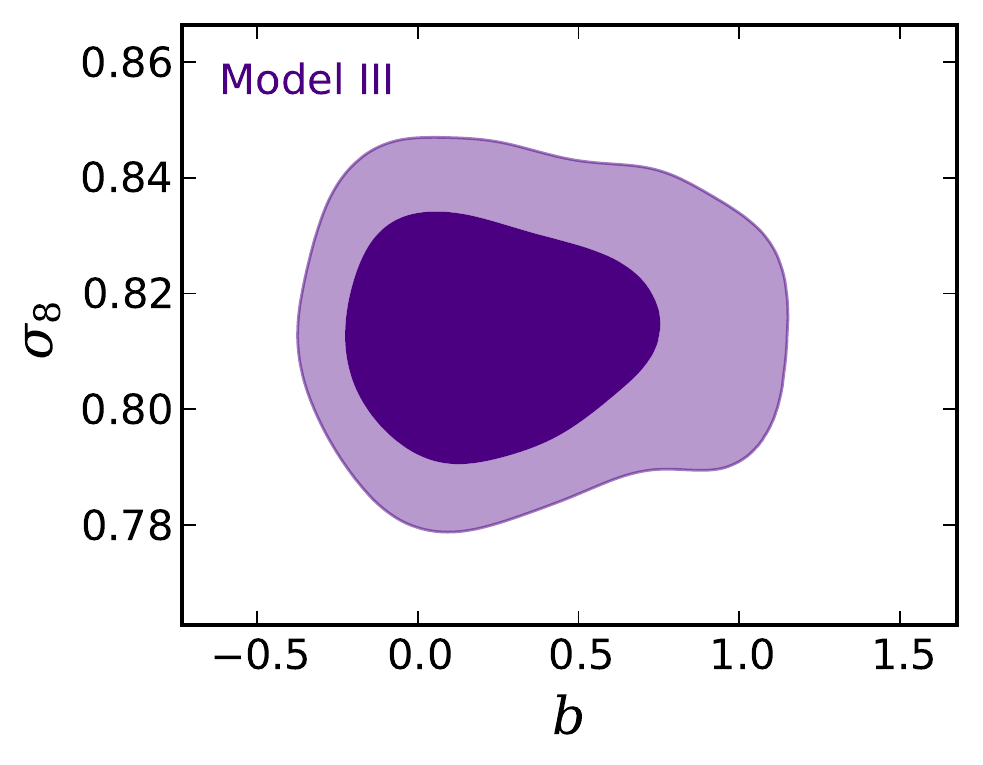}\\
\includegraphics[width=0.24\textwidth]{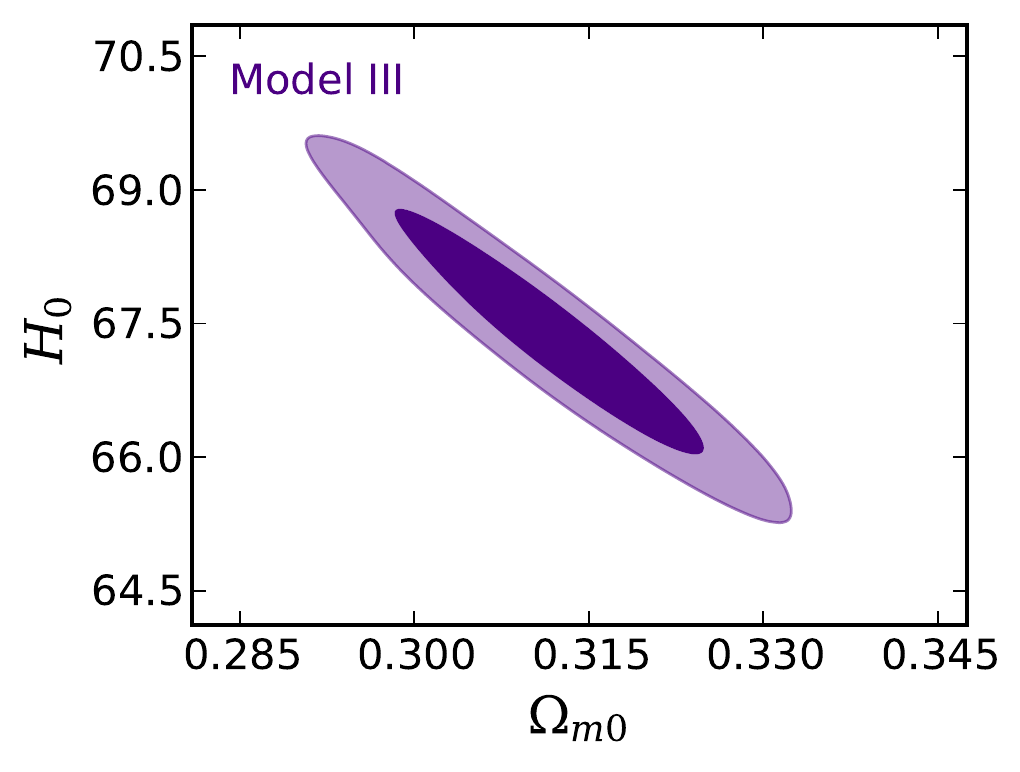}
\includegraphics[width=0.24\textwidth]{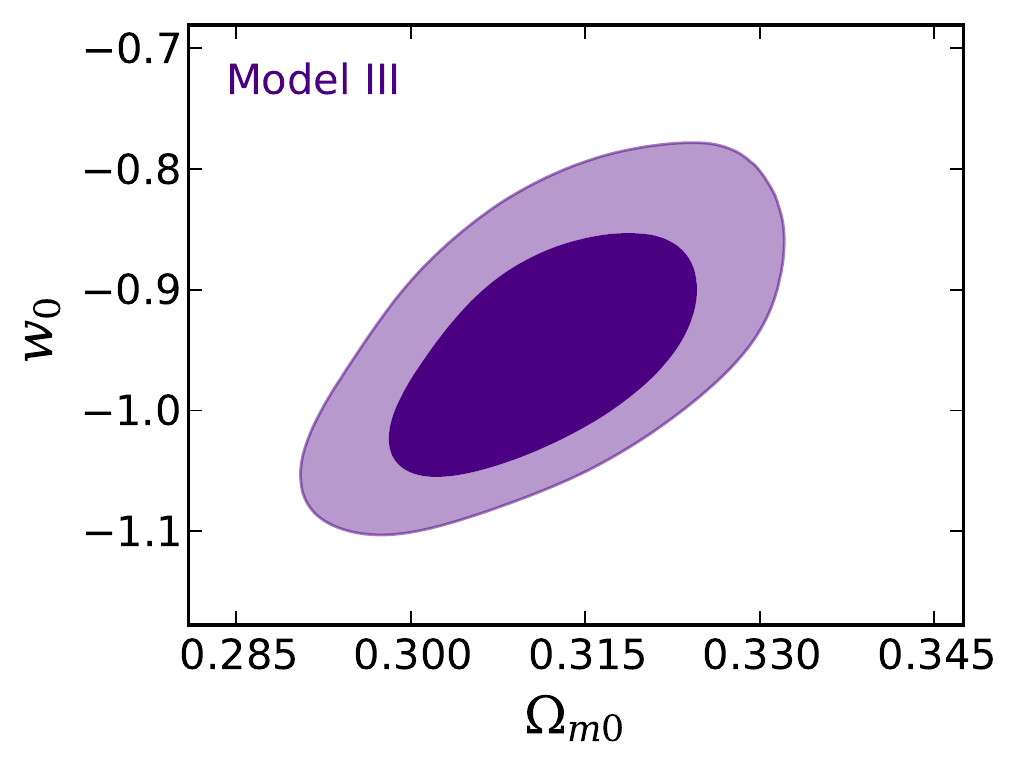}
\includegraphics[width=0.24\textwidth]{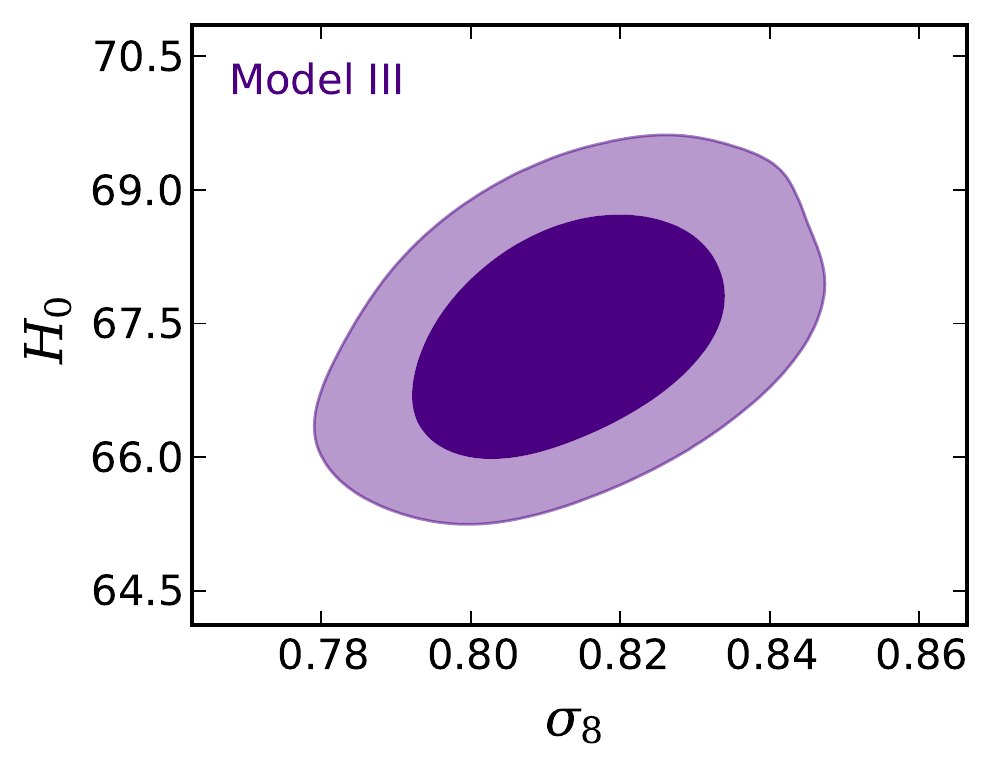}
\includegraphics[width=0.24\textwidth]{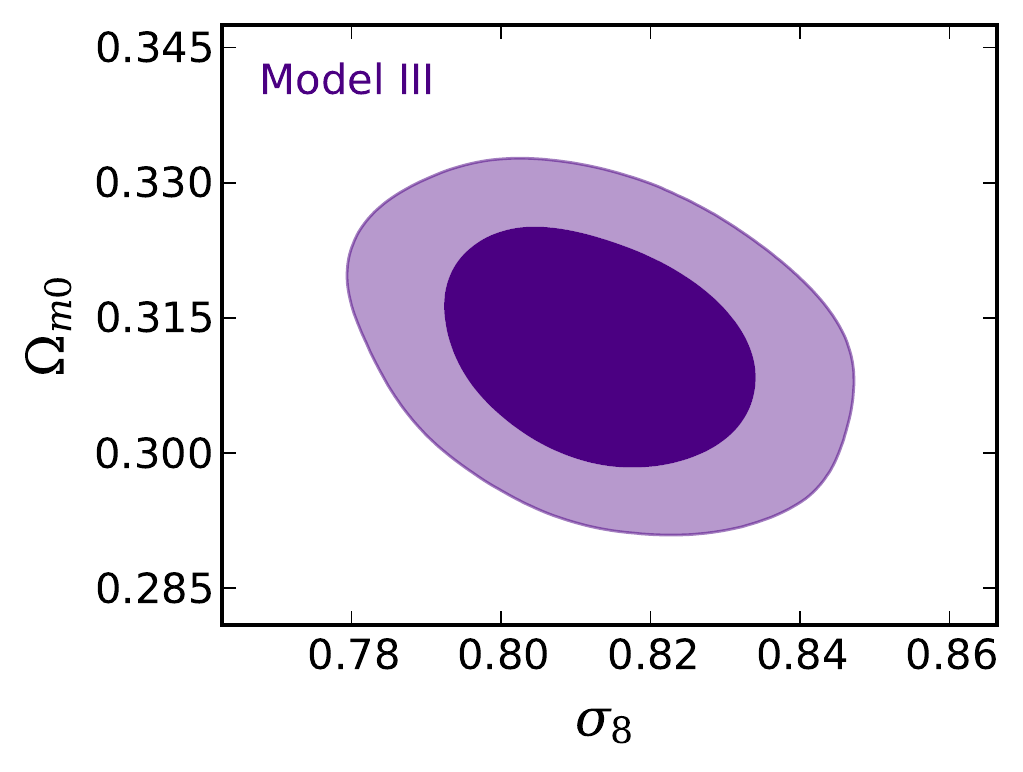}
\caption{\textit{1$\sigma$ (68.3\%) and 2$\sigma$ (95.4\%) confidence level contour plots 
for different combinations of the model parameters of Model III of 
(\ref{model3}), for the combined observational data JLA $+$ BAO $+$ Planck TT, 
TE, EE $+$ LowTEB 
$+$ RSD $+$ WL$+$ CC.   }}
\label{fig:contour-ModelIII}
\end{figure*}
\begin{figure*}
\includegraphics[width=0.19\textwidth]{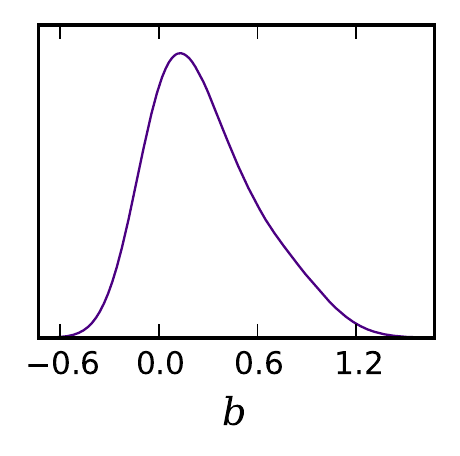}
\includegraphics[width=0.19\textwidth]{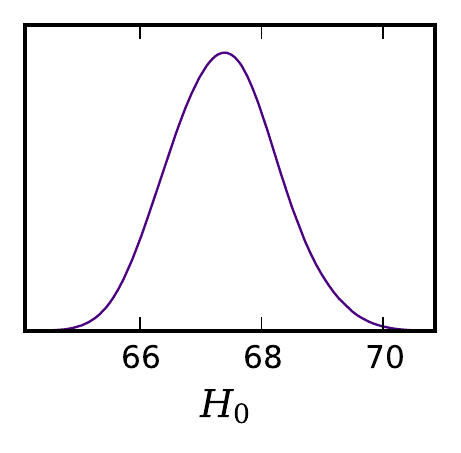}
\includegraphics[width=0.19\textwidth]{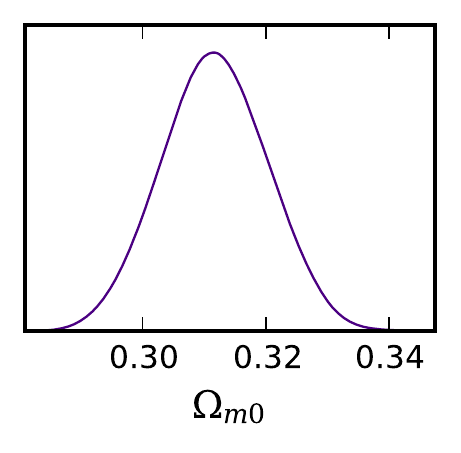}
\includegraphics[width=0.19\textwidth]{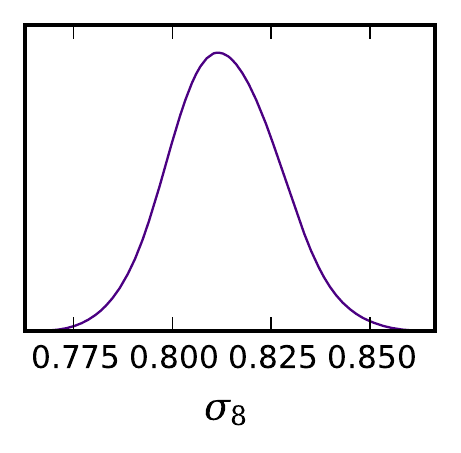}
\includegraphics[width=0.19\textwidth]{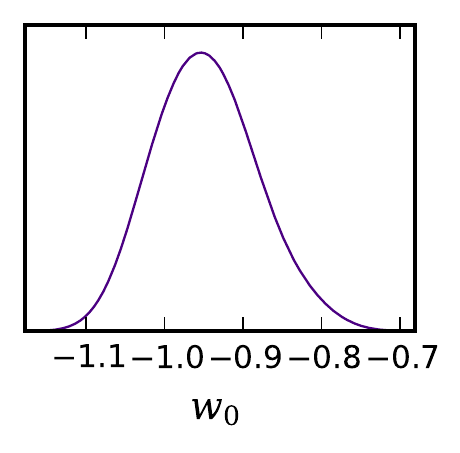}
\caption{\textit{The  marginalized 1-dimensional posterior distributions for the model 
parameters of Model III of (\ref{model3}), for the combined observational data 
JLA 
$+$ BAO $+$ Planck TT, TE, EE $+$ LowTEB 
$+$ RSD $+$ WL$+$ CC.  
}}
\label{fig:ModelIII-posterior}
\end{figure*}
\begin{figure*}[!]
\includegraphics[width=0.5\textwidth]{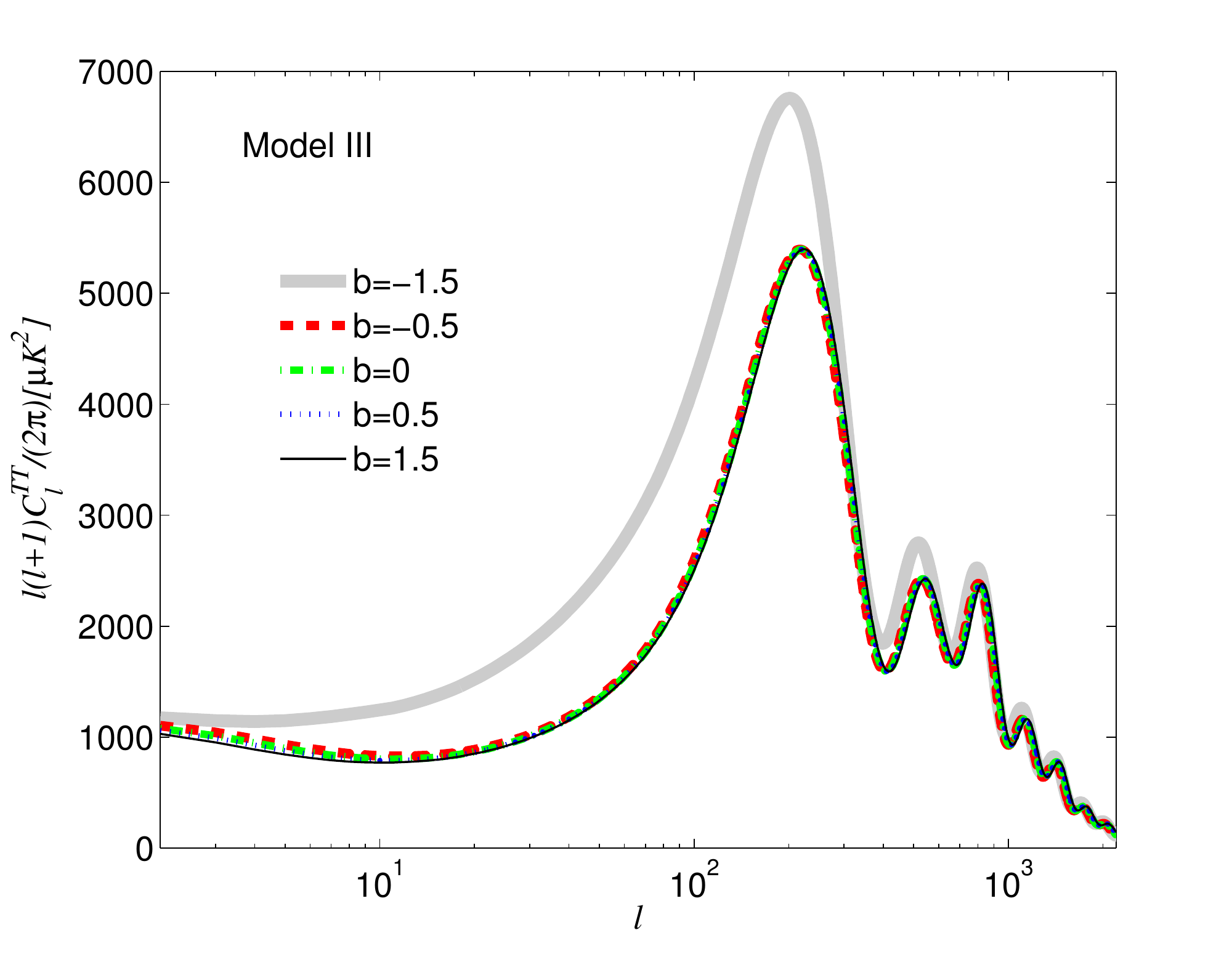}
\includegraphics[width=0.5\textwidth]{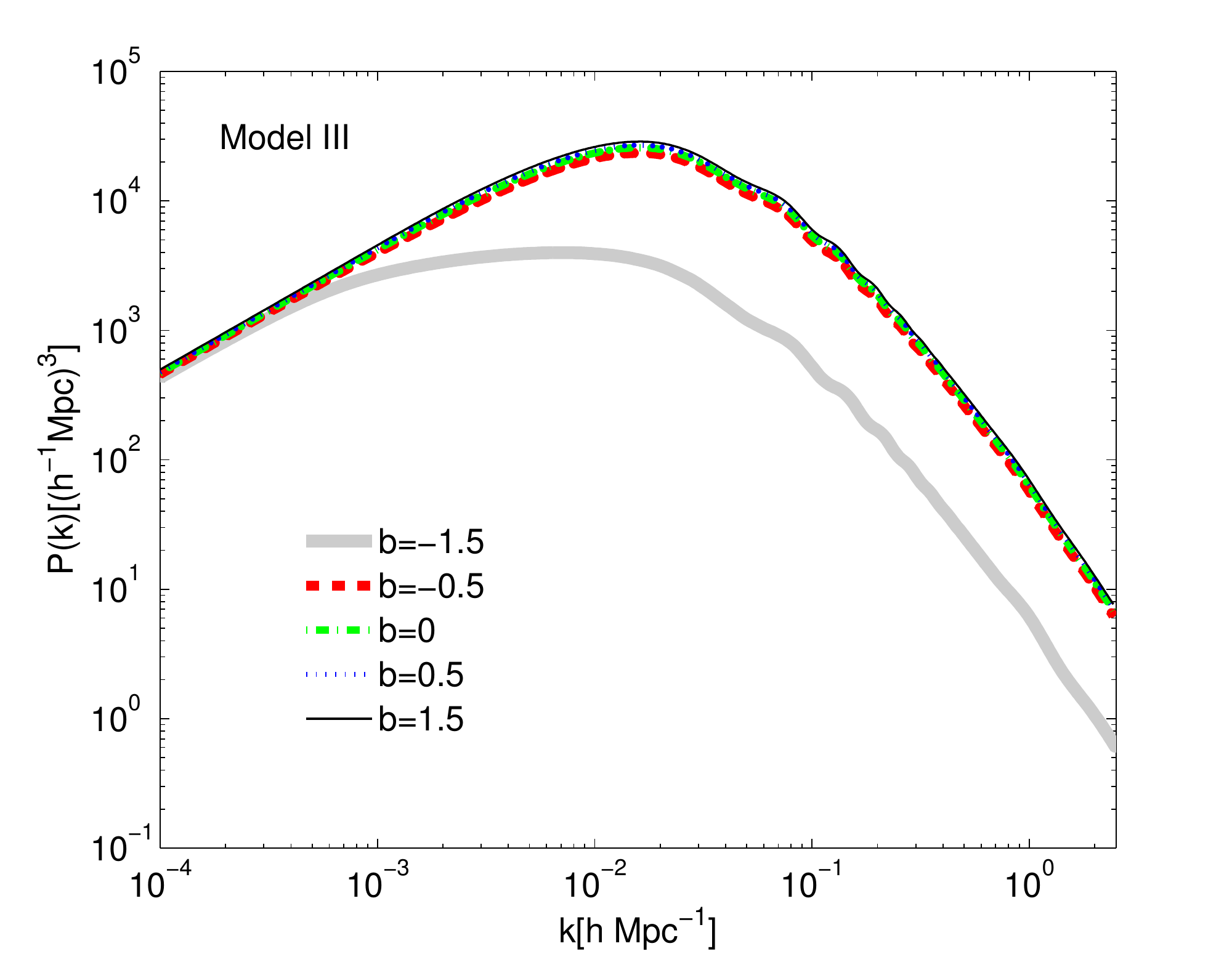}
\caption{\textit{The temperature anisotropy 
in the CMB spectra (left panel) and the matter power spectra (right panel),
for Model III of (\ref{model3}), for different values of 
the parameter $b$.  }}
\label{fig:CMB-Matter-ModelIII}
\end{figure*}   
\begin{table}[ht]
\centering                                                      
\begin{tabular}{ccc}                                                            
\hline\hline                                                                             
 Parameters & Mean $\pm$  1$\sigma$ $\pm$ 2$\sigma$ $\pm$ 3$\sigma$ & Best fit \\ \hline
$\Omega_c h^2$ & $    0.1185_{-    0.0013-    0.0025-    0.0033}^{+    0.0013+    0.0025+ 
   0.0033}
$ & $    0.1182$\\
$\Omega_b h^2$ & $    0.02228_{-    0.00015-    0.00028-    0.00036}^{+    0.00014+    
0.00030+    
0.00038}$ & $    0.02229$\\
$100\theta_{MC}$ & $    1.04060_{-    0.00032-    0.00062-    0.00077}^{+    0.00031+    
0.00062+   
 0.00079}$ & $    1.04065$\\
$\tau$ & $    0.067_{-    0.017-    0.035-    0.047}^{+    0.017+    0.034+    0.045}$ & 
$    0.083$\\
$n_s$ & $    0.9754_{-    0.0042-    0.0083-    0.0105}^{+    0.0042+    0.0080+    
0.0108}$ & $    
0.9777$\\
${\rm{ln}}(10^{10} A_s)$ & $    3.075_{-    0.032-    0.069-    0.093}^{+    0.033+    
0.066+    0.
088}$ & $    3.102$\\
$w_0$ & $   -0.9480_{-    0.0749-    0.1300-    0.1523}^{+    0.0631+    0.1389+    
0.1761}$ & $   -
0.8720$\\
$b$ & $    0.2700_{-    0.4373-    0.5691-    0.6634}^{+    0.2205+    0.7261+    
0.9531}$ 
& $    0.5788$\\
$\Omega_{m0}$ & $    0.312_{-    0.009-    0.017-    0.021}^{+    0.009+    0.017+    
0.021}$ & $   
 0.316$\\
$\sigma_8$ & $    0.813_{-    0.015-    0.027-    0.035}^{+    0.014+    0.027+    
0.037}$ 
& $    0.820$\\
$H_0$ & $   67.38_{-    0.91-    1.77-    2.11}^{+    0.88+    1.80+    2.31}$ & $   
66.85$\\
\hline\hline                                                                          
\end{tabular}
\caption{Summary of the observational constraints on Model III of (\ref{model3}) 
using the 
observational data JLA $+$ BAO $+$ Planck TT, TE, EE $+$ LowTEB $+$ RSD $+$ WL$+$ CC. 
 }\label{table-ModelIII}                                                 
\end{table}

We perform the combined analysis described above, for the Model III of (\ref{model3}), 
and in Table \ref{table-ModelIII} we give the  summary of the main observational 
constraints. Furthermore, in Fig.  \ref{fig:contour-ModelIII} we present the 1$\sigma$ 
and 2$\sigma$ confidence-level contour plots for several combinations of the model 
parameters and of the derived parameters. Additionally,  in Fig. 
\ref{fig:ModelIII-posterior} 
we display the 
corresponding marginalized one-dimensional posterior distributions for the involved 
quantities.  
 
According to the joint analysis we find that for   Model III, both the best fit and the 
mean values of the dark-energy equation-of-state parameter at present  exhibit 
quintessential behaviour. However,  the $1\sigma$ 
lower confidence level may allow for phantom behavior too, although 
only slightly.
Moreover, from the temperature anisotropy in the CMB spectra and the matter 
power spectra depicted in Fig. \ref{fig:CMB-Matter-ModelIII}, we can see that at large 
scales, and for large negative values of $b$ (different from Model I and II), the changes 
in both 
CMB spectra and matter power spectra, are huge. This implies that 
this model might exhibit a non-zero deviation from $w$CDM cosmology and hence from   
$\Lambda$CDM 
cosmology too.  
 However, from Table 
\ref{table-ModelIII} we can see  that $-0.0721 < b$ at  3$\sigma$  confidence level, and 
thus this exotic behaviour is practically not observable. Hence, Model III is 
practically close to 
$w$CDM cosmology, and thus to   $\Lambda$CDM cosmology too.

\subsection{Model IV}

\begin{table}[ht]
\centering           
\begin{tabular}{ccc}  
\hline\hline                          
Parameters & Mean $\pm$  1$\sigma$ $\pm$ 2$\sigma$ $\pm$ 3$\sigma$ & Best fit \\ \hline

$\Omega_c h^2$ & $    0.1179_{-    0.0012-    0.0023-    0.0029}^{+    0.0011+    0.0022+ 
   0.0031}
$ & $    0.1182$\\
$\Omega_b h^2$ & $    0.02233_{-    0.00014-    0.00030-    0.00035}^{+    0.00013+    
0.00029+    
0.00040}$ & $    0.02216$\\
$100\theta_{MC}$ & $    1.04065_{-    0.00035-    0.00056-    0.00074}^{+    0.00030+    
0.00062+   
 0.00081}$ & $    1.04056$\\
$\tau$ & $    0.074_{-    0.017-    0.032-    0.043}^{+    0.017+    0.032+    0.042}$ & 
$    0.056$\\
$n_s$ & $    0.9770_{-    0.0041-    0.0081-    0.0110}^{+    0.0040+    0.0081+    
0.0111}$ & $    
0.9763$\\
${\rm{ln}}(10^{10} A_s)$ & $    3.087_{-    0.032-    0.060-    0.084}^{+    0.033+    
0.064+    0.
082}$ & $    3.046$\\
$w_0$ & $   -0.9807_{-    0.0360-    0.0569-    0.0756}^{+    0.0333+    0.0570+    
0.0761}$ & $   -
1.0382$\\
$b$ & $    0.0544_{-    0.0887-    0.2142-    0.3038}^{+    0.0876+    0.2324+    
0.2798}$ & $    0.
0461$\\
$\Omega_{m0}$ & $    0.312_{-    0.008-    0.017-    0.020}^{+    0.009+    0.016+    
0.020}$ & $   
 0.299$\\
$\sigma_8$ & $    0.811_{-    0.015-    0.028-    0.038}^{+    0.014+    0.029+    
0.038}$ & $    0.
812$\\
$H_0$ & $   67.26_{-    1.02-    1.53-    1.97}^{+    0.83+    1.65+    2.05}$ & $   
68.73$\\
\hline\hline 
\end{tabular}  
\caption{Summary of the observational constraints on Model IV of (\ref{model4}) 
using the 
observational data JLA $+$ BAO $+$ Planck TT, TE, EE $+$ LowTEB $+$ RSD $+$ WL$+$ CC. }
\label{table-ModelIV}                                                    
\end{table}                                                                               
 
\begin{figure*}
\includegraphics[width=0.24\textwidth]{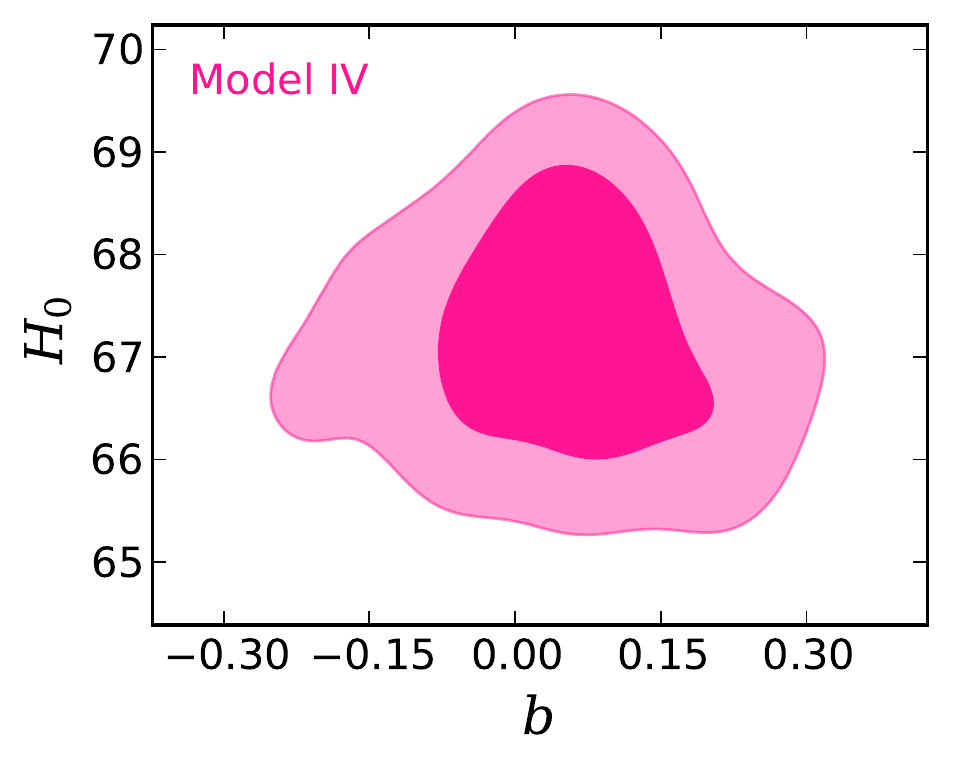}
\includegraphics[width=0.24\textwidth]{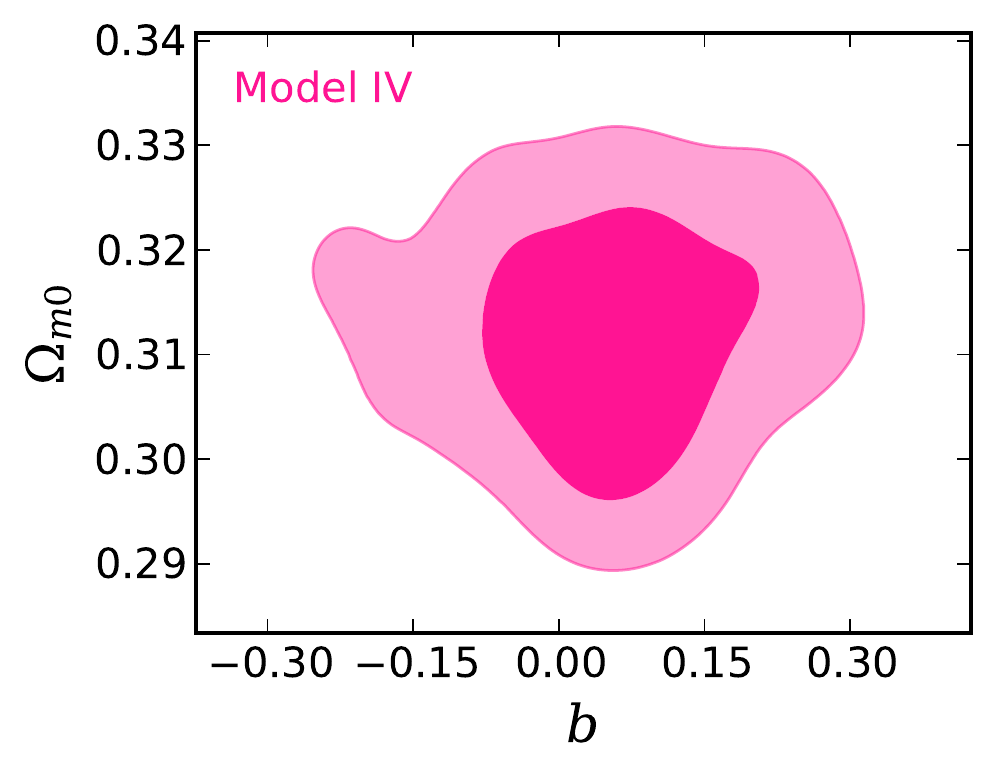}
\includegraphics[width=0.24\textwidth]{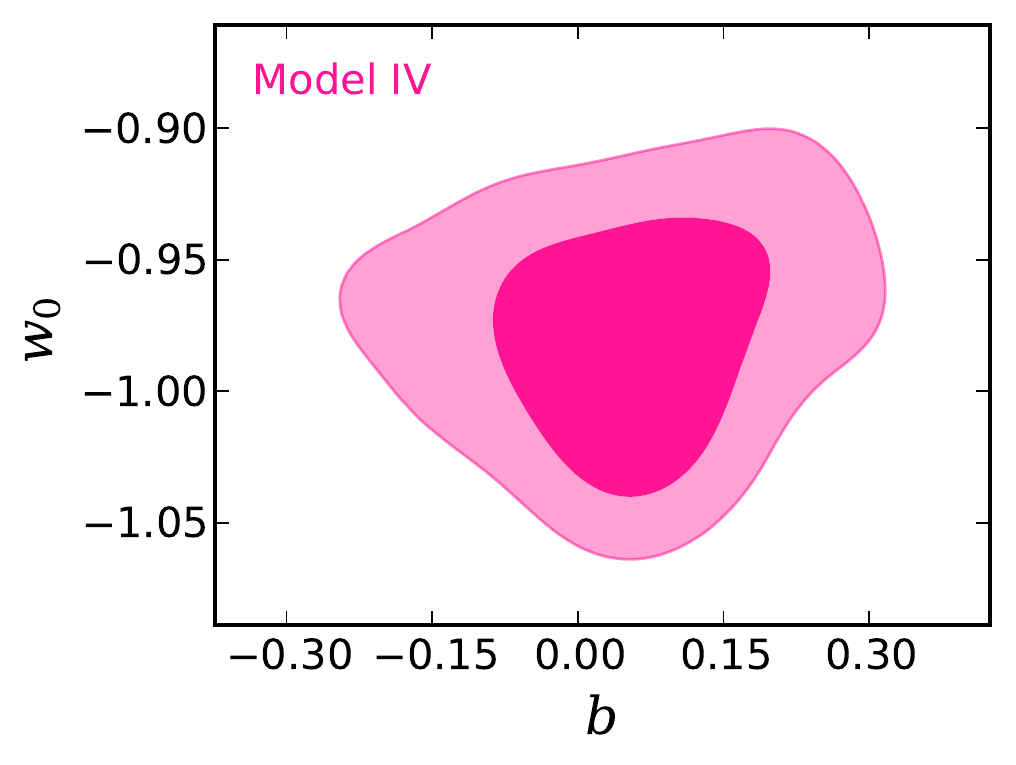}
\includegraphics[width=0.24\textwidth]{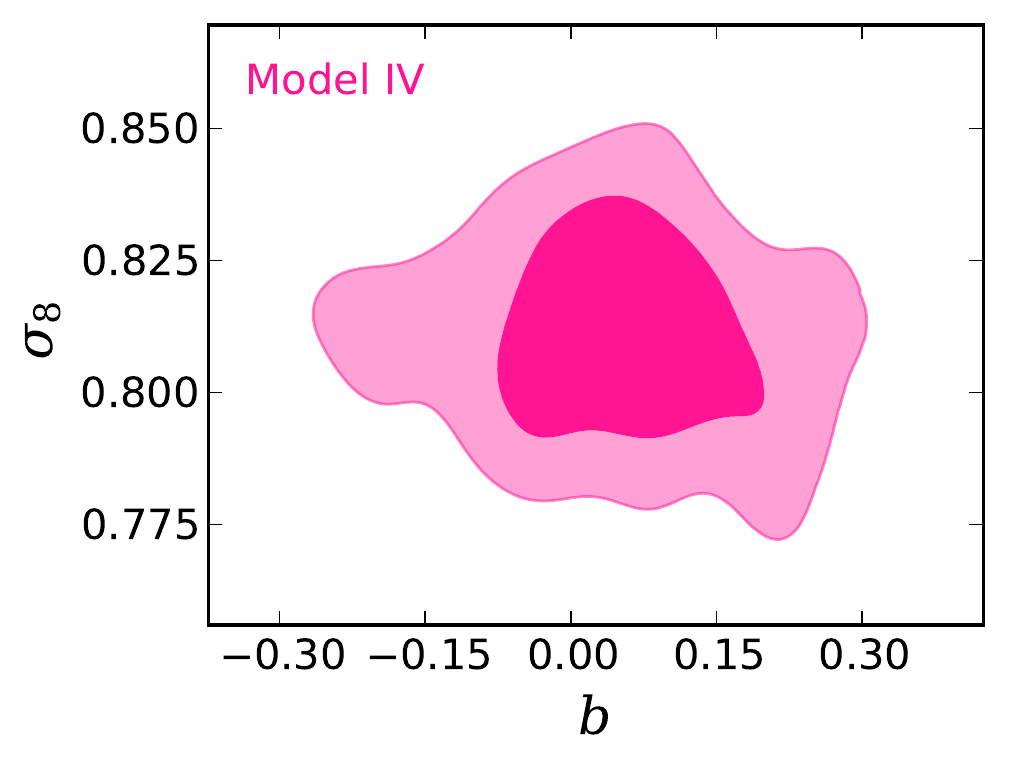}\\
\includegraphics[width=0.24\textwidth]{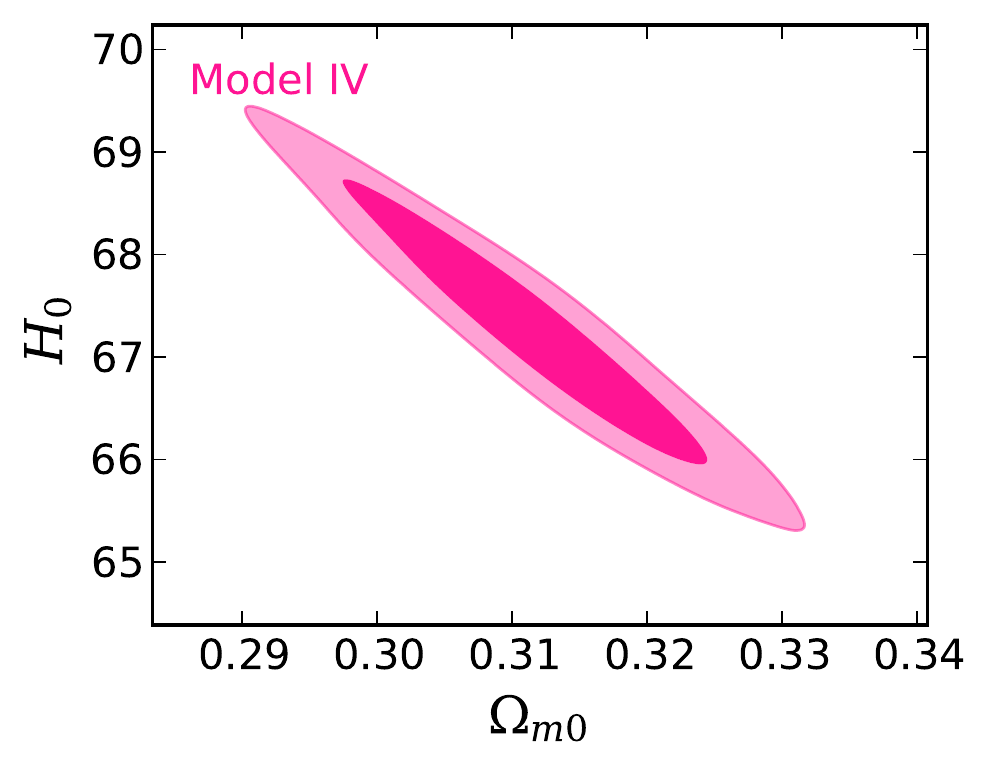}
\includegraphics[width=0.24\textwidth]{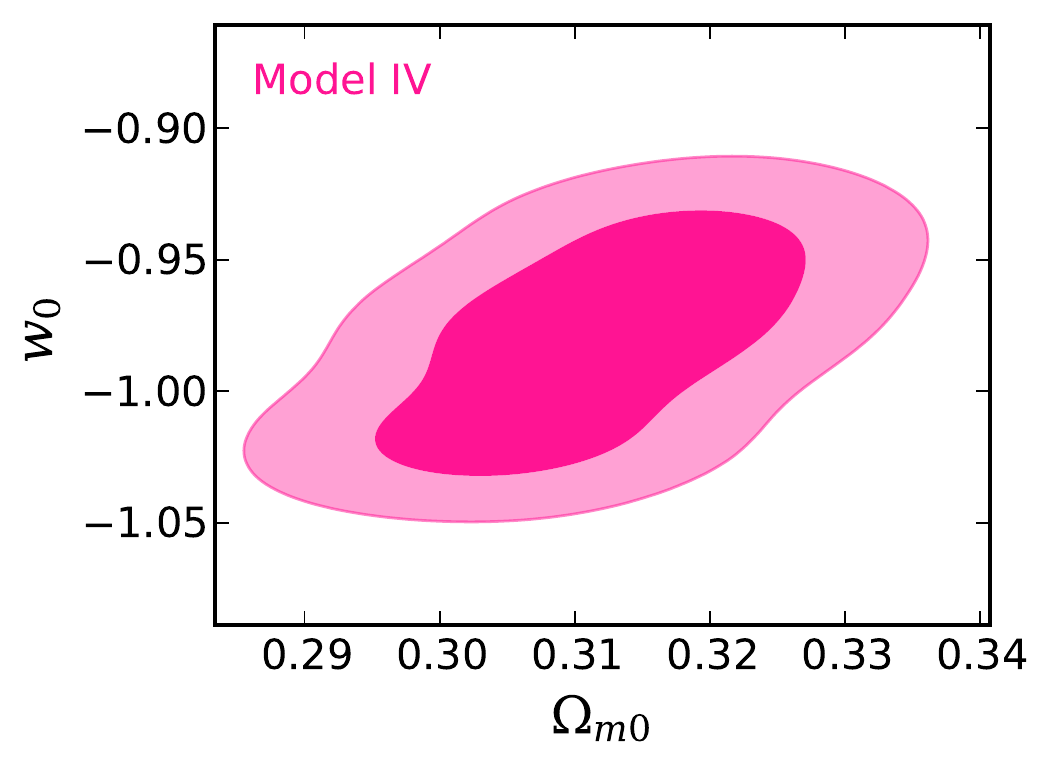}
\includegraphics[width=0.24\textwidth]{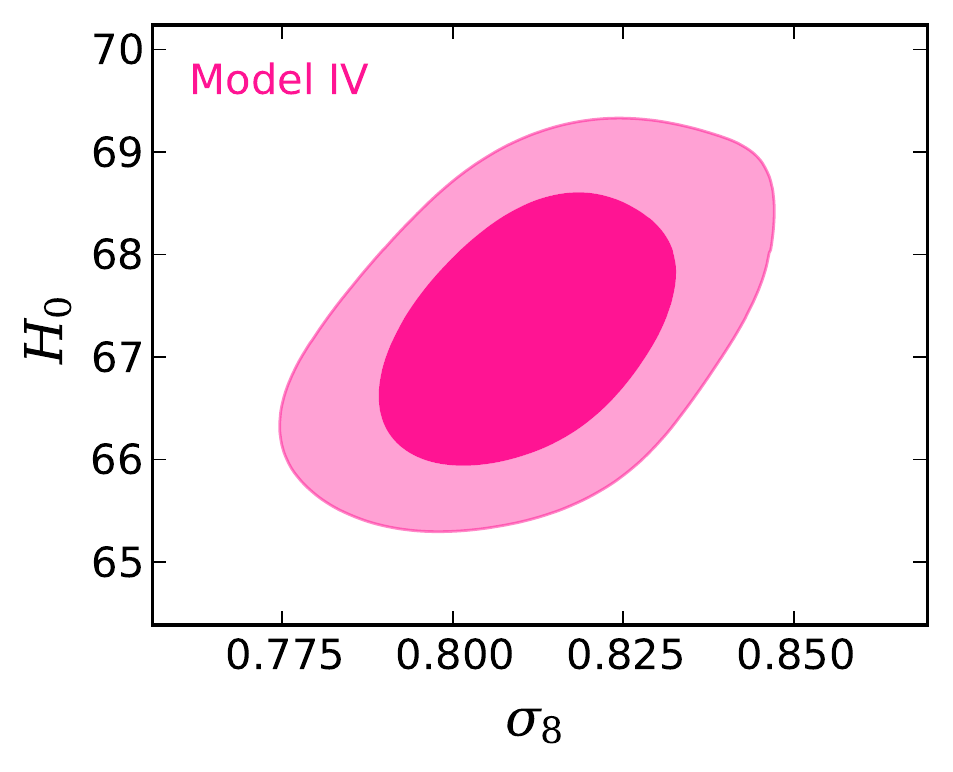}
\includegraphics[width=0.24\textwidth]{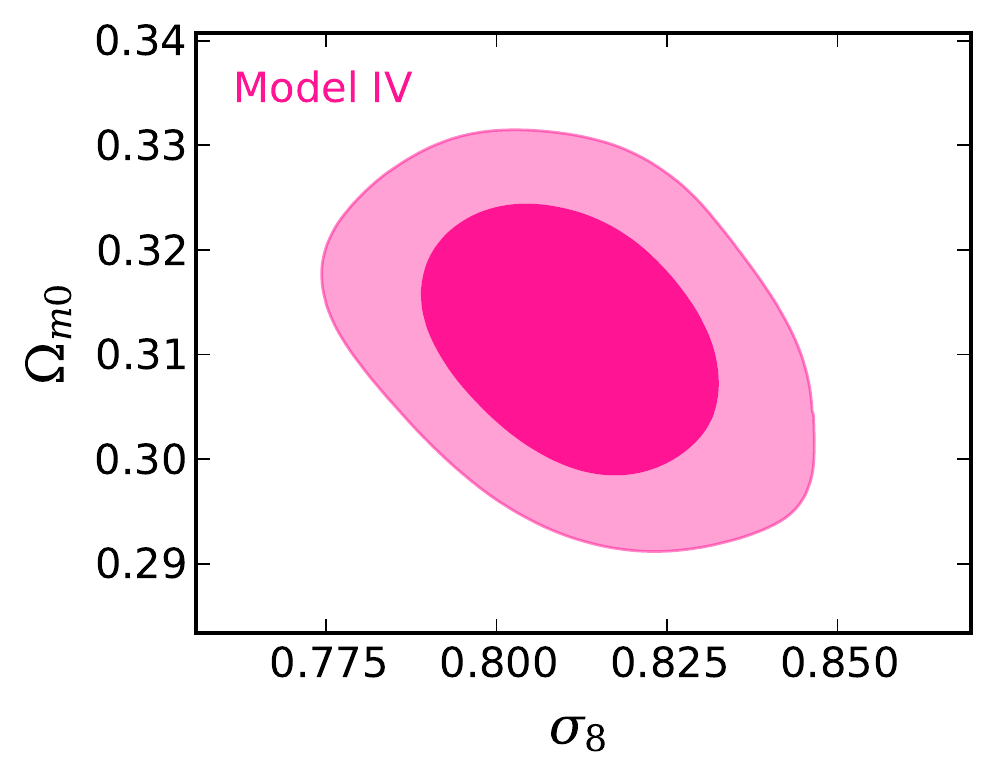}
\caption{\textit{1$\sigma$ (68.3\%) and 2$\sigma$ (95.4\%) confidence level contour plots 
for different combinations of the model parameters of Model IV of 
(\ref{model4}), for the combined observational data JLA $+$ BAO $+$ Planck TT, 
TE, EE $+$ LowTEB 
$+$ RSD $+$ WL$+$ CC. 
}}
\label{fig:contour-ModelIV}
\end{figure*}

\begin{figure*}
\includegraphics[width=0.19\textwidth]{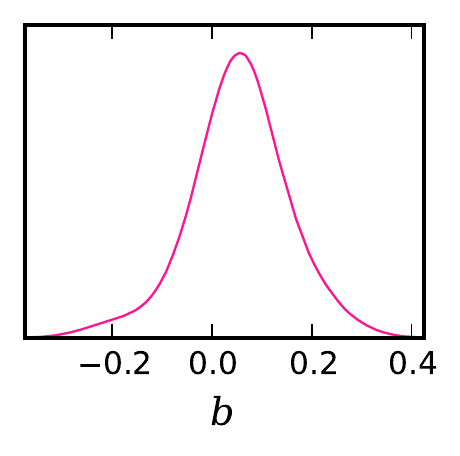}
\includegraphics[width=0.19\textwidth]{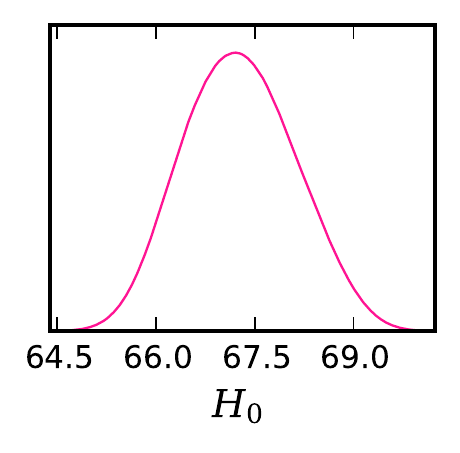}
\includegraphics[width=0.19\textwidth]{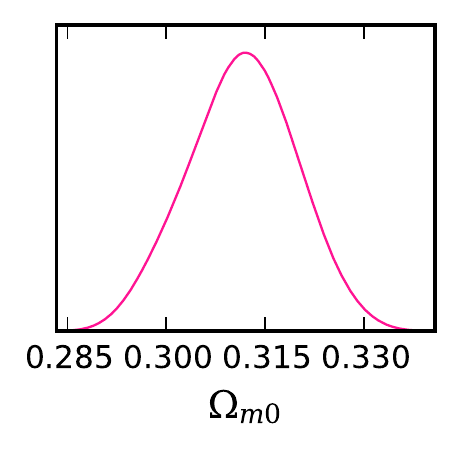}
\includegraphics[width=0.19\textwidth]{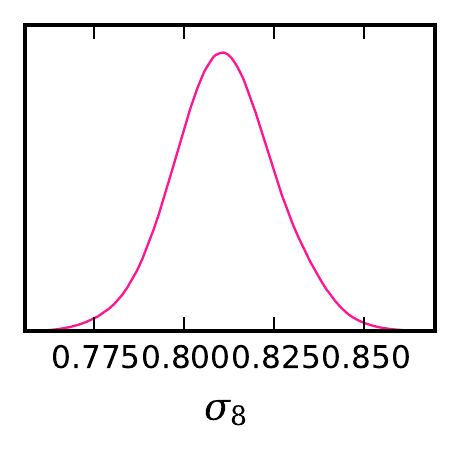}
\includegraphics[width=0.19\textwidth]{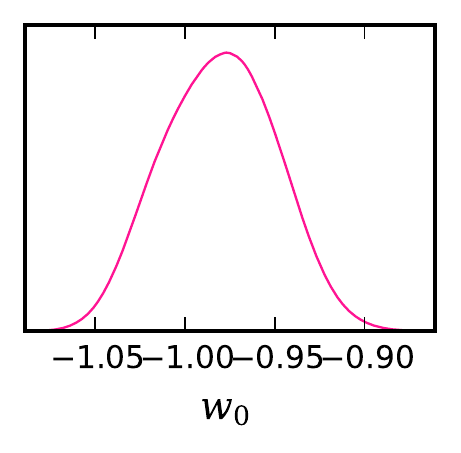}
\caption{\textit{The  marginalized 1-dimensional posterior distributions for the model 
parameters of Model IV of (\ref{model4}), for the combined observational data 
JLA $+$ BAO $+$ Planck TT, TE, EE $+$ LowTEB 
$+$ RSD $+$ WL$+$ CC.   }}
\label{fig:ModelIV-posterior}
\end{figure*}

 Finally, for the Model IV of (\ref{model4}) we perform the joint analysis 
and in Table \ref{table-ModelIV} we display the  summary of the main observational 
constraints. 
Moreover, in Fig. \ref{fig:contour-ModelIV} we show the 1$\sigma$ and 2$\sigma$ 
confidence-level contour plots for several combinations of the model parameters and the 
derived parameters. Additionally, in Fig. \ref{fig:ModelIV-posterior} we present the 
corresponding marginalized one-dimensional posterior distributions for the involved 
quantities.              
\begin{figure*}[!]
\includegraphics[width=0.5\textwidth]{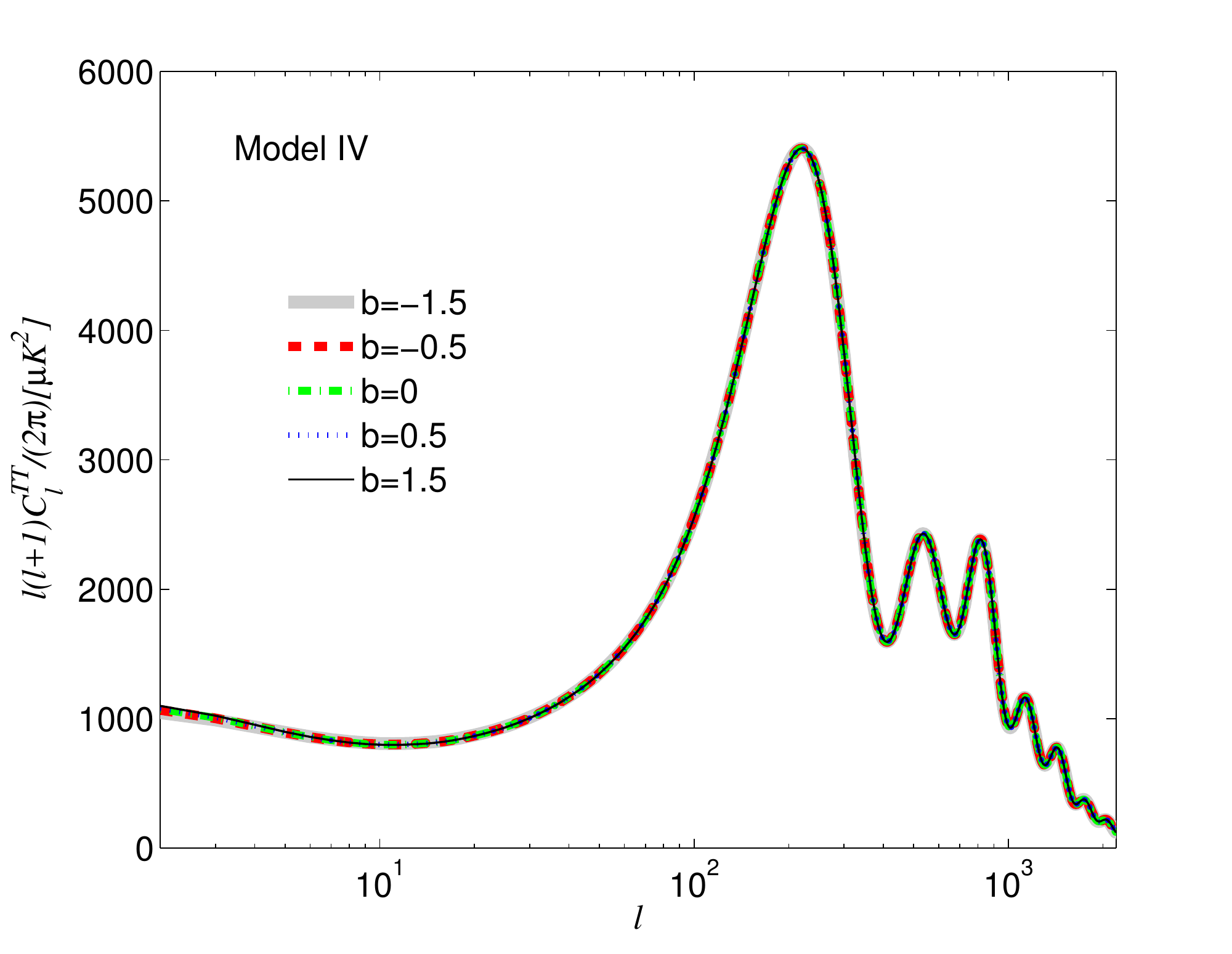}
\includegraphics[width=0.5\textwidth]{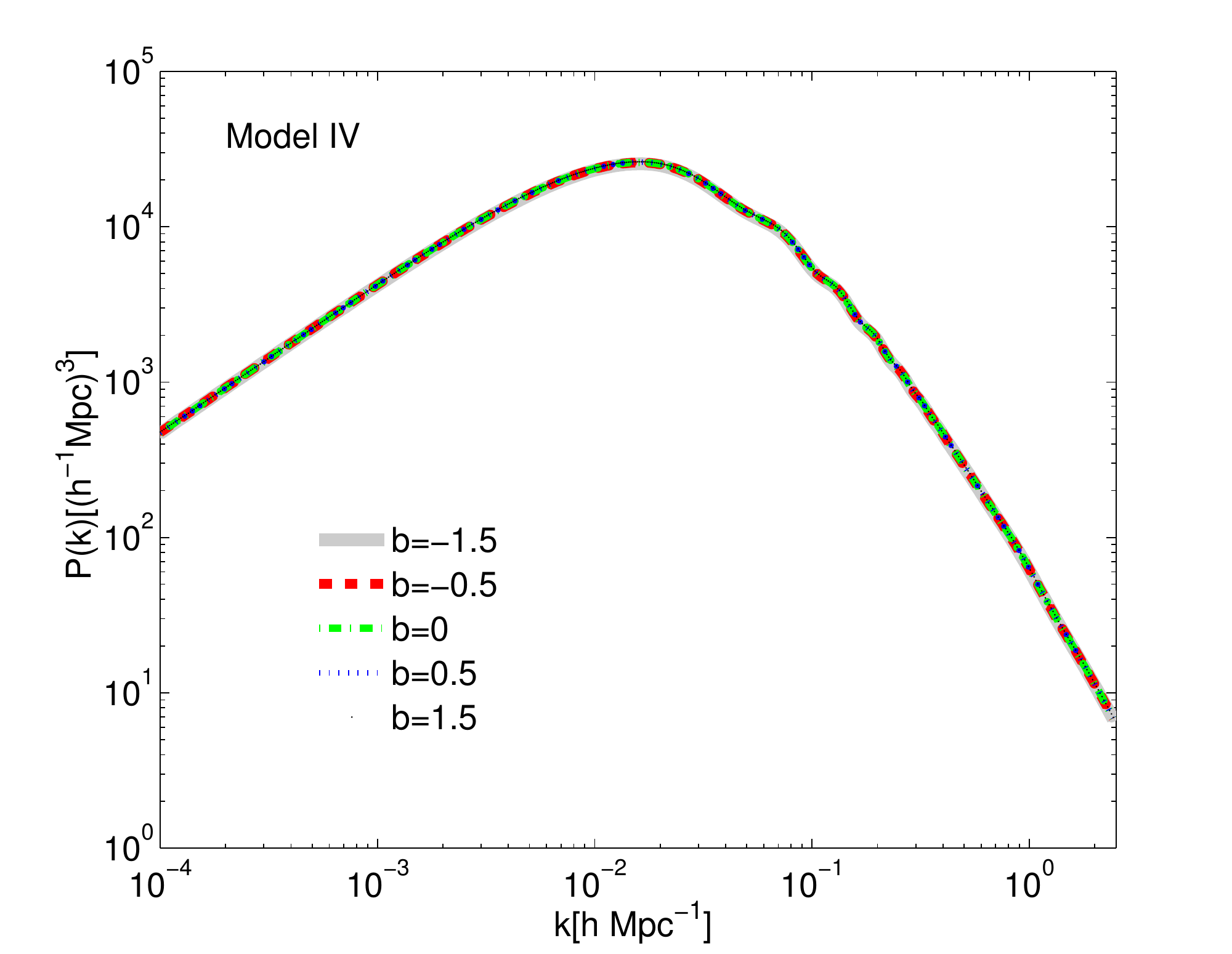}
\caption{\textit{The temperature anisotropy 
in the CMB spectra (left panel) and the matter power spectra (right panel),
for Model IV of (\ref{model4}), for different values of 
the parameter $b$.  
}}
\label{fig:CMB-Matter-ModelIV}
\end{figure*}

As we can observe, the joint analysis reveals  that for   Model IV,  the 
best-fit value 
of the dark-energy equation-of-state parameter at present  exhibits 
phantom behaviour, although very close to $-1$, while the mean value of 
$w_0$ lies in the   quintessence regime. 
However, as one can see from Table \ref{table-ModelIV}, within 1$\sigma$ 
confidence-region the phantom character of $w_0$ is not excluded.
Additionally, from the temperature anisotropy in the CMB spectra and 
the matter power spectra depicted in Fig. \ref{fig:CMB-Matter-ModelIV}, we can see 
that we do not find any significant variation from $w$CDM cosmology, and thus from
$\Lambda$CDM paradigm.
 
\begin{figure}[ht]
\begin{center}
\includegraphics[width=0.7\textwidth]{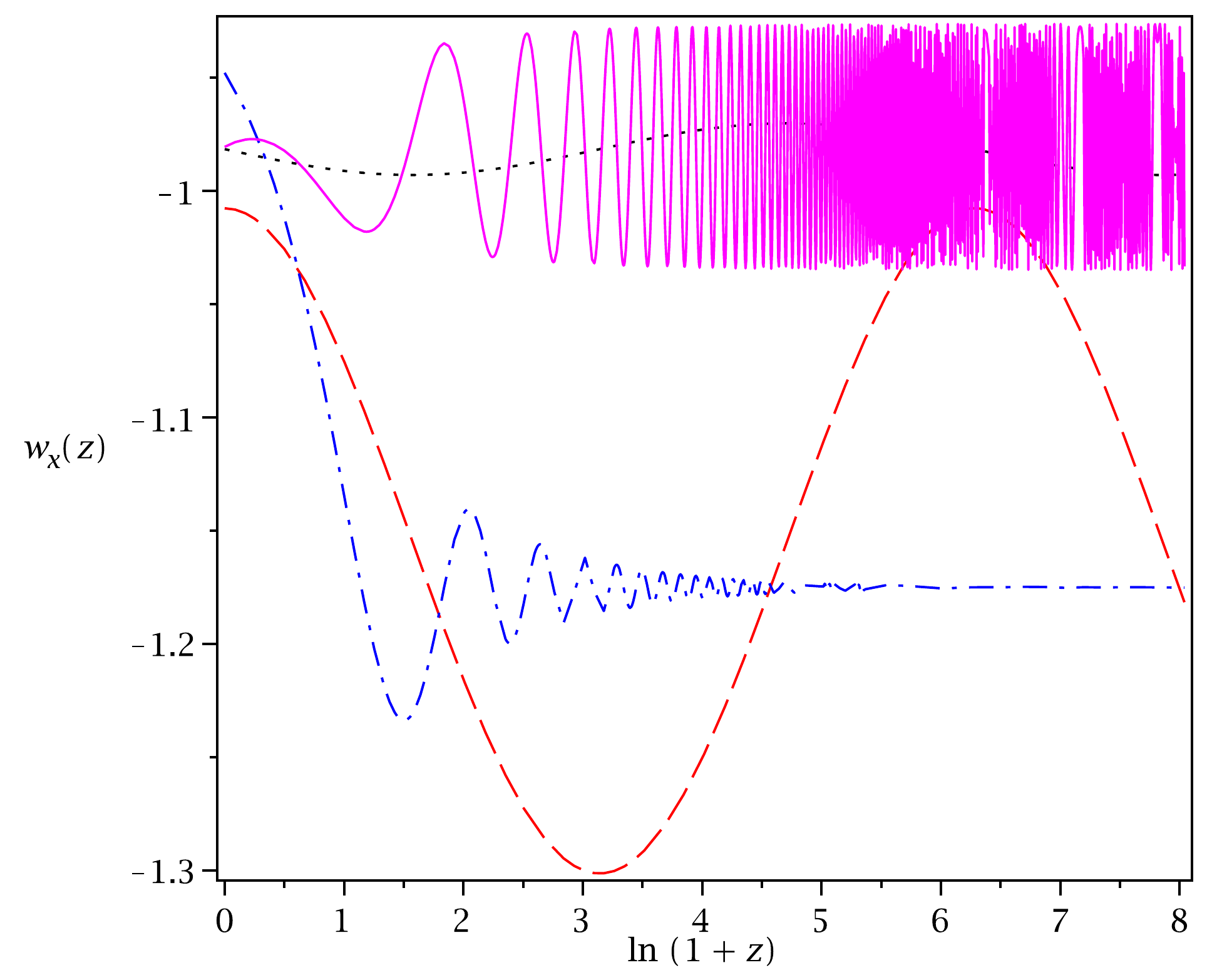}
\end{center}
\caption{\textit{The evolution of the dark-energy equation-of-state parameter $w_x (z)$,
 has been shown using the mean 
values of $(w_0, b)$ that arise from the combined analysis JLA $+$  
BAO $+$ Planck TT, TE, EE $+$ LowTEB $+$ RSD $+$ WL$+$ CC for
Model I of (\ref{model1-current}) (red-dashed), for 
Model II of (\ref{model2-current}) (black-dotted), 
for Model III of (\ref{model3}) (blue-dash-dotted), and for Model IV of (\ref{model4})  
(magenta-solid). }}
\label{fig:eos-all}
\end{figure}

We close this section displaying the evolution of the dark-energy 
equation-of-state parameter of all Models in Fig. \ref{fig:eos-all}. 
This figure shows the 
qualitative differences between the oscillating  dark energy models, both at 
low- and high- redshifts. From the evolutions of all oscillating  dark energy models we 
find 
that Model I gives a phantom  dark energy during the entire evolution of the universe, 
while  Model 
II always exhibits a 
quintessential character. Interestingly, Model III presents a different behaviour: at 
high redshifts it exhibits a phantom behaviour, while very recently the dark energy 
equation of state transits from the phantom regime to quintessence one. Concerning Model 
IV, we find that it exhibits both quintessence and phantom behaviour. In fact, unlike 
other oscillating  dark energy models, it periodically   enters into 
quintessence and phantom region.   

\section{Statistical Model Comparison: Bayesian Evidence}
\label{sec-compare}

In this section we perform a statistical comparison of the oscillating dark 
energy models through the Bayesian evidence, also known as marginal likelihood or  
model likelihood. The Bayesian
evidence plays an important role to compare 
different cosmological models based on their performance with   observational data.   

In the Bayesian analysis we need to determine the posterior probability of the model 
parameters 
(denoted by $\theta$), given a data set $x$, any prior information, and a model $M$. 
In particular, using 
Bayes theorem, one can write 
\begin{eqnarray}\label{BE}
p(\theta|x, M) = \frac{p(x|\theta, M)\,\pi(\theta|M)}{p(x|M)},
\end{eqnarray}
where $p(x|\theta, M)$ is the likelihood which is considered to be function of 
the model 
parameters $\theta$ with the data set fixed, and $\pi(\theta|M)$ is the prior. The 
denominator $p(x|M)
$ in  (\ref{BE}) 
is the Bayesian evidence used for the model comparison, and it is the integral 
over the unnormalised posterior $\tilde{p} (\theta|x, M) \equiv 
p(x|\theta,M)\,\pi(\theta|M)$:

\begin{eqnarray}
E \equiv p(x|M) = \int d\theta\, p(x|\theta,M)\,\pi(\theta|M),
\end{eqnarray}
and thus it is also referred to as the marginal likelihood. Considering 
any particular model $M_i$  and the reference model $M_j$,  
the posterior probability is thus given by the product of the ratio of the model priors 
with
the ratio of evidences, namely 
\begin{eqnarray}
\frac{p(M_i|x)}{p(M_j|x)} = \frac{\pi(M_i)}{\pi(M_j)}\,\frac{p(x| M_i)}{p(x|M_j)} = 
\frac{\pi(M_i)}{
\pi(M_j)}\, B_{ij},
\end{eqnarray}
where $B_{ij} = \frac{p(x| M_i)}{p(x|M_j)}$ is called the Bayes factor of the 
model $M_i$ 
relative to the reference model $M_j$.

\begin{table}[!h]     
\centering   
\begin{tabular}{cc}                 
\hline\hline               
$\ln B_{ij}$ & Strength of evidence for model ${M}_i$ \\ \hline
$0 \leq \ln B_{ij} < 1$ & Weak \\
$1 \leq \ln B_{ij} < 3$ & Definite/Positive \\
$3 \leq \ln B_{ij} < 5$ & Strong \\
$\ln B_{ij} \geq 5$ & Very strong \\
\hline\hline              
\end{tabular}             
\caption{Summary of the revised Jeffreys scale, to quantify the observational 
support 
of   
model $M_i$ with respect to model $M_j$. }\label{tab:jeffreys}         
\end{table}                                   
\begin{table}[!h]    
\centering  
\begin{tabular}{cccc}                              
\hline\hline                              
Model & $\ln B_{ij}$ &~ Strength of evidence for model $\Lambda$CDM \\ \hline
Model I &    $-9.4$  & Very Strong \\
Model II &   $-4.6$  & Strong \\
Model III &  $-6.2$  & Very Strong  \\
Model IV  &  $-8.4$  & Very Strong \\
\hline\hline                                             
\end{tabular}                    
\caption{Summary of the values of $\ln B_{ij}$, calculated for the 
oscillating 
dark energy 
models with respect to the reference $\Lambda$CDM paradigm.  
According to the Bayesian point of view, the negative values of $\ln B_{ij}$ indicate 
that 
$\Lambda$CDM is favored over the oscillating dark energy models. }\label{tab:bayesian}  
\end{table}                   
\begin{figure*}
\includegraphics[width=0.24\textwidth]{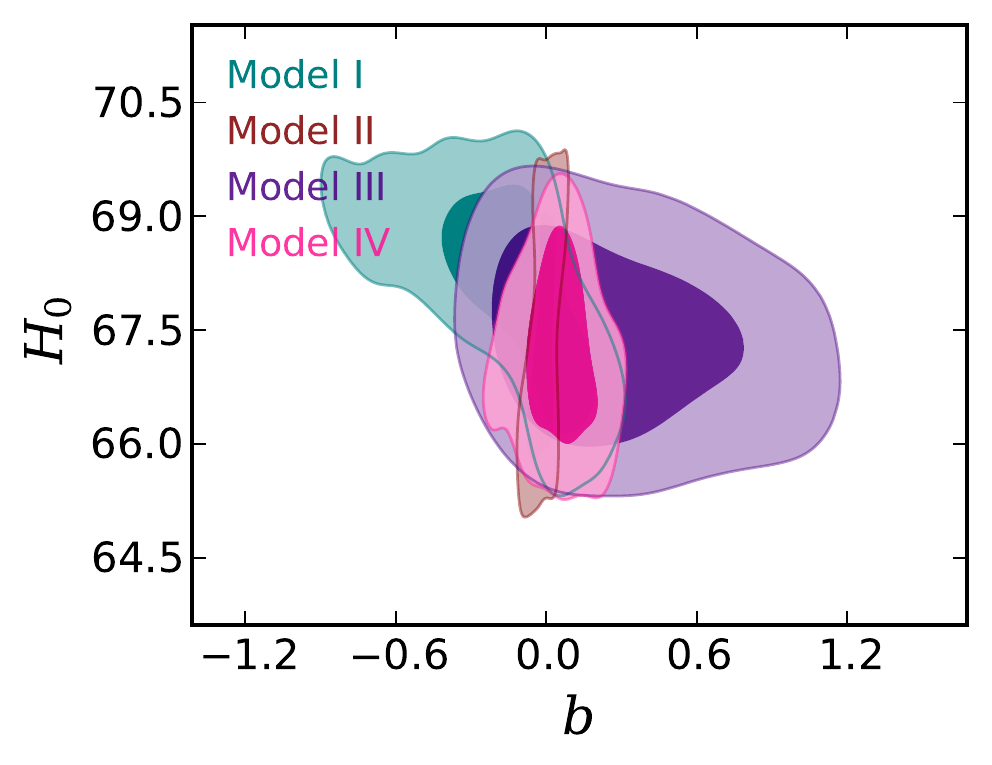}
\includegraphics[width=0.24\textwidth]{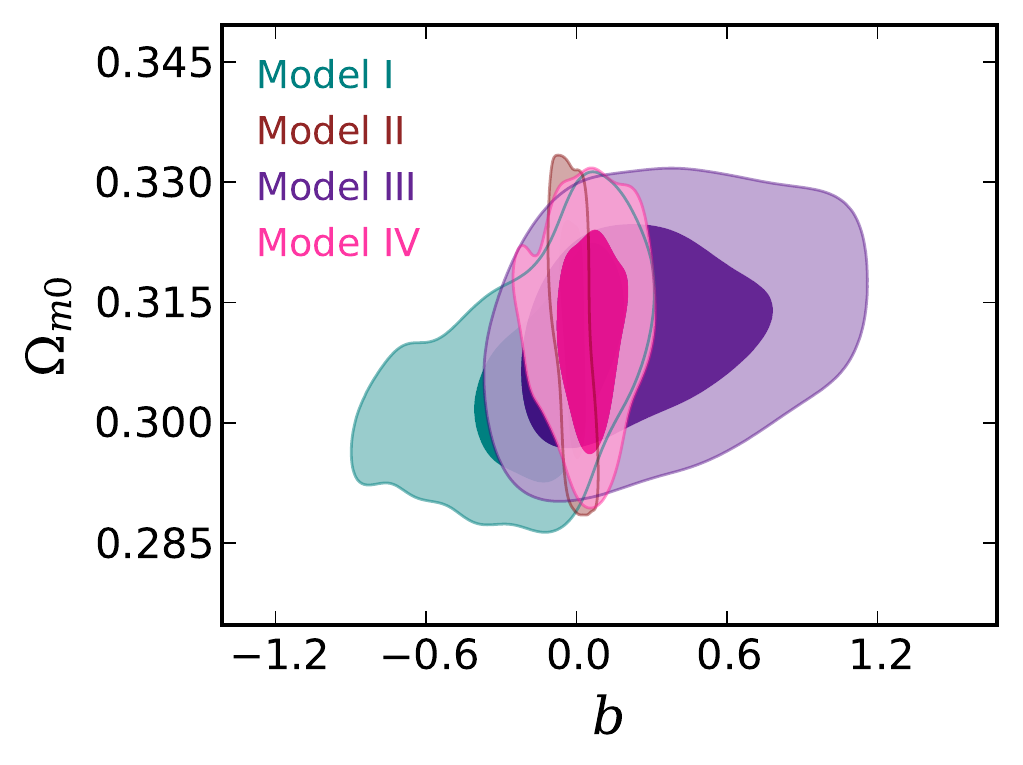}
\includegraphics[width=0.24\textwidth]{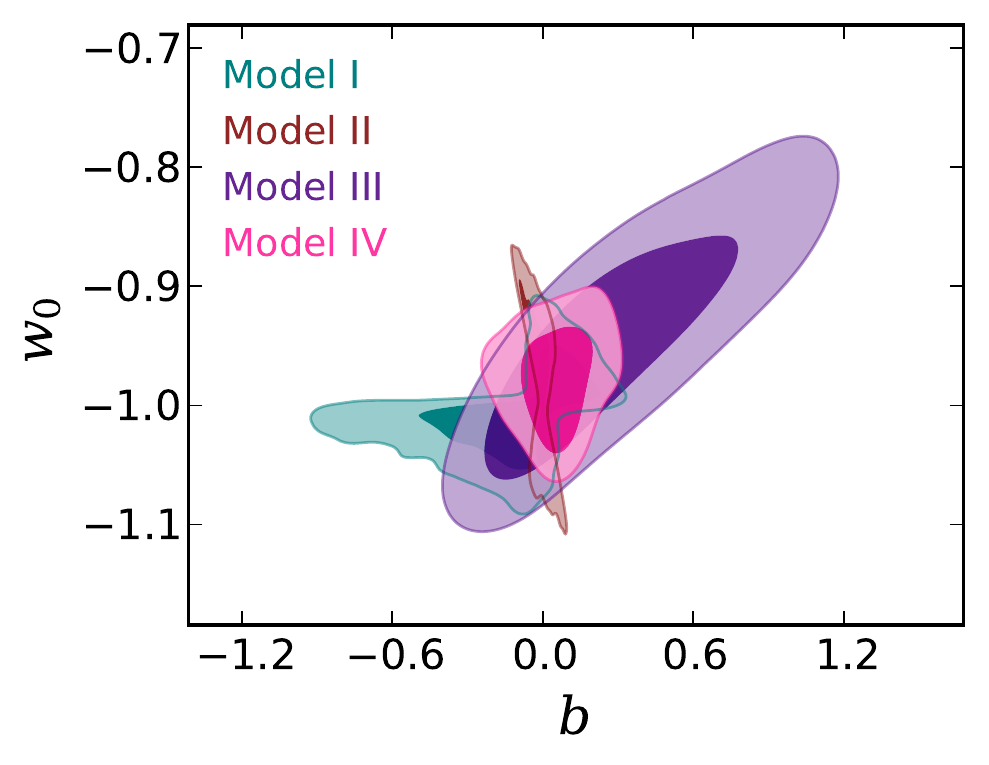}
\includegraphics[width=0.24\textwidth]{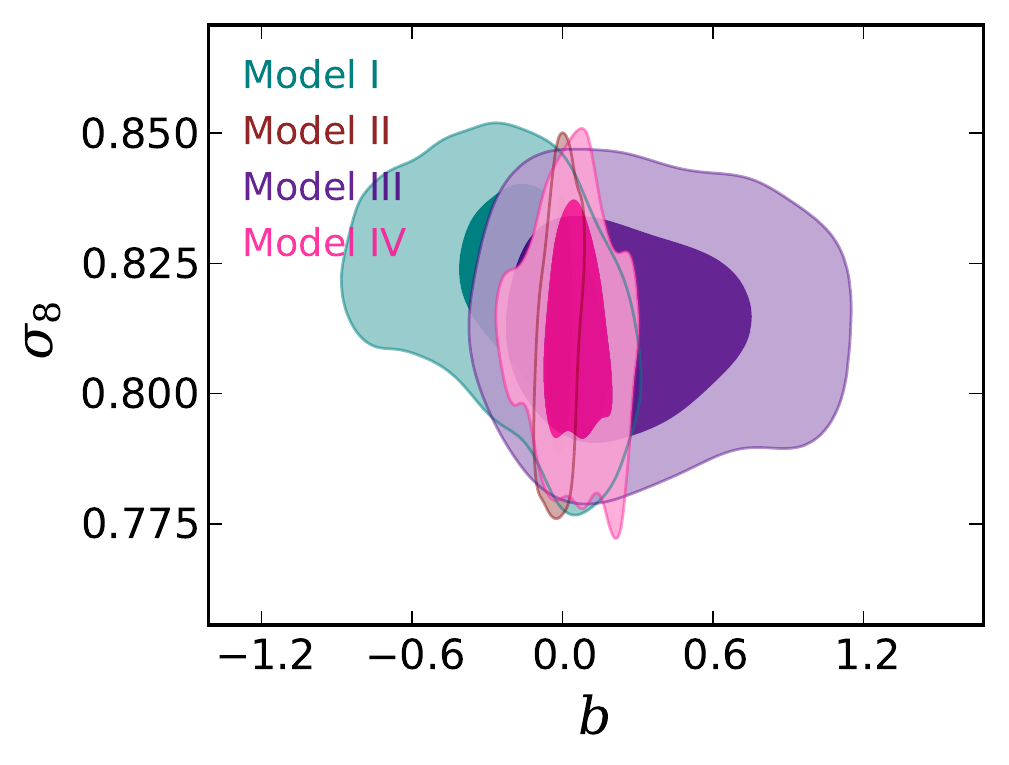}\\
\includegraphics[width=0.24\textwidth]{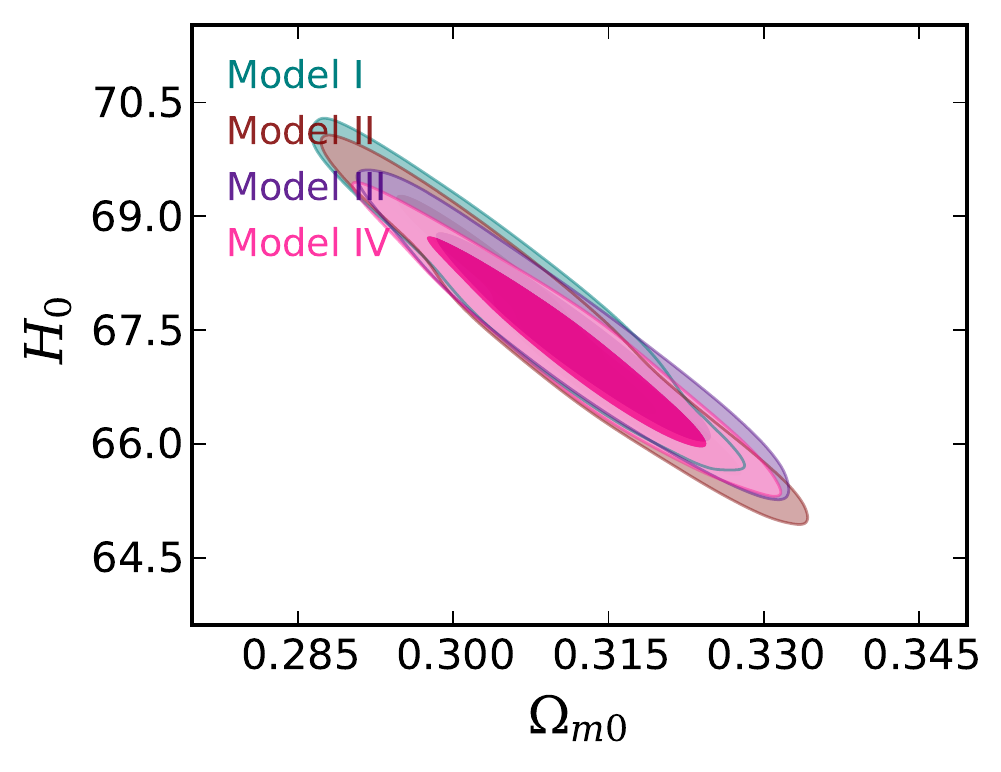}
\includegraphics[width=0.24\textwidth]{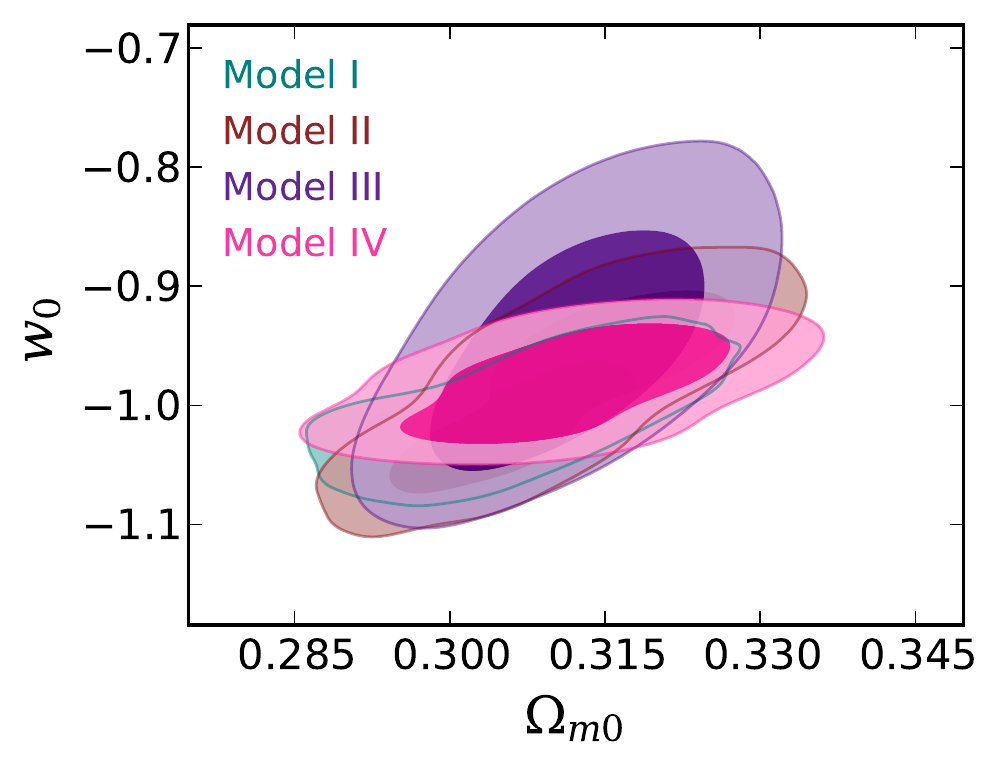}
\includegraphics[width=0.24\textwidth]{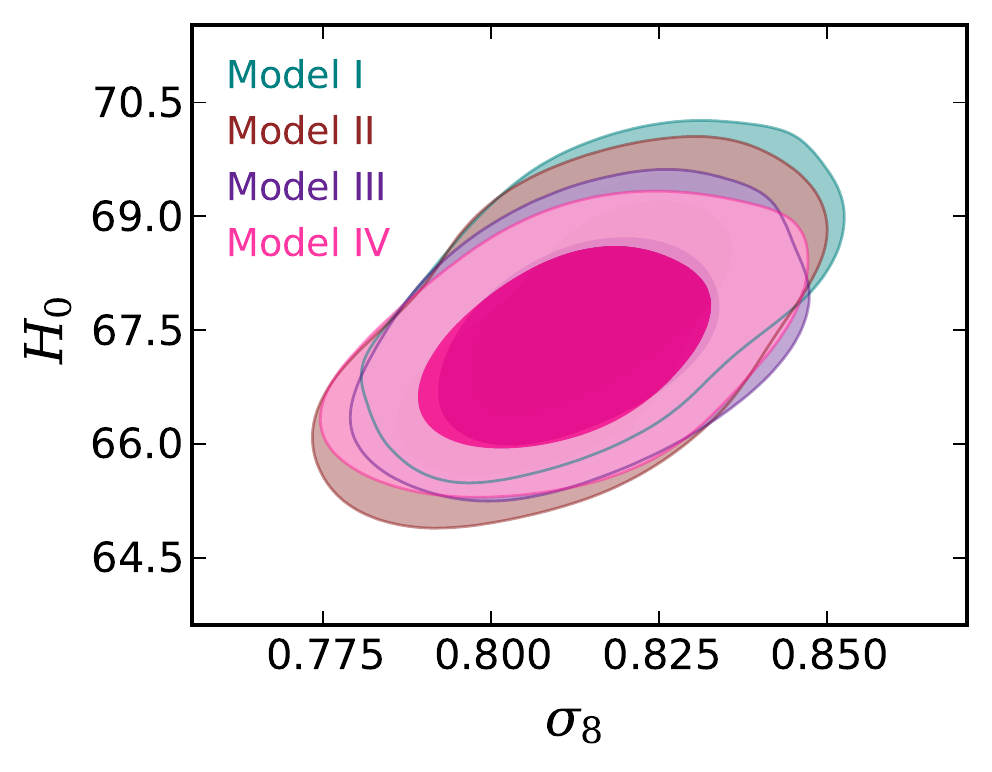}
\includegraphics[width=0.24\textwidth]{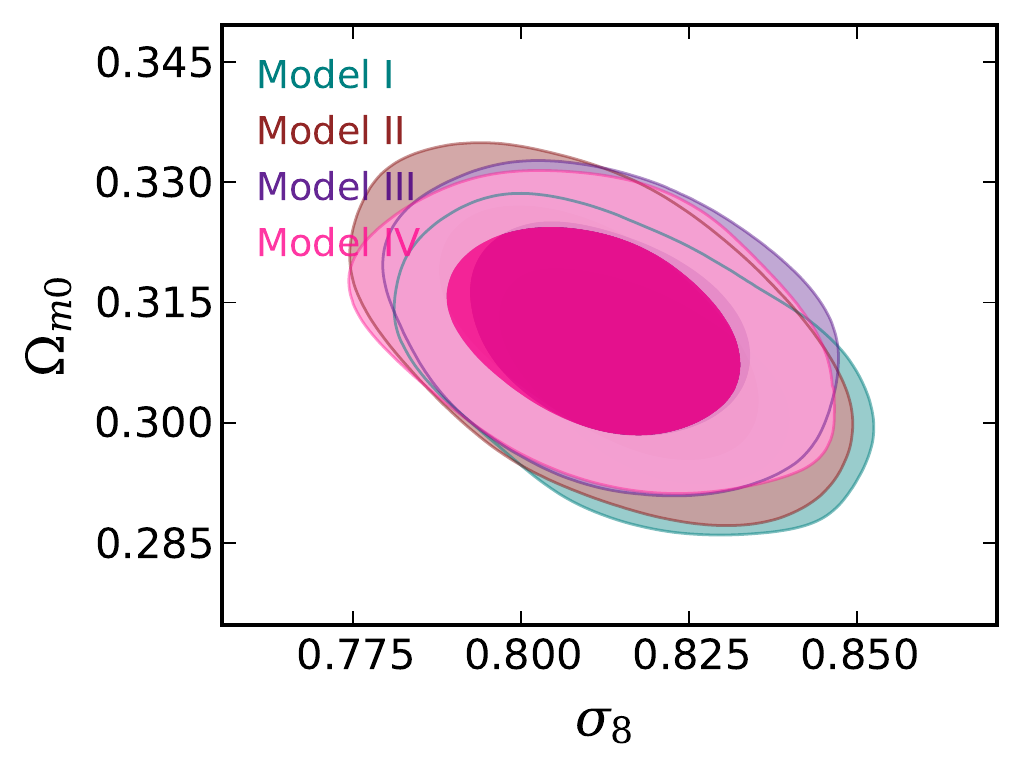}
\caption{\textit{
1$\sigma$ (68.3\%) and 2$\sigma$ (95.4\%) confidence level contour plots 
for different combinations of the model parameters,  for all Models I-IV of 
(\ref{model1-current})-(\ref{model4}) simultaneously, for the combined observational data 
JLA $+$ 
BAO $+$ Planck TT, TE, EE $+$ LowTEB 
$+$ RSD $+$ WL$+$ CC.  }}
\label{fig:All-contours}
\end{figure*}
\begin{figure*}
\includegraphics[width=0.5\textwidth]{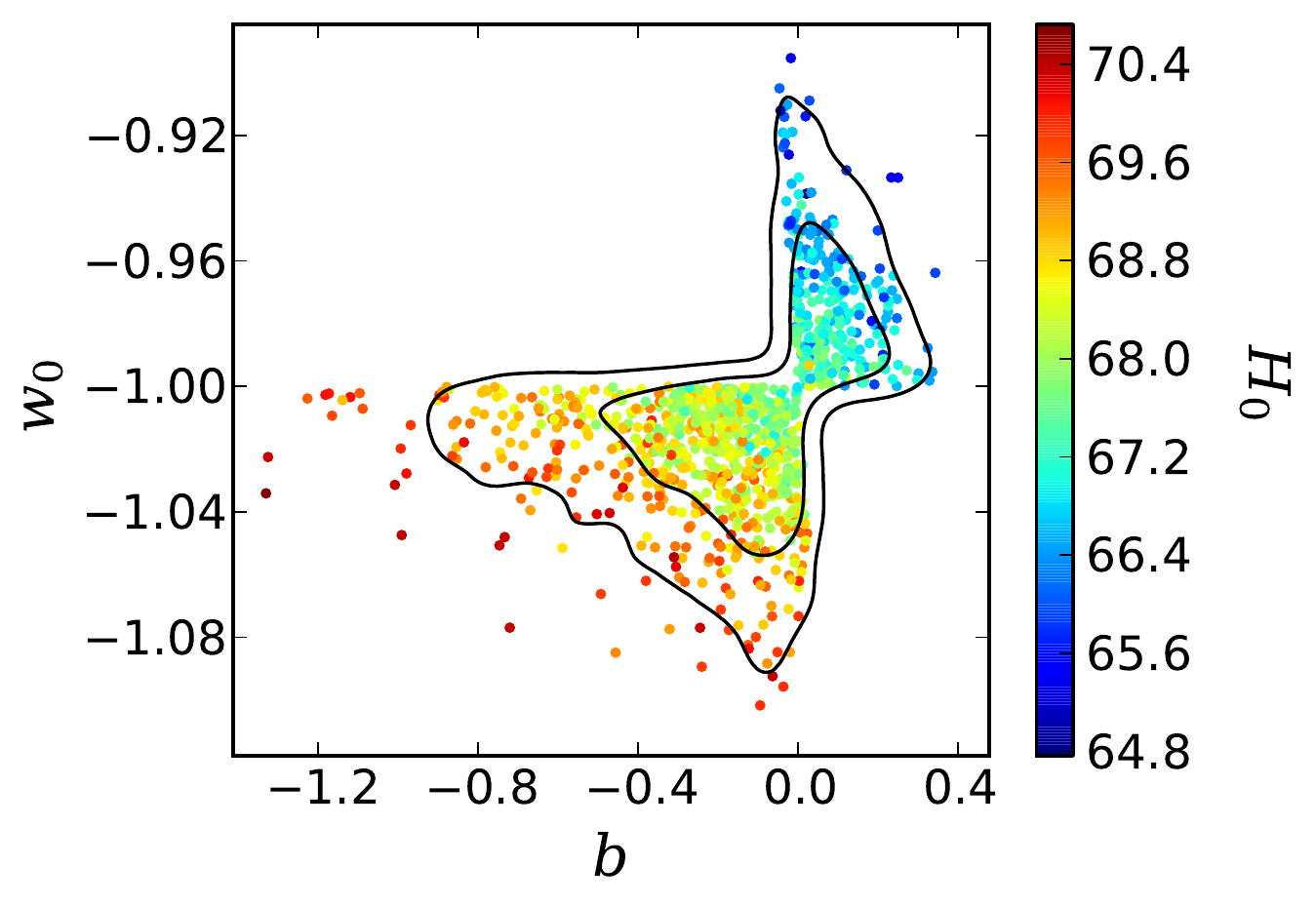}
\includegraphics[width=0.5\textwidth]{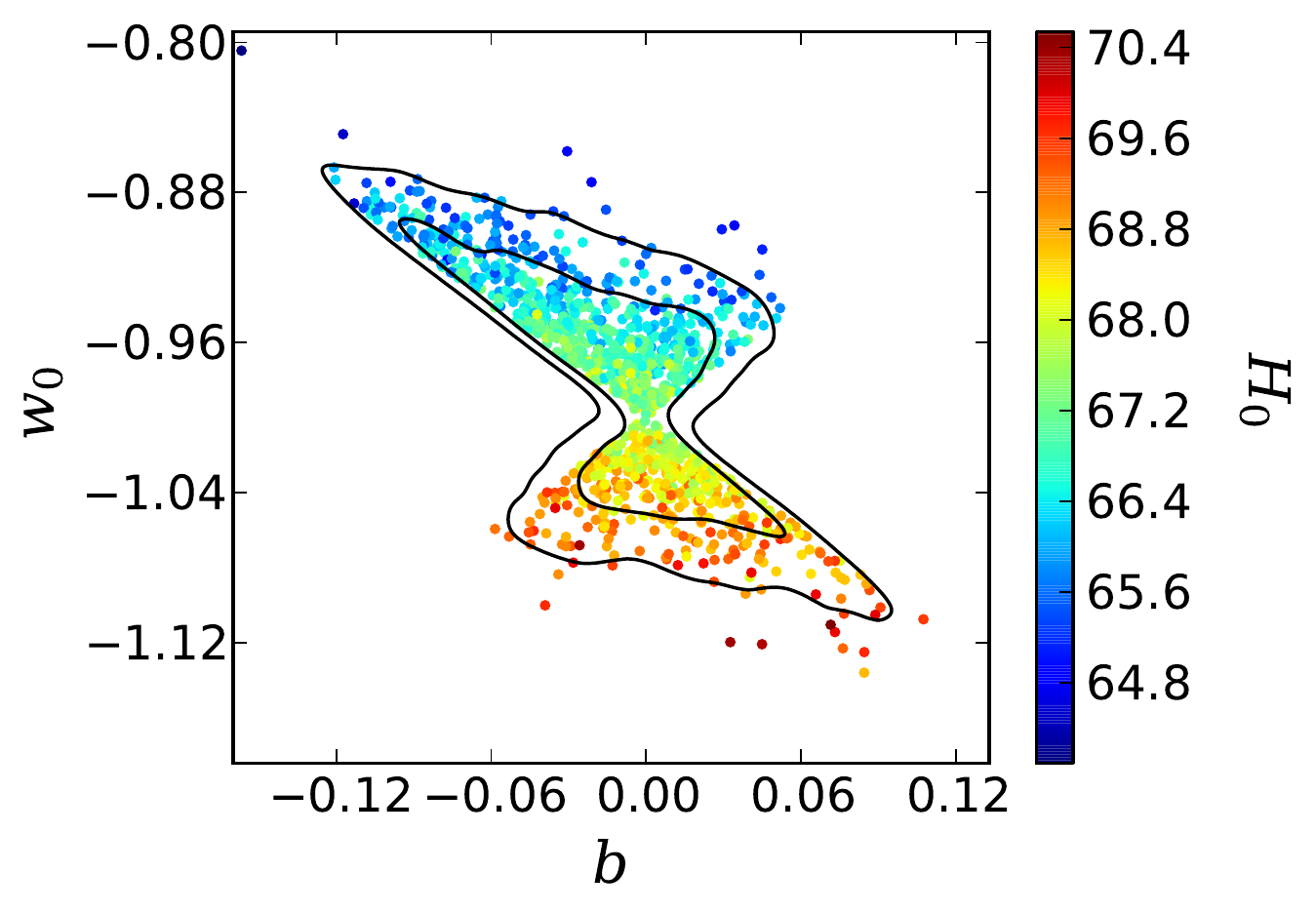}
\includegraphics[width=0.5\textwidth]{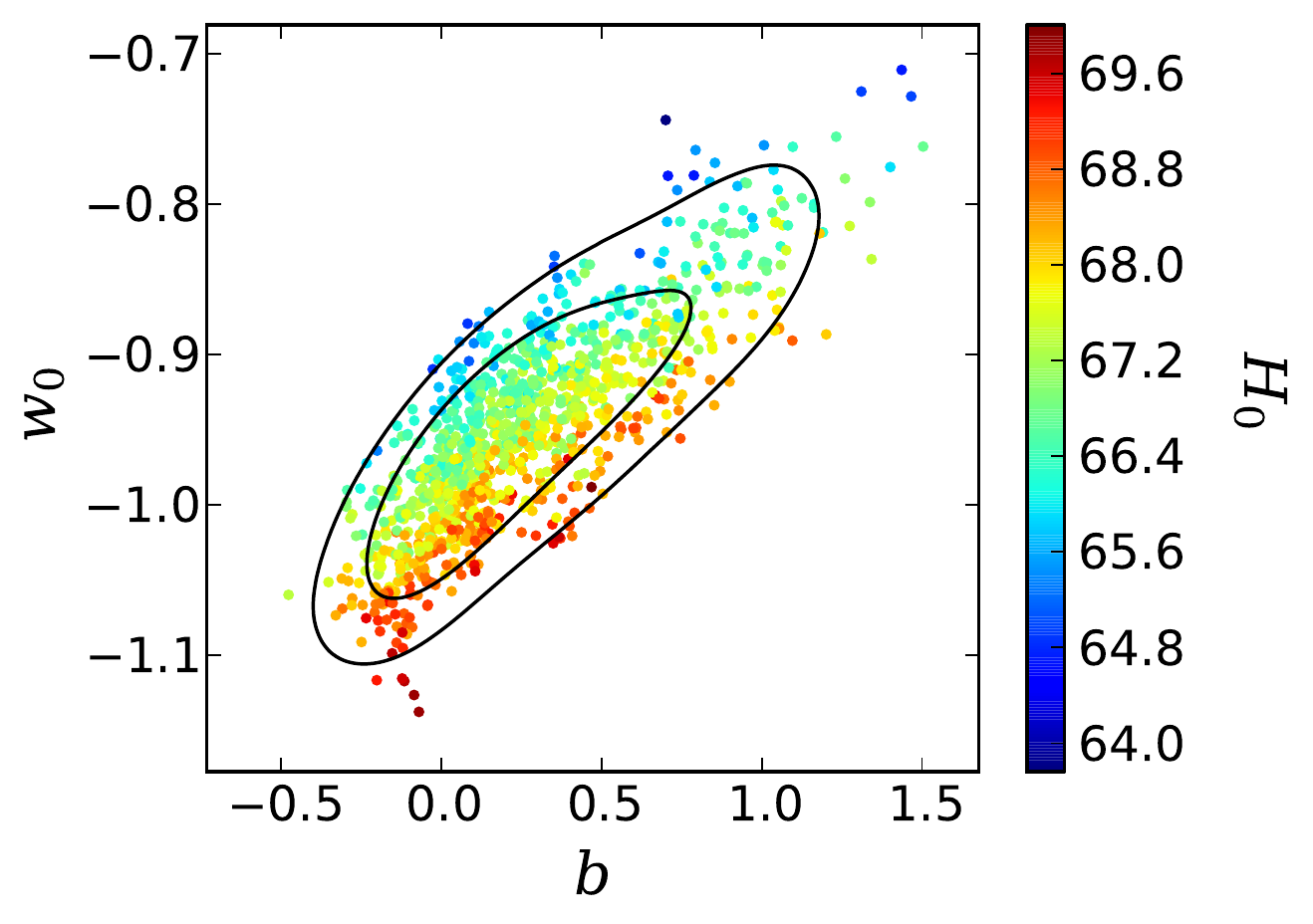}
\includegraphics[width=0.5\textwidth]{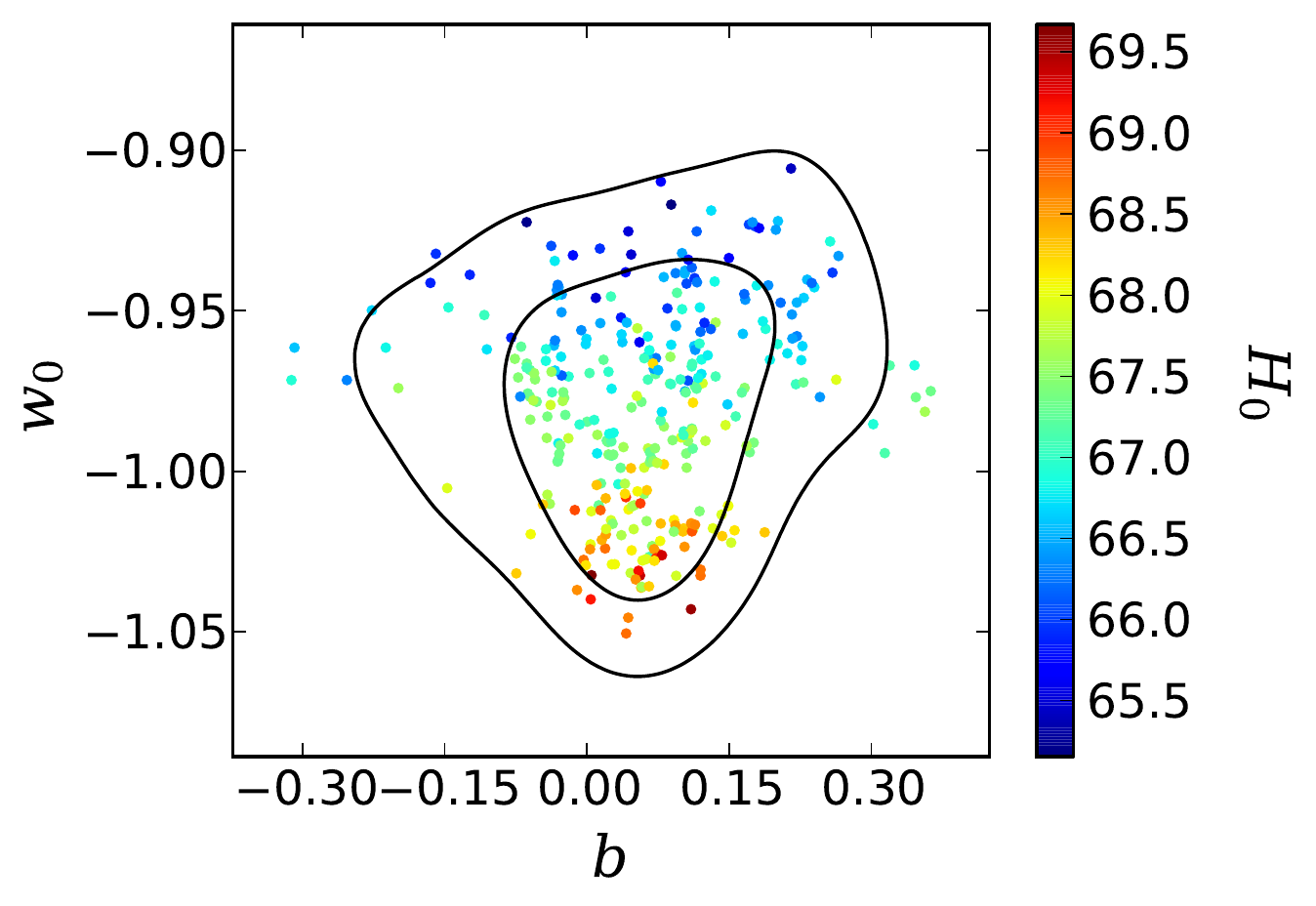}
\caption{ \textit{The trend of the key parameters $(b, w_0)$ of the oscillating  dark 
energy 
models, namely for Model I of (\ref{model1-current}) (upper left graph), for Model II of 
(\ref{model2-current}) (upper right graph), for Model III of (\ref{model3}) 
(lower left graph) and for Model IV of (\ref{model4}) 
(lower right graph), for different values of $H_0$, from the MCMC chain of the combined 
analysis JLA $+$ BAO $+$ Planck TT, TE, EE $+$ LowTEB $+$ RSD $+$ WL$+$ CC. }} 
\label{fig:scattered-plots}
\end{figure*}

For $B_{ij} > 1 $ we deduce that the cosmological data employed in the analysis support 
  model $M_i$ more strongly than   model $M_j$. The behaviour 
of the models can be quantified using different values of $B_{ij}$ (or  equivalently
$\ln B_{ij}$).
 In this work, we shall use the widely accepted Jeffreys scales \cite{Kass:1995loi}, 
summarized in 
Table \ref{tab:jeffreys}, that quantify the viabilities of the models under 
consideration.

The Bayesian evidence for the scenarios at hand can be easily calculated, since only 
the MCMC chains, 
which are used for parameters estimation, are needed. A detailed explanation can be found 
in two recent articles \cite{Heavens:2017hkr,Heavens:2017afc}, where the algorithm 
that calculates the Bayesian evidence is known as the \texttt{MCEvidence} 
code\footnote{This code is publicly
available  at 
\href{https://github.com/yabebalFantaye/MCEvidence}{github.com/yabebalFantaye/
MCEvidence}.}.

\begin{figure*}
\includegraphics[width=0.5\textwidth]{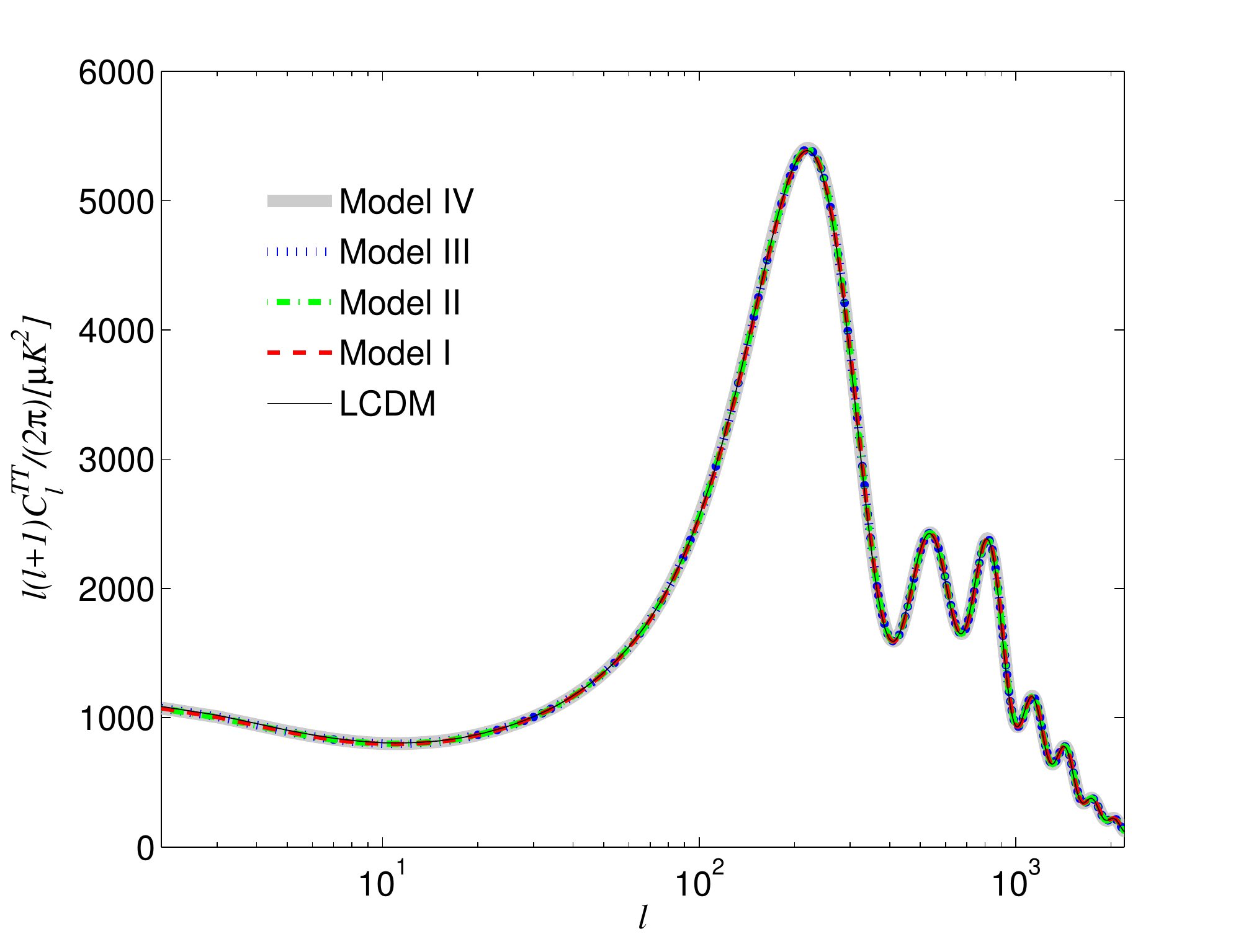}
\includegraphics[width=0.5\textwidth]{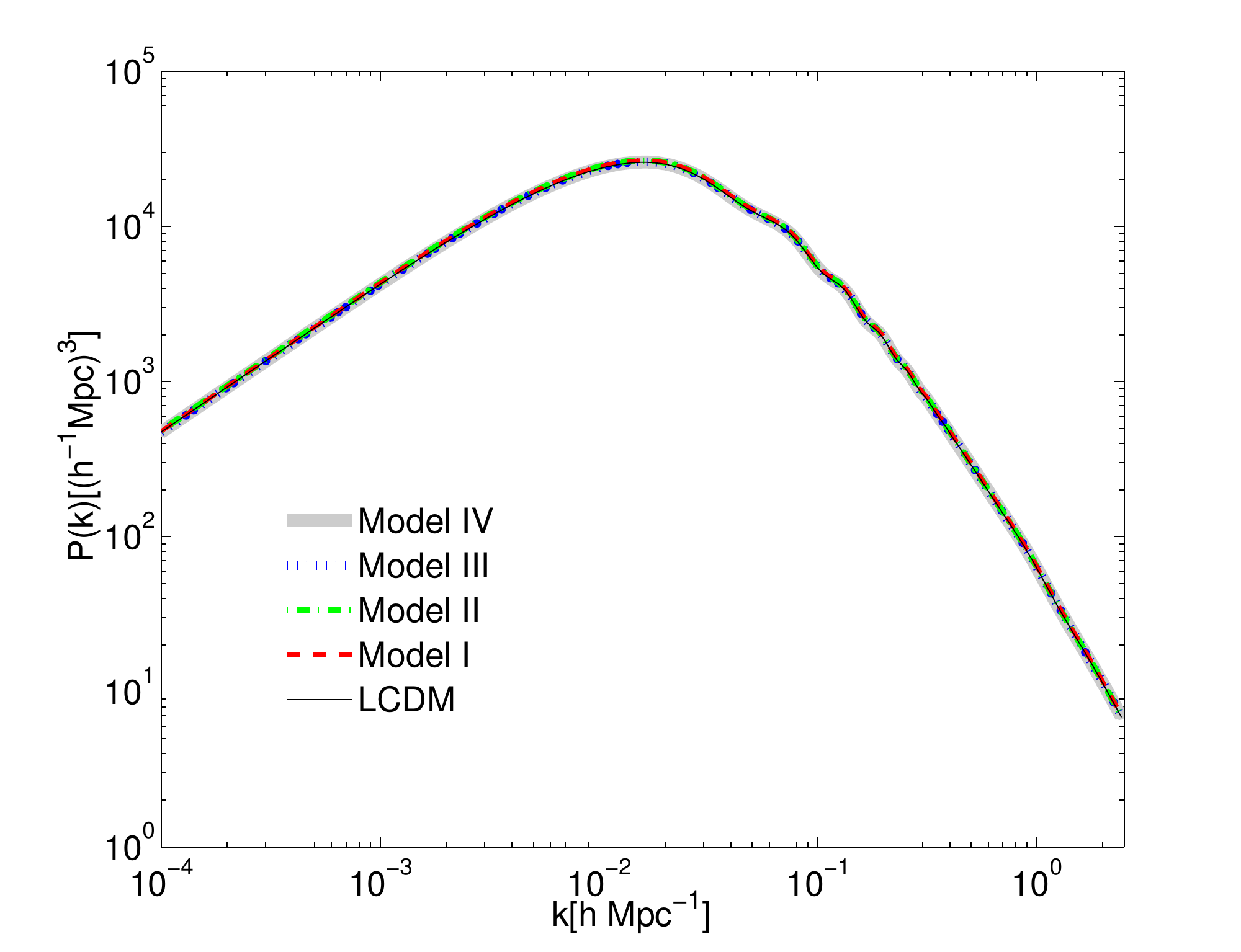}
\caption{\textit{The temperature anisotropy in the CMB spectra (left panel) and the 
matter power spectra (right panel), for all Models I-IV of 
(\ref{model1-current})-(\ref{model4}) simultaneously, for the mean values of $(w_0, b)$ 
that arise from the combined analysis JLA $+$ BAO $+$ Planck TT, TE, EE $+$ LowTEB $+$ 
RSD $+$ WL$+$ CC, and the corresponding curves of $\Lambda$CDM cosmology.}}
\label{fig:cmb}
\end{figure*} 
\begin{figure*}
\includegraphics[width=0.5\textwidth]{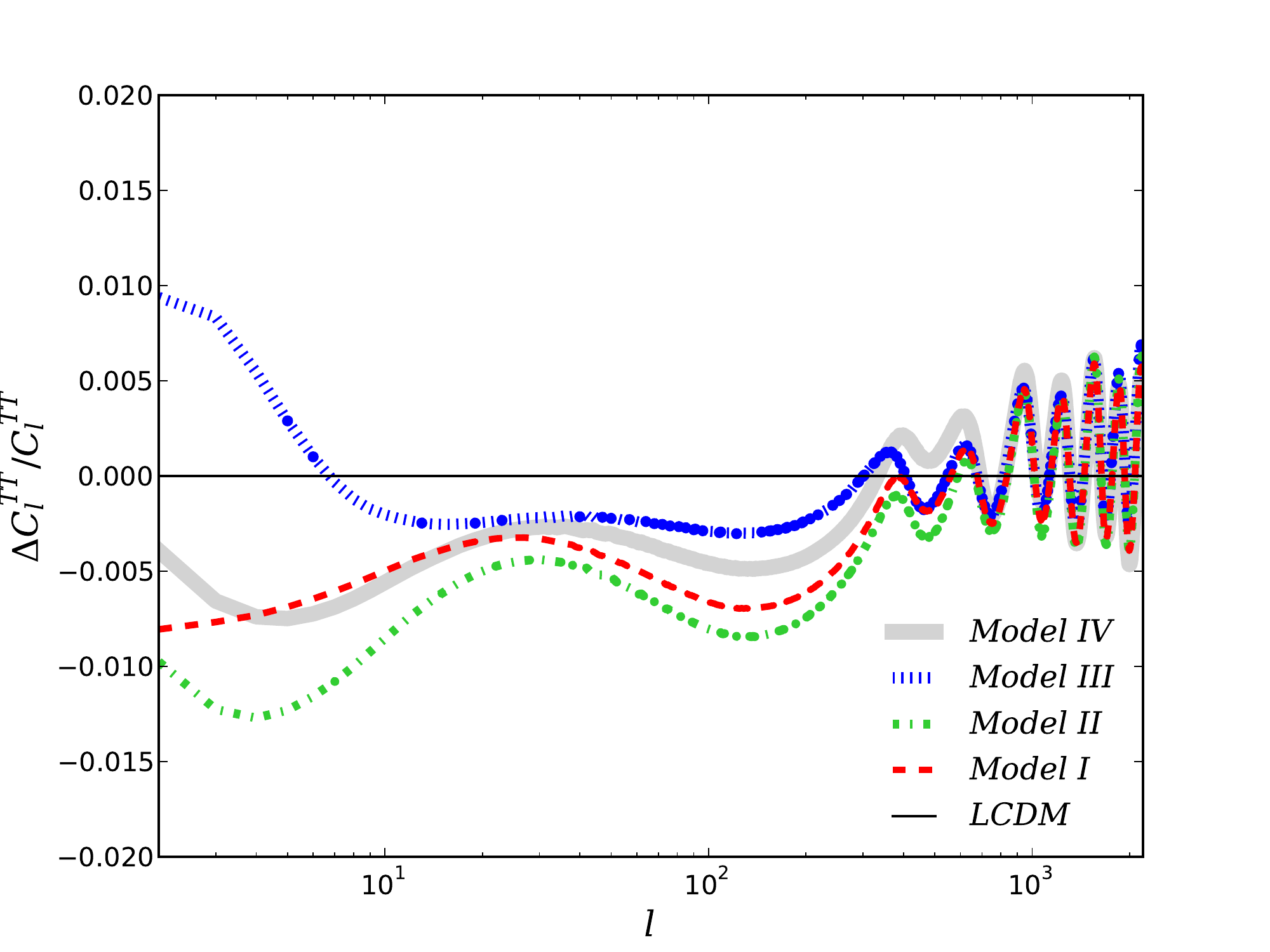}
\includegraphics[width=0.5\textwidth]{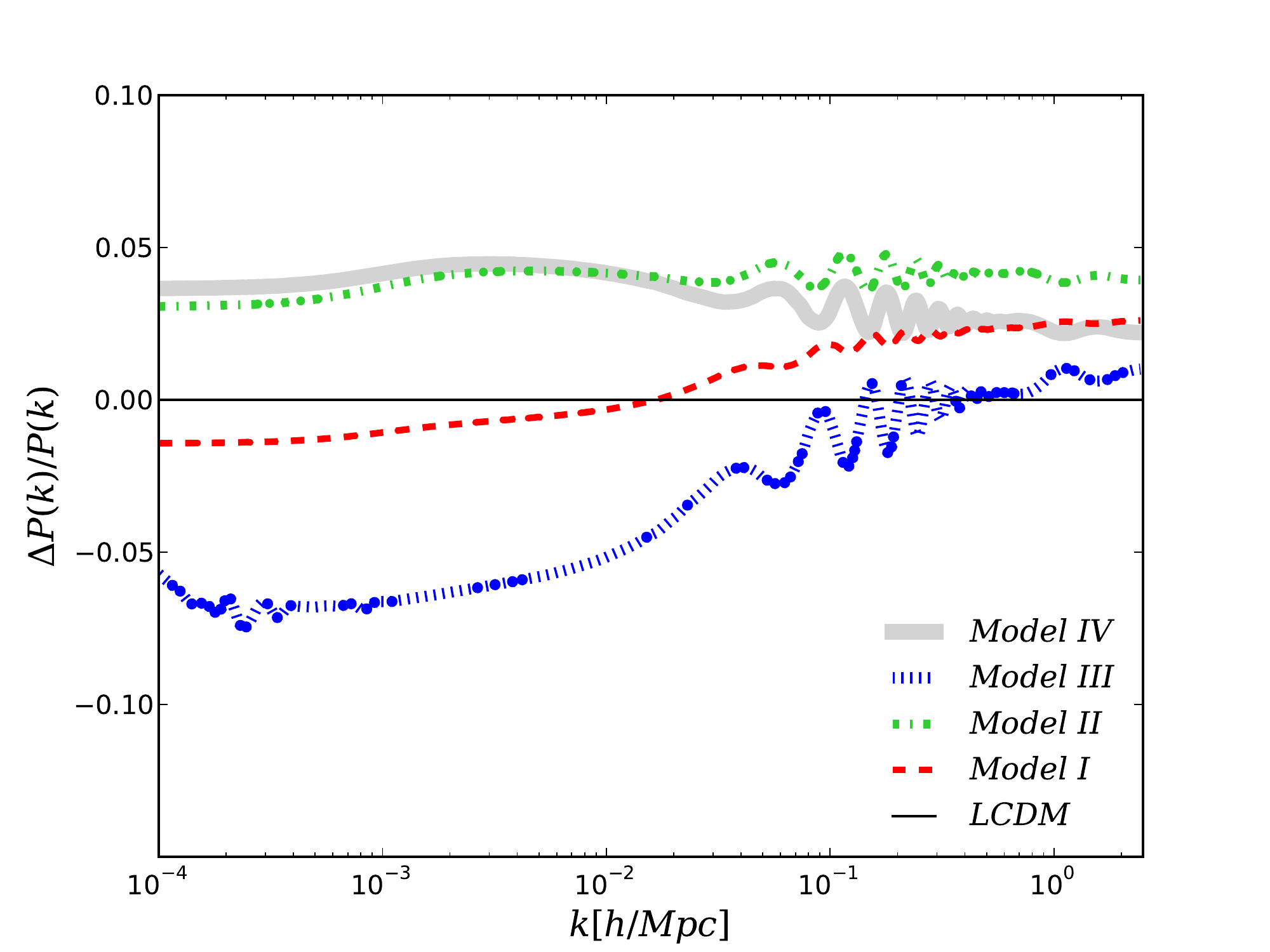}
\caption{ \textit{Relative deviation of the CMB TT spectra (left panel) and of the matter 
power 
spectra (right panel) from the $\Lambda$CDM paradigm ($b = 0$, $w_0 = -1$) for Model I of 
(\ref{model1-current}), for Model II of (\ref{model2-current}), for Model III of 
(\ref{model3}), and  for Model IV of (\ref{model4}), shown for the mean values of 
$b$ obtained from the combined analysis JLA $+$ 
BAO $+$ Planck TT, TE, EE $+$ LowTEB $+$ RSD $+$ WL$+$ CC. } }
\label{fig:ratio}
\end{figure*}
Using the \texttt{MCEvidence} code we calculate the logarithm of the Bayes 
factor, i.e. $\ln B_{ij}$, where $i=$ Model I $-$ Model IV   and $j=\Lambda$CDM. In Table 
\ref{tab:bayesian} we summarize the calculated values of $\ln B_{ij}$ for all 
  oscillating dark energy models with respect to the base $\Lambda$CDM.  For Models I, 
III 
and IV we find that    $|\ln B_{ij}| > 5$. In particular, we see
that $\ln B_{ij} = -9.4$ (Model I), $\ln B_{ij} = -6.2$ (Model III), $\ln B_{ij} = -8.4$ 
(Model IV), which implies that for all these three models  we obtain a very 
strong preference for $\Lambda$CDM. Additionally,  for Model II we acquire $\ln B_{ij} = 
-4.6$,  which indicates the strong preference of  $\Lambda$CDM over Model II. Overall, we 
 find that $\Lambda$CDM cosmology is significantly favored    compared to the examined 
oscillating dark 
energy models.

\section{Conclusions}
\label{sec-discuss}

Since the nature of the dark energy sector is unknown, one can incorporate its effect in 
a phenomenological way, i.e. introducing various parametrizations of the  dark energy 
equation-of-state parameter. One interesting parametrization class  is the case where 
$w_x(z)$ exhibits oscillating behaviour \cite{Sahni:1999qe, Dodelson:2001fq, 
Feng:2004ff,Xia:2004rw,Jain:2007fa, Lazkoz:2010gz, Ma:2011nc}, since it may lead to 
interesting cosmology. 

In order to thoroughly examine whether oscillating dark-energy models are in agreement 
with the latest observational data, we have performed a complete 
observational confrontation using the latest data, namely: Joint Light Curve analysis 
(JLA) sample from Supernoave Type Ia, Baryon Acoustic Oscillations (BAO) distance 
measurements, Cosmic Microwave Background (CMB) observations, redshift space distortion, 
weak gravitational lensing, Hubble parameter measurements from 
cosmic chronometers, and the local Hubble constant value.

We considered four oscillating dark energy models, namely, Model I of 
(\ref{model1-current}), Model II of (\ref{model2-current}), Model III of (\ref{model3}) 
and Model IV of (\ref{model4}). Our analysis shows that for Model I, Model II and 
Model IV, 
the best fit values of the dark energy equation-of-state parameter $w_0$ lies in the  
phantom regime, nevertheless 
in all models  the quintessential regime is 
also allowed within 1$\sigma$ confidence-level. The models indicate  
deviations from   $\Lambda$CDM cosmology, although such deviations are small.  The 
fittings suggest that in all viable oscillating dark-energy  models, the parameter $b$ 
that quantifies the deviation from  $w$CDM and $\Lambda$CDM cosmology is relatively small 
for three 
models namely Model I, Model II and Model IV, while for Model III  
$b$ is larger. Thus, effectively, Model III exhibits 
a non-zero 
deviation from   $w$CDM   as well as $\Lambda$CDM cosmology, however the deviation 
is not   significant. 

As a next step we analyzed the behaviour of the oscillating models at large scales, 
through the impact on the temperature anisotropy of the CMB spectra and on the matter 
power spectra. Moreover, we compared the results  with the $w$CDM and $\Lambda$CDM
scenarios, examining the corresponding deviations. As we showed, for  Models II and III 
the deviation from  $w$CDM and $\Lambda$CDM models is clear  for large negative values of 
the parameter $b$. On the other hand,  Model I exhibits a slight deviation, while  for 
Model IV the deviation is non-significant. 

Furthermore, we presented the Bayesian evidences for all oscillating dark-energy 
models with respect to the reference $\Lambda$CDM scenario. The results have been 
summarized in Table \ref{tab:bayesian}, from which we found that  according to 
the present observational data    $\Lambda$CDM cosmology is favored  over all considered  
  models.

We close this work with a short statistical comparison of the models, both at 
background  and perturbative levels. 
In Fig. \ref{fig:All-contours} we present the 1$\sigma$ and 2$\sigma$ confidence-level 
contour plots for several combinations of the free parameters and of the derived 
parameters, for all Models I-IV simultaneously.
As we can observe, Model I is slightly different compared to the other three models, 
although not
significantly. Moreover, we analyze the trend of the two main parameters of the  
oscillating models, namely $b$ and $w_0$, 
for different values of $H_0$, using the MCMC 
chain of the combined analysis JLA $+$ BAO $+$ Planck TT, TE, EE $+$ LowTEB $+$ RSD $+$ 
WL$+$ CC, and in Fig. \ref{fig:scattered-plots} we present the results for all  
models. From the analysis of the MCMC chain we can clearly 
notice that higher values of $H_0$ (the red sample points in Fig. 
\ref{fig:scattered-plots}) favour the phantom 
behaviour of dark energy, while for low values of the Hubble 
constant $H_0$ (the blue sample points in Fig. \ref{fig:scattered-plots}) a 
quintessence-like dark energy is favored, 
however within $1\sigma$, $w_0$ is close to $-1$.

In order to examine whether these differences can be observed at large scales, in Fig. 
\ref{fig:cmb} we depict the temperature anisotropy in the CMB spectra (left 
graph)  and the matter power spectra (right graph), for all models simultaneously, using 
for each model the corresponding mean value for the parameter $b$. From both graphs  
we deduce that we cannot distinguish the various models, and moreover all models are 
found to exhibit a behaviour close to that of the flat $\Lambda$CDM scenario.  
However, a 
slight difference is expected as the estimated value of $b$ for all oscillating  dark 
energy models 
is non-null. This difference can be seen in  Fig. \ref{fig:ratio},
in which we show the relative deviations in the CMB TT spectra (left panel) 
and in the
matter power spectra (right panel).

In summary, the analysis of the present work reveals that the oscillating dark energy 
models can be 
in agreement with observations. However, according to the Bayesian analysis, 
$\Lambda$CDM cosmology is 
 favored   compared to them. 
One interesting extension of the above investigation would be to proceed to a more 
general 
formalism where the sound speed of 
the dark energy could be variable, instead of constant. This study could enlighten the   
intrinsic nature of the oscillating dark energy models, especially in comparison with 
non-oscillating models. Such an investigation is left for a future project.

\begin{acknowledgments}
We thank an anonymous referee for   essential and enlighting comments that helped to 
improve 
the manuscript in a singificant manner. Additionally, we are also grateful to Rafael C. 
Nunes and 
Eleonora Di Valentino for many fruitful discussions for this article. 
The  research of SP was supported by the SERB-NPDF grant (File No. PDF/2015/000640). SP 
also thanks 
the DPS, IISER Kolkata, India, where a part of the work was finished. W. 
Yang's work is supported by the National Natural Science Foundation of China under Grants 
No. 11705079 and No. 11647153.
This article is based upon work from COST Action ``Cosmology and Astrophysics Network
for Theoretical Advances and Training Actions'', supported by COST (European Cooperation
in Science and Technology).
 \end{acknowledgments}


\begin{thebibliography}{99}

 

\bibitem{Copeland:2006wr}
  E.~J.~Copeland, M.~Sami and S.~Tsujikawa,
  {\it{Dynamics of dark energy}},
  Int.\ J.\ Mod.\ Phys.\  D {\bf 15}, 1753 (2006),
     [\href{http://xxx.lanl.gov/abs/hep-th/0603057}
{{\tt arXiv:hep-th/0603057}}].


\bibitem{Cai:2009zp}
  Y.~-F.~Cai, E.~N.~Saridakis, M.~R.~Setare and J.~-Q.~Xia,
  {\it{Quintom Cosmology: Theoretical implications and observations}},
  Phys.\ Rept.\  {\bf 493}, 1 (2010),
       [\href{http://xxx.lanl.gov/abs/0909.2776}
{{\tt arXiv:0909.2776}}].



 
\bibitem{Nojiri:2006ri} 
  S.~Nojiri and S.~D.~Odintsov,
  {\it{Introduction to modified gravity and gravitational alternative for dark energy}},
  eConf C {\bf 0602061}, 06 (2006)
  [Int.\ J.\ Geom.\ Meth.\ Mod.\ Phys.\  {\bf 04}, 115 (2007)]
       [\href{http://xxx.lanl.gov/abs/hep-th/0601213}
{{\tt arXiv:hep-th/0601213}}].



 

 
\bibitem{Capozziello:2011et}
  S.~Capozziello and M.~De Laurentis,
  {\it{Extended Theories of Gravity}},
  Phys.\ Rept.\  {\bf 509}, 167 (2011),
  [\href{http://xxx.lanl.gov/abs/1108.6266}
{{\tt arXiv:1108.6266}}].



\bibitem{Cai:2015emx}
  Y.~F.~Cai, S.~Capozziello, M.~De Laurentis and E.~N.~Saridakis,
  {\it{f(T) Teleparallel Gravity and Cosmology}},
  Rept.\ Prog.\ Phys.\  {\bf 79}, no. 10, 106901 (2016),
  [\href{http://xxx.lanl.gov/abs/1511.07586}
{{\tt arXiv:1511.07586}}].

  
  
  

\bibitem{Gong:2005de} 
  Y.~g.~Gong and Y.~Z.~Zhang,
    {\it{Probing the curvature and dark energy}},
  Phys.\ Rev.\ D {\bf 72}, 043518 (2005),
    [\href{http://xxx.lanl.gov/abs/astro-ph/0502262}
{{\tt arXiv:astro-ph/0502262}}].

  
  
   
  

\bibitem{Chevallier:2000qy} 
  M.~Chevallier and D.~Polarski,
    {\it{Accelerating universes with scaling dark matter}},
  Int.\ J.\ Mod.\ Phys.\ D {\bf 10}, 213 (2001),
      [\href{http://xxx.lanl.gov/abs/gr-qc/0009008}
{{\tt arXiv:gr-qc/0009008}}].

  
  
   
\bibitem{Linder:2002et} 
  E.~V.~Linder,
    {\it{Exploring the expansion history of the universe}},
  Phys.\ Rev.\ Lett.\  {\bf 90}, 091301 (2003),
        [\href{http://xxx.lanl.gov/abs/astro-ph/0208512}
{{\tt arXiv:astro-ph/0208512}}].

  
   
  
\bibitem{Cooray:1999da} 
  A.~R.~Cooray and D.~Huterer,
    {\it{Gravitational lensing as a probe of quintessence}},
  Astrophys.\ J.\  {\bf 513}, L95 (1999),
          [\href{http://xxx.lanl.gov/abs/astro-ph/9901097}
{{\tt arXiv:astro-ph/9901097}}].

   
  
\bibitem{Astier:2000as} 
  P.~Astier,
    {\it{Can luminosity distance measurements probe the equation of state of dark 
energy}},
  Phys.\ Lett.\ B {\bf 500}, 8 (2001),
            [\href{http://xxx.lanl.gov/abs/astro-ph/0008306}
{{\tt arXiv:astro-ph/0008306}}].

    
  
\bibitem{Weller:2001gf} 
  J.~Weller and A.~Albrecht,
    {\it{Future supernovae observations as a probe of dark energy}},
  Phys.\ Rev.\ D {\bf 65}, 103512 (2002),
              [\href{http://xxx.lanl.gov/abs/astro-ph/0106079}
{{\tt arXiv:astro-ph/0106079}}].


 

\bibitem{Efstathiou:1999tm} 
  G.~Efstathiou,
    {\it{Constraining the equation of state of the universe from distant type Ia 
supernovae 
and cosmic 
microwave background anisotropies}},
  Mon.\ Not.\ Roy.\ Astron.\ Soc.\  {\bf 310}, 842 (1999),
                [\href{http://xxx.lanl.gov/abs/astro-ph/9904356}
{{\tt arXiv:astro-ph/9904356}}].

 


\bibitem {Jassal:2005qc}
H.~K.~Jassal, J.~S.~Bagla and T.~Padmanabhan,
  {\it{Observational constraints on low redshift evolution of dark energy: How consistent 
are 
different 
observations?}},
Phys.\ Rev.\ D \textbf{72}, 103503 (2005),
  [\href{http://xxx.lanl.gov/abs/astro-ph/0506748}
{{\tt arXiv:astro-ph/0506748}}].

 


\bibitem {Barboza:2008rh}E.~M.~Barboza, Jr. and J.~S.~Alcaniz,
  {\it{A parametric model for dark energy}},
Phys.\ Lett.\ B \textbf{666}, 415 (2008),
  [\href{http://xxx.lanl.gov/abs/0805.1713}
{{\tt arXiv:0805.1713}}].

 
 


\bibitem{Feng:2012gf} 
  C.~J.~Feng, X.~Y.~Shen, P.~Li and X.~Z.~Li,
    {\it{A New Class of Parametrization for Dark Energy without Divergence}},
  JCAP {\bf 1209}, 023 (2012),
    [\href{http://xxx.lanl.gov/abs/1206.0063}
{{\tt arXiv:1206.0063}}].

 
  

\bibitem{Feng:2011zzo} 
  L.~Feng and T.~Lu,
    {\it{A new equation of state for dark energy model}},
  JCAP {\bf 1111}, 034 (2011),
      [\href{http://xxx.lanl.gov/abs/1203.1784}
{{\tt arXiv:1203.1784}}].

 
   

  
\bibitem{Pantazis:2016nky} 
  G.~Pantazis, S.~Nesseris and L.~Perivolaropoulos,
    {\it{Comparison of thawing and freezing dark energy parametrizations}},
  Phys.\ Rev.\ D {\bf 93}, no. 10, 103503 (2016),
        [\href{http://xxx.lanl.gov/abs/1603.02164}
{{\tt arXiv:1603.02164}}].

 
 
\bibitem{DiValentino:2016hlg} 
  E.~Di Valentino, A.~Melchiorri and J.~Silk,
  {\it{Reconciling Planck with the local value of $H_0$ in extended parameter space}},
  Phys.\ Lett.\ B {\bf 761}, 242 (2016),
  [\href{http://xxx.lanl.gov/abs/1606.00634}
{{\tt arXiv:1606.00634}}].
  
 
\bibitem{Escamilla-Rivera:2016qwv} 
  C.~Escamilla-Rivera,
  {\it{Status on bidimensional dark energy parameterizations using SNe Ia JLA and BAO 
datasets}},
  Galaxies {\bf 4}, no. 3, 8 (2016),
  [\href{http://xxx.lanl.gov/abs/1605.02702}
{{\tt arXiv:1605.02702}}].

  
  
\bibitem{Zhao:2017cud}
  G.~B.~Zhao {\it et al.},
  {\it{``Dynamical dark energy in light of the latest observations''}},
  Nat.\ Astron.\  {\bf 1}, 627 (2017),
  [\href{http://xxx.lanl.gov/abs/1701.08165}
{{\tt arXiv:1701.08165}}].
  
  
\bibitem{Yang:2017amu} 
  W.~Yang, R.~C.~Nunes, S.~Pan and D.~F.~Mota,
  {\it{Effects of neutrino mass hierarchies on dynamical dark energy models}},
  Phys.\ Rev.\ D {\bf 95}, 103522 (2017),
  [\href{http://xxx.lanl.gov/abs/1703.02556}
{{\tt arXiv:1703.02556}}].
  
  
\bibitem{DiValentino:2017zyq} 
  E.~Di Valentino, A.~Melchiorri, E.~V.~Linder and J.~Silk,
  {\it{Constraining Dark Energy Dynamics in Extended Parameter Space}},
  Phys.\ Rev.\ D {\bf 96},  023523 (2017),
  [\href{http://xxx.lanl.gov/abs/1704.00762}
{{\tt arXiv:1704.00762}}].
 
\bibitem{DiValentino:2017gzb} 
  E.~Di Valentino,
  {\it{Crack in the cosmological paradigm}},
  Nat.\ Astron.\  {\bf 1}, 569 (2017),
  [\href{http://xxx.lanl.gov/abs/1709.04046}
{{\tt arXiv:1709.04046}}].
  

\bibitem{Yang:2017alx} 
  W.~Yang, S.~Pan and A.~Paliathanasis,
    {\it{Latest astronomical constraints on some nonlinear parametric dark energy 
models}}, Mon.\ Not.\ Roy.\ Astron.\ Soc.\  {\bf 475}, 2605 (2018),
[\href{http://xxx.lanl.gov/abs/1708.01717}
{{\tt arXiv:1708.01717}}].
 
  
\bibitem{Ma:2011nc} 
  J.~Z.~Ma and X.~Zhang,
    {\it{Probing the dynamics of dark energy with novel parametrizations}},
  Phys.\ Lett.\ B {\bf 699}, 233 (2011),
  [\href{http://xxx.lanl.gov/abs/1102.2671}
{{\tt arXiv:1102.2671}}].
 
   
    
  
  
\bibitem{Linder:2005ne} 
  E.~V.~Linder and D.~Huterer,
    {\it{How many dark energy parameters?}},
  Phys.\ Rev.\ D {\bf 72}, 043509 (2005),
  [\href{http://xxx.lanl.gov/abs/astro-ph/0505330}
{{\tt arXiv:astro-ph/0505330}}].
 
 
  
\bibitem{DeFelice:2012vd} 
  A.~De Felice, S.~Nesseris and S.~Tsujikawa,
    {\it{Observational constraints on dark energy with a fast varying equation of state}},
  JCAP {\bf 1205}, 029 (2012),
  [\href{http://xxx.lanl.gov/abs/1203.6760}
{{\tt arXiv:1203.6760}}].
 
 
\bibitem{Marcondes:2017vjw} 
  R.~J.~F.~Marcondes and S.~Pan,
  {\it{Cosmic chronometers constraints on some fast-varying dark energy equations of 
state}}, [\href{http://xxx.lanl.gov/abs/1711.06157}
{{\tt arXiv:1711.06157}}].



\bibitem{Sahni:1999qe} 
  V.~Sahni and L.~M.~Wang,
    {\it{A New cosmological model of quintessence and dark matter}},
  Phys.\ Rev.\ D {\bf 62}, 103517 (2000),
  [\href{http://xxx.lanl.gov/abs/astro-ph/9910097}
{{\tt arXiv:astro-ph/9910097}}].


\bibitem{Dodelson:2001fq} 
  S.~Dodelson, M.~Kaplinghat and E.~Stewart,
    {\it{Solving the coincidence problem: Tracking oscillating energy}},
  Phys.\ Rev.\ Lett.\  {\bf 85}, 5276 (2000),
  [\href{http://xxx.lanl.gov/abs/astro-ph/0002360}
{{\tt arXiv:astro-ph/0002360}}].
 

  \bibitem{Feng:2004ff} 
  B.~Feng, M.~Li, Y.~S.~Piao and X.~Zhang,
    {\it{Oscillating quintom and the recurrent universe}},
  Phys.\ Lett.\ B {\bf 634}, 101 (2006),
  [\href{http://xxx.lanl.gov/abs/astro-ph/0407432}
{{\tt arXiv:astro-ph/0407432}}].
 
 
  
  
  \bibitem{Xia:2004rw} 
  J.~Q.~Xia, B.~Feng and X.~M.~Zhang,
    {\it{Constraints on oscillating quintom from supernova, microwave background and 
galaxy  clustering}},
  Mod.\ Phys.\ Lett.\ A {\bf 20}, 2409 (2005),
  [\href{http://xxx.lanl.gov/abs/astro-ph/0411501}
{{\tt arXiv:astro-ph/0411501}}].
  
\bibitem{Xia:2006rr} 
  J.~Q.~Xia, G.~B.~Zhao, H.~Li, B.~Feng and X.~Zhang,
  {\it{Features in Dark Energy Equation of State and Modulations in the Hubble Diagram}},
  Phys.\ Rev.\ D {\bf 74}, 083521 (2006),
  [\href{http://xxx.lanl.gov/abs/astro-ph/0605366}
{{\tt arXiv:astro-ph/0605366}}].
  
 
\bibitem{Zhao:2006qg}
  G.~B.~Zhao, J.~Q.~Xia, H.~Li, C.~Tao, J.~M.~Virey, Z.~H.~Zhu and X.~Zhang,
  {\it{Probing for dynamics of dark energy and curvature of universe with latest 
cosmological 
observations}},
  Phys.\ Lett.\ B {\bf 648}, 8 (2007),
  [\href{http://xxx.lanl.gov/abs/astro-ph/0612728}
{{\tt arXiv:astro-ph/0612728}}].
    
 

\bibitem{Nojiri:2006ww} 
  S.~Nojiri and S.~D.~Odintsov,
   {\it{The Oscillating dark energy: Future singularity and coincidence problem}},
  Phys.\ Lett.\ B {\bf 637}, 139 (2006),
  [\href{http://xxx.lanl.gov/abs/hep-th/0603062}
{{\tt arXiv:hep-th/0603062}}].


\bibitem{Jain:2007fa} 
  D.~Jain, A.~Dev and J.~S.~Alcaniz,
    {\it{Cosmological bounds on oscillating dark energy models}},
  Phys.\ Lett.\ B {\bf 656}, 15 (2007),
  [\href{http://xxx.lanl.gov/abs/0709.4234}
{{\tt arXiv:0709.4234}}].
  
 
\bibitem{Lazkoz:2010gz} 
  R.~Lazkoz, V.~Salzano and I.~Sendra,
    {\it{Oscillations in the dark energy EoS: new MCMC lessons}},
  Phys.\ Lett.\ B {\bf 694}, 198 (2010),
  [\href{http://xxx.lanl.gov/abs/1003.6084}
{{\tt arXiv:1003.6084}}].
 
   

\bibitem{Pace:2011kb} 
  F.~Pace, C.~Fedeli, L.~Moscardini and M.~Bartelmann,
  {\it{Structure formation in cosmologies with oscillating dark energy}},
  Mon.\ Not.\ Roy.\ Astron.\ Soc.\  {\bf 422}, 1186 (2012),
  [\href{http://xxx.lanl.gov/abs/1111.1556}
{{\tt arXiv:1111.1556}}].
  
  
\bibitem{Zhang:2017idq} 
  Y.~Zhang, H.~Zhang, D.~Wang, Y.~Qi, Y.~Wang and G.~B.~Zhao,
  {\it{Probing dynamics of dark energy with latest observations}},
  Res.\ Astron.\ Astrophys.\  {\bf 17}, 050 (2017),
  [\href{http://xxx.lanl.gov/abs/1703.08293}
{{\tt arXiv:1703.08293}}].
  
  
\bibitem{Nesseris:2004wj} 
  S.~Nesseris and L.~Perivolaropoulos,
    {\it{A Comparison of cosmological models using recent supernova data}},
  Phys.\ Rev.\ D {\bf 70}, 043531 (2004),
  [\href{http://xxx.lanl.gov/abs/astro-ph/0401556}
{{\tt arXiv:astro-ph/0401556}}].
 
  
  
  
  
  \bibitem {Mukhanov}V. F. Mukhanov, H. A. Feldman and R. H. Brandenberger, {\it Theory 
of cosmological perturbations},
Phys. Rept. \textbf{215}, 203 (1992).



\bibitem {Ma:1995ey}C.~P.~Ma and E.~Bertschinger,
  {\it{Cosmological perturbation theory in the synchronous and conformal Newtonian 
gauges}},
Astrophys.\ J.\ \textbf{455}, 7 (1995),
[\href{http://xxx.lanl.gov/abs/astro-ph/9506072}
{{\tt arXiv:astro-ph/9506072}}].
 
 

\bibitem{Malik:2008im} 
  K.~A.~Malik and D.~Wands,
    {\it{Cosmological perturbations}},
  Phys.\ Rept.\  {\bf 475}, 1 (2009),
  [\href{http://xxx.lanl.gov/abs/0809.4944}
{{\tt arXiv:0809.4944}}].
 
  

\bibitem{Erickson:2001bq} 
  J.~K.~Erickson, R.~R.~Caldwell, P.~J.~Steinhardt, C.~Armendariz-Picon and 
V.~F.~Mukhanov,
    {\it{Measuring the speed of sound of quintessence}},
  Phys.\ Rev.\ Lett.\  {\bf 88}, 121301 (2002),
  [\href{http://xxx.lanl.gov/abs/astro-ph/0112438}
{{\tt arXiv:astro-ph/0112438}}].
  
    
  
  
  
    
\bibitem{Weller:2003hw} 
  J.~Weller and A.~M.~Lewis,
    {\it{Large scale cosmic microwave background anisotropies and dark energy}},
  Mon.\ Not.\ Roy.\ Astron.\ Soc.\  {\bf 346}, 987 (2003),
  [\href{http://xxx.lanl.gov/abs/astro-ph/0307104}
{{\tt arXiv:astro-ph/0307104}}].

 
 
  
 
\bibitem{Hannestad:2005ak} 
  S.~Hannestad,
    {\it{Constraints on the sound speed of dark energy}},
  Phys.\ Rev.\ D {\bf 71}, 103519 (2005),
  [\href{http://xxx.lanl.gov/abs/astro-ph/0504017}
{{\tt arXiv:astro-ph/0504017}}].

 
  
  
  


\bibitem{Betoule:2014frx} 
  M.~Betoule {\it et al.} [SDSS Collaboration],
    {\it{Improved cosmological constraints from a joint analysis of the SDSS-II and SNLS 
supernova  samples}},
  Astron.\ Astrophys.\  {\bf 568}, A22 (2014),
  [\href{http://xxx.lanl.gov/abs/1401.4064}
{{\tt arXiv:1401.4064}}].
  
  \bibitem{Riess:1998cb} 
  A.~G.~Riess {\it et al.} [Supernova Search Team],
  {\it ``Observational evidence from supernovae for an accelerating universe and a 
cosmological 
constant,''}
  Astron.\ J.\  {\bf 116}, 1009 (1998),
  [\href{http://xxx.lanl.gov/abs/astro-ph/9805201}
{{\tt arXiv:astro-ph/9805201}}].

  
  \bibitem{Eisenstein:2005su} 
  D.~J.~Eisenstein {\it et al.} [SDSS Collaboration],
  {\it ``Detection of the Baryon Acoustic Peak in the Large-Scale Correlation Function of 
SDSS 
Luminous Red Galaxies,''}
  Astrophys.\ J.\  {\bf 633}, 560 (2005),
  [\href{http://xxx.lanl.gov/abs/astro-ph/0501171}
{{\tt arXiv:astro-ph/0501171}}].
       
  

\bibitem{Beutler:2011hx} 
  F.~Beutler {\it et al.},
    {\it{The 6dF Galaxy Survey: Baryon Acoustic Oscillations and the Local Hubble 
Constant}},
  Mon.\ Not.\ Roy.\ Astron.\ Soc.\  {\bf 416}, 3017 (2011),
   [\href{http://xxx.lanl.gov/abs/1106.3366}
{{\tt arXiv:1106.3366}}].
 
  
  
\bibitem{Ross:2014qpa} 
  A.~J.~Ross, L.~Samushia, C.~Howlett, W.~J.~Percival, A.~Burden and M.~Manera,
    {\it{The clustering of the SDSS DR7 main Galaxy sample $-$ I. A 4 per cent distance 
measure at $z = 0.15$}},
  Mon.\ Not.\ Roy.\ Astron.\ Soc.\  {\bf 449}, no. 1, 835 (2015),
  [\href{http://xxx.lanl.gov/abs/1409.3242}
{{\tt arXiv:1409.3242}}].
 
  
\bibitem{Gil-Marin:2015nqa} 
  H.~Gil-Mar\'{i}n {\it et al.},
    {\it{The clustering of galaxies in the SDSS-III Baryon Oscillation Spectroscopic 
Survey: BAO  measurement from the LOS-dependent power spectrum of DR12 BOSS galaxies}},
  Mon.\ Not.\ Roy.\ Astron.\ Soc.\  {\bf 460}, no. 4, 4210 (2016),
  [\href{http://xxx.lanl.gov/abs/1509.06373}
{{\tt arXiv:1509.06373}}].
  
  




\bibitem{Adam:2015rua} 
  R.~Adam {\it et al.} [Planck Collaboration],
    {\it{Planck 2015 results. I. Overview of products and scientific results}},
  Astron.\ Astrophys.\  {\bf 594}, A1 (2016),
  [\href{http://xxx.lanl.gov/abs/1502.01582}
{{\tt arXiv:1502.01582}}].
  
   
\bibitem{Aghanim:2015xee} 
  N.~Aghanim {\it et al.} [Planck Collaboration],
    {\it{Planck 2015 results. XI. CMB power spectra, likelihoods, and robustness of 
parameters}},
  Astron.\ Astrophys.\  {\bf 594}, A11 (2016),
  [\href{http://xxx.lanl.gov/abs/1507.02704}
{{\tt arXiv:1507.02704}}].
   
 \bibitem{Gil-Marin:2016wya}
  H.~Gil-Mar\'{i}n {\it et al.},
 {\it{``The clustering of galaxies in the SDSS-III Baryon Oscillation Spectroscopic 
Survey: RSD 
measurement from the power spectrum and bispectrum of the DR12 BOSS galaxies''}},
  Mon.\ Not.\ Roy.\ Astron.\ Soc.\  {\bf 465}, no.2,  1757 (2017),
  [\href{http://xxx.lanl.gov/abs/1606.00439}
{{\tt arXiv:1606.00439}}].

\bibitem {Heymans:2013fya}
C.~Heymans \textit{et al.},
  {\it{CFHTLenS tomographic weak lensing cosmological parameter constraints: Mitigating 
the impact of intrinsic galaxy alignments}},
Mon.\ Not.\ Roy.\ Astron.\ Soc.\ \textbf{432}, 2433 (2013),
[\href{http://xxx.lanl.gov/abs/1303.1808}
{{\tt arXiv:1303.1808}}].
   


\bibitem {Asgari:2016xuw}
M.~Asgari, C.~Heymans, C.~Blake, J.~Harnois-Deraps,
P.~Schneider and L.~Van Waerbeke,
  {\it{Revisiting CFHTLenS cosmic shear: Optimal E/B mode decomposition using COSEBIs and 
 
compressed COSEBIs}},
Mon.\ Not.\ Roy.\ Astron.\ Soc.\ \textbf{464}, 1676 (2017),
[\href{http://xxx.lanl.gov/abs/1601.00115}
{{\tt arXiv:1601.00115}}].
 


\bibitem{Nunes:2016dlj} 
  R.~C.~Nunes, S.~Pan and E.~N.~Saridakis,
   {\it{New constraints on interacting dark energy from cosmic chronometers}},
Phys.\ Rev.\ D {\bf 94}, no. 2, 023508 (2016),
[\href{http://xxx.lanl.gov/abs/1605.01712}
{{\tt arXiv:1605.01712}}].
 
  
   
   
\bibitem{Anagnostopoulos:2017iao} 
  F.~K.~Anagnostopoulos and S.~Basilakos,
  {\it ``Constraining the dark energy models with $H(z)$ data: An approach independent of 
$H_0$,''}
  Phys.\ Rev.\ D {\bf 97}, no. 6, 063503 (2018),
  [\href{http://xxx.lanl.gov/abs/1709.02356}
{{\tt arXiv:1709.02356}}]. 
  
 
  
  
  
\bibitem {Moresco:2016mzx}
M.~Moresco \textit{et al.},
  {\it{A 6\% measurement of the Hubble parameter at $z\sim0.45$: direct evidence of the 
epoch of cosmic  re-acceleration}},
JCAP \textbf{1605}, 014 (2016),
[\href{http://xxx.lanl.gov/abs/1601.01701}
{{\tt arXiv:1601.01701}}]. 
   
     
  
 \bibitem{Lewis:2002ah} 
  A.~Lewis and S.~Bridle,
    {\it{Cosmological parameters from CMB and other data: A Monte Carlo approach}},
  Phys.\ Rev.\ D {\bf 66}, 103511 (2002).
  [\href{http://xxx.lanl.gov/abs/astro-ph/0205436}
{{\tt arXiv:astro-ph/0205436}}].


  \bibitem{Kass:1995loi}
  R.~E.~Kass and A.~E.~Raftery,
  {\it ``Bayes Factors,''}
  J.\ Am.\ Statist.\ Assoc.\  {\bf 90}, no.430,  773 (1995).

\bibitem{Heavens:2017afc} 
  A.~Heavens, Y.~Fantaye, A.~Mootoovaloo, H.~Eggers, Z.~Hosenie, S.~Kroon and 
E.~Sellentin,
  {\it ``Marginal Likelihoods from Monte Carlo Markov Chains,''},
  [\href{http://xxx.lanl.gov/abs/1704.03472}
{{\tt arXiv:1704.03472}}].
  
  
  \bibitem{Heavens:2017hkr} 
  A.~Heavens, Y.~Fantaye, E.~Sellentin, H.~Eggers, Z.~Hosenie, S.~Kroon and 
A.~Mootoovaloo,
  {\it ``No evidence for extensions to the standard cosmological model,''}
  Phys.\ Rev.\ Lett.\  {\bf 119}, no. 10, 101301 (2017),
  [\href{http://xxx.lanl.gov/abs/1704.03467}
{{\tt arXiv:1704.03467}}].
  
  
\end{thebibliography}
\end{document}